%% file: main.tex
\RequirePackage{fix-cm}
\documentclass[natbib]{svjour3}                     
\smartqed  
\usepackage[utf8]{inputenc}
\usepackage[pdftex]{graphicx}
\usepackage[english]{babel}
\usepackage{pgfplots}
\pgfplotsset{compat=1.12}
\usepackage{multirow}
\usepackage{multicol}
\usepackage{url}
\usepackage{amsmath}
\usepackage{listings}
\usepackage{natbib}
\input{files}
\begin{document}

\title{Evaluating the robustness of source code plagiarism detection tools to pervasive plagiarism-hiding modifications}
\titlerunning{Evaluating the robustness of source code plagiarism detection tools}

\author{Hayden Cheers \and Yuqing Lin \and Shamus P. Smith}

\institute{
Hayden Cheers \at 
The School of Electrical Engineering \& Computing,\\
The University of Newcastle, Callaghan, NSW, Australia\\
\email{hayden.cheers@newcastle.edu.au}
\and
Yuqing Lin \at
The School of Electrical Engineering \& Computing\\
The University of Newcastle, Callaghan, NSW, Australia\\
\email{yuqing.lin@newcastle.edu.au}
\and
Shamus P. Smith \at
The School of Electrical Engineering \& Computing\\
The University of Newcastle, Callaghan, NSW, Australia\\
\email{shamus.smith@newcastle.edu.au}
}


\date{Received: date / Accepted: date}

\maketitle

\begin{abstract}
Source code plagiarism is a common occurrence in undergraduate computer science education. In order to identify such cases, many source code plagiarism detection tools have been proposed. A source code plagiarism detection tool evaluates pairs of assignment submissions to detect indications of plagiarism. However, a plagiarising student will commonly apply plagiarism-hiding modifications to source code in an attempt to evade detection. Subsequently, prior work has implied that currently available source code plagiarism detection tools are not robust to the application of pervasive plagiarism-hiding modifications. In this article, 11 source code plagiarism detection tools are evaluated for robustness against plagiarism-hiding modifications. The tools are evaluated with data sets of simulated undergraduate plagiarism, constructed with source code modifications representative of undergraduate students. The results of the performed evaluations indicate that currently available source code plagiarism detection tools are not robust against modifications which apply fine-grained transformations to the source code structure. Of the evaluated tools, JPlag and Plaggie demonstrates the greatest robustness to different types of plagiarism-hiding modifications. However, the results also indicate that graph-based tools (specifically those that compare programs as program dependence graphs) show potentially greater robustness to pervasive plagiarism-hiding modifications.

\end{abstract}

\input{main_body}

\bibliography{references}

\end{document}

%% file: main_body.tex
\section{Introduction}

Plagiarism is a common occurrence in tertiary education \citep{joy1999, sheard2003, yeo2007, sraka2009, curtis2011, pierce2017}. In computing education, plagiarism is often encountered as source code plagiarism. This is where one student has appropriated the source code of another student (either as files or fragments), and submitted it as their own work \citep{parker1989, joy1999, cosma2008, sraka2009}. In order to identify cases of source code plagiarism, many automated tools and techniques have been proposed in the form of Source Code Plagiarism Detection Tools (SCPDTs) \citep{joy1999, martins2014, novak2019}. A SCPDT evaluates assignment submissions for similarity in order to identify suspiciously similar assignment pairs. A high similarity implies plagiarism has occurred, mid-range similarity can imply students have collaborated on an assignment, while low similarity does not raise suspicion of plagiarism.

In order to hide plagiarism, a plagiarising student may apply source code modifications to reduce the similarity of the plagiarised work to its original. Such modifications serve to differentiate the source code such that it evades detection by a human reviewer or automated detection tool. In this work, source code modifications used to hide plagiarism are collectively referred to as \textit{plagiarism-hiding modifications}. Plagiarism-hiding modifications are commonly applied by either \textit{transforming} the structure and appearance of the source code (e.g. by renaming identifiers, shuffling statements, or modifying comments); or by \textit{injecting} spurious fragments of source code to give the appearance of distinct implementations \citep{faidhi1987, whale1990, prechelt2002, freire2007, granzer2013, novak2019}. 

In order to accurately detect plagiarism, a SCPDT must be robust against plagiarism-hiding modifications. \textit{Robustness} is considered to be the ability of a SCPDT to withstand plagiarism-hiding modifications without decrease in the measurement of similarity. The robustness of SCPDTs can be compared relatively by the impact plagiarism-hiding modifications have upon the measurement of similarity between a plagiarised work and its original. A SCPDT with greater robustness to plagiarism-hiding modifications will evaluate a lesser decrease in similarity as a result of the modifications; while a SCPDT that is vulnerable to plagiarism-hiding modifications will evaluate a greater decrease in similarity. 

Robustness and accuracy are related but distinct qualities of a SCPDT. A SCPDT is robust when it can accommodate for plagiarism-hiding modifications. This will result in a SCPDT reporting a high similarity between a plagiarised assignment with applied source code modifications and its source. Similarly, a SCPDT is accurate when it measures a high level of similarity between a plagiarised assignment and its source to imply plagiarism is present; while also measuring a low similarity between unrelated works, implying that plagiarism is not present. 

Prior work has indicated that currently available SCPDTs are not robust to certain plagiarism-hiding modifications, especially when pervasively applied \citep{schulze2013, cheers2020}. \textit{Pervasive plagiarism-hiding modifications} occur when plagiarism-hiding modifications are applied throughout the body of plagiarised source code. This can result in many fine-grained modifications to the structure of the source code, overall resulting in a large decrease in measured similarity by a SCPDT. When pervasive plagiarism-hiding modifications are applied, it has been indicated that the similarity of plagiarised submission pairs can drop into a range that does not warrant suspicion of plagiarism \citep{cheers2020}.

The research presented here evaluates the robustness of 11 SCPDTs against pervasively-applied plagiarism-hiding modifications. This is to identify vulnerabilities of available SCPDTs, as well as to identify future potential directions of work for the development of SCPDTs. Robustness is evaluated upon data sets of simulated plagiarism representative of undergraduate students. Cases of simulated plagiarism are generated with different selections of source code modifications, with increasing pervasiveness of modification (i.e. being increasingly modified). This allows for the evaluation of SCPDT robustness against diverse selections of source code modifications, along a sliding scale of pervasiveness of modification. 

The remainder of this work is structured as followed. Section 2 presents background on existing SCPDTs and plagiarism-hiding modifications. Section 3 presents the design of software tools used in this evaluation to accommodate for known deficiencies in the evaluation of SCPDTs. Section 4 presents the experimental design and describes the purpose of the evaluation performed in this work. Sections 5 and 6 describe the setup of each evaluation and their performed experiments. Section 7 discusses the results of the evaluations and reflects upon the robustness of the evaluated SCPDTs. Section 8 discusses threats and vulnerabilities from the results identified in this work. Section 9 identifies related works. And finally, Section 10 concludes this research, and identifies future directions of work.

\subsection{Research Questions \& Contributions}

This work is guided by three research questions:

\textbf{RQ1:} \textit{What are the impacts of source code transformations on SCPDTs?} Source code transformations typically apply cosmetic and structural changes to the source code. However, the impact of specific source code transformations upon the measurement of similarity is unclear. Certain SCPDTs and techniques may be robust to specific transformations, but significantly impacted upon others. Hence, how are the SCPDTs affected by specific source code transformations?

\textbf{RQ2:} \textit{What is the impact of source code injection on SCPDTs?} Source code injection adds new fragments of source code to a program. This effectively changes the `size' of the source code by introducing new elements for a SCPDT to compare. Injecting source code will undoubtedly lower the similarity scores of the compared submissions. However, it is unclear if this is by a meaningful amount such that it can hide indications of plagiarism. Furthermore, it is unknown if certain SCPDTs and techniques are more robust to the addition of source code fragments than others. Hence, how are SCPDTs affected by source code injection? 

\textbf{RQ3:} \textit{What SCPDT is most robust to plagiarism-hiding modifications?} When a program is modified with plagiarism-hiding source code modifications, it will contain many changes to the structure of the source code. Hence, what SCPDT shows the least impact upon the evaluation of similarity in the presence of pervasive plagiarism-hiding modifications?

In the exploration of these research questions, four contributions are made:
\begin{itemize}
    \item A comprehensive evaluation of existing SCPDTs for robustness to plagiarism-hiding modifications.
    \item The identification of 5 specific source code transformations that have a large impact on the evaluation of source code similarity with existing SCPDTs.
    \item The implementation of 6 simple SCPDTs used in this evaluation, 2 that are modelled after unavailable SCPDTs.
    \item A toolset for the generation and evaluation of simulated undergraduate source code plagiarism data sets for similarity.
\end{itemize}

\section{Background}

This section will firstly, provide background on existing SCPDTs, and secondly, identify commonly encountered examples of plagiarism-hiding source code modifications.

\subsection{Source Code Plagiarism Detection Tools}

Many SCPDTs have been proposed to identify indications of plagiarism in pairs of undergraduate assignment submissions. This is typically thorough the evaluation of source code similarity. Approaches to Source Code Plagiarism Detection (SCPD) can be broadly categorised by the aspects of source code compared to evaluate similarity:
\begin{itemize}
    \item Metric-based,
    \item Text-based,
    \item Token-based,
    \item Tree-base, 
    \item Graph-based, or
    \item Behavioural.
\end{itemize}

Metric-based approaches count attributes of source code elements to identify similar documents. Metrics may include (but are not limited to) the number of operands, operators, declared variables or literals in code. Metric-based approaches have been shown to be less effective in comparison to structural approaches \citep{whale1990, whale1990a, kapser2003}, and as such are mostly considered historic. \cite{faidhi1987} identified suspected plagiarism by counting 24 distinct code metrics. \cite{ottenstein1976} compared source code with Halstead complexity measures \citep{halstead1977} to identify programs with similar attribute counts. More recently, \cite{shan2014} applied attribute counting and the chi-squared test method to evaluate program similarity.

Text-based approaches analyse textual character strings extracted from source code documents. Such approaches are typically applied to identify documents with short edit distances or those with large quantities of overlapping sub-strings. Identifying such cases imply the documents have a high degree of similarity, providing an indication of plagiarism. For example, two documents with a relatively short edit distance imply they have the same origin, but have been modified. Likewise, documents with many overlapping sub-strings share content in common. Sim-Gitchell \citep{gitchell1999, gitchell1999a} applies string alignment to measure similarity (similar in concept to overlapping sub-strings). Sherlock-Sydney \citep{pikeunknown} compares text files through the extraction of digital signatures. Such signatures are simply hashed word sequences extracted from the source documents. These signatures are then compared for similarity. \cite{rani2018} proposed an extended Levenshtein String Edit Distance for measuring similarity. 

Token-based approaches represent a source code document as a stream of tokens. A token is a lexically significant term in a programming language. Tokens may include (but are not limited to) identifiers, keywords, grammatical delimiters, and literal values. Streams of tokens can be compared for similarity using techniques similar to text-based approaches. It is also common to see approaches utilise a technique referred to as token tiling, where tokens strings from one program are placed over another to identify the coverage of token sequences. Two notable token-based plagiarism detection tools include MOSS and JPlag. MOSS \citep{schleimer2003} implements a winnowing algorithm to find overlapping sub-strings of token hashes. JPlag \citep{prechelt2002} implements a greedy token tiling algorithm that covers one source document with token sub-strings from another document. Sim-Grune \citep{grune1989} is another similar token-based tool that identifies similarity though common token sub-strings. This allows for the identification of similar segments of code that differ in terms of layout, comments, identifiers and literal values. More recently, \cite{anzai2019} proposed an extended edit distance for the calculation of program similarity that takes into consideration some of the common modifications applied by plagiarisers (e.g. changing order of blocks and statements).

Tree-based approaches use a language-specific parser to construct Abstract Syntax Trees (AST). An AST represents the syntax of source code in a hierarchical manner showing the grammatical structure of the source code. The AST is constructed with a stream of lexical tokens extracted from a source code document. The structure of this tree can is compared for similarity. For example, by finding isomorphic or similar sub-trees. \cite{li2010} identified plagiarism by comparing AST structures. \cite{zhao2015} identified plagiarism by comparing the hash values of AST nodes. Tree-based approaches are known to suffer from a high computational complexity due to the complexity of tree comparison algorithms \citep{baxter1998}.

Graph-based approaches developed to represent the semantics of source code. This is commonly through the use of Program Dependence Graphs (PDG) \citep{ferrante1987}. Like tree-based approaches, graph-based approaches also suffer from high computational complexity in analysing similarity \citep{baxter1998}. \cite{liu2006}, and \cite{chen2010} implement such methods by evaluating the similarity of PDGs. These methods are claimed to be immune to plagiarism-hiding modifications such as statement reordering, and mapping statements to semantic equivalents. Alternatively, \cite{chae2013} use an API-labelled control flow graph to identify similar sequences of API calls. 

In recent years there have been a few notable approaches that perform dynamic or symbolic analysis on programs to identify similarity. These approaches consider the execution behaviour of a program in attempts to be robust to obfuscations. JIVE \citep{anjali2015} analyses the call tree of a program to find similarity in method call sequences. The authors claim that the call tree is robust to obfuscations such as renaming and statement reordering. LoPD \citep{zhang2014} \& Cop \citep{luo2017} attempt to identify the similarity of programs based on their implemented program logics to determine if they are semantically equivalent. VaPD \citep{jhi2011} identifies program similarity through runtime execution analysis by identifying identical values stored in memory during execution. The foundation of VaPD is that from observation, certain runtime values of a program cannot be changed through semantics-preserving obfuscations.

There also exist hybrid tools that implement one or more of these approaches. Sherlock-Warwick \citep{joy1999} implements both text and token-based similarity measurements, combined with normalisation of the source code to reduce any variation. Furthermore, there are also tools that implement other novel techniques to identify similarity. \cite{chen2004} evaluate program similarity through approximations of Kolmogorov Complexity \citep{kolmogorov1998}. \cite{cosma2012} apply latent semantic analysis to match source code documents with similar terms with PlaGate. While \cite{karnalim2016} identified plagiarism though the analysis of Java Bytecode sequences.

\subsection{Plagiarism-Hiding Modifications}\label{s:modifications}

Plagiarism-hiding modifications differentiate a plagiarised program from its original in an attempt to hide the committed plagiarism. This work explores two types of plagiarism-hiding modifications: 
\begin{itemize}
    \item Source code transformation
    \item Source code injection
\end{itemize}

Source code transformation occurs when the original source code is modified to appear different. Such transformations are typically cosmetic or structural in nature, and have no impact on the operation of the program. Behaviourally the plagiarised work will remain the same, however, the transformed program will appear different to a human reviewer. In general, a source code transformation will not introduce new statements into the source code, however it may split existing statements where appropriate. Overall, the original source code is present, however it may take a structurally or cosmetically distinct form. Examples of source code transformations used to hide plagiarism include: modifying comments, reordering statements or members, replacing control structures with equivalents, mapping expressions to semantic equivalents, or renaming identifiers \citep{joy1999, jones2001, cosma2008, allyson2019}. 

Source code injection refers to the addition of new or unrelated source code to a program. In minor cases, this source code can be non-functional `junk' \citep{jhi2011}, and only serves to make a plagiarised program appear different. For example, injected source code can consist of simple `print' statements, or unused variable declarations \cite{joy1999}. However, in more advanced cases the injected source code itself may be plagiarised. For example, a plagiariser may appropriate whole source files, classes, methods, or continuous blocks of code, and integrate them into their own work. This constitutes `partial plagiarism', where only fragments of a plagiarisers work is inappropriately sourced.


A plagiarising student may apply plagiarism-hiding modifications with different intensity. This is subject to the skill of the plagiariser, and subsequently the effort they apply to evade detection. In basic cases of plagiarism, an undergraduate plagiariser will have little understanding of a program or fragment of source code that they have appropriated \citep{joy1999}. Hence, applied plagiarism-hiding modifications can be expected to be simple. For example, applying minor cosmetic changes such as reformatting code or modifying comments. However, there are potential cases where a more advanced plagiariser will begin to modify the structure of the source code throughout the entire program. This can be by reordering declarations and statements in the source code, or potentially creating new classes and methods. Furthermore, plagiarism can also be committed by more advanced students with greater programming skills \citep{cheers2020}. For example, consider a time-poor student who is proficient at programming. They may apply many in-depth modifications to the source code, potentially pervasively modifying it such that the plagiarised code no longer bares resemblance to the original. In this case, the plagiariser has effectively paraphrased (rewritten) an other's work to evade detection.

Other works term plagiarism-hiding modifications as source code obfuscation (for example, \cite{jhi2011,zhang2014,luo2017,ko2017}). Plagiarism-hiding modifications are a form of source code obfuscation. However, obfuscation is typically applied to reduce the comprehension or understandability of source code. Plagiarism-hiding modifications are not applied to reduce the comprehension of source code. The modified source code will still be understandable by a reviewer, but is expected to be superficially distinct compared to the original. Hence, plagiarism-hiding modifications can be seen as `lesser' source code obfuscations. In this work, the term plagiarism-hiding modification is used to refer to source code obfuscations that are representative of those applied by undergraduate plagiarisers to attempt evading detection. 

\section{Evaluation Tooling}

From a review of existing evaluations of SCPDTs \citep{novak2019}, two common deficiencies can be identified:
\begin{enumerate}
    \item Tool availability - many evaluated SCPDTs are not made available for reuse.
    \item Data set availability - evaluations do not use reproducible or comprehensive data sets.
\end{enumerate}

The first deficiency of tool availability of is a significant problem in the evaluation of SCPDTs. Many proposed SCPDTs are simply not made available by their authors for reuse after initial publication \citep{novak2019, cheers2020}; and without access to the proposed SCPDTs, it is difficult to determine if newer approaches are suitable for use in the detection of plagiarism. For use in this evaluation, only 6 SCPDTs were identified to be available for reuse:
\begin{itemize}
    \item MOSS \citep{schleimer2003}
    \item JPlag \citep{prechelt2002}
    \item Plaggie \citep{ahtiainen2006}
    \item Sim-Grune \citep{grune1989}
    \item Sherlock-Warwick \citep{joy1999}
    \item Sherlock-Sydney \citep{pikeunknown}
\end{itemize}

From a recent study of SCPDTs, \cite{novak2019} confirmed that these 6 SCPDTs are commonly used in comparative evaluation of SCPDTs. Hence, it can be assumed that these 6 tools have potential to be used in the detection of plagiarism at academic institutions (that do not otherwise have their own internal tools, or use commercial alternatives). However, these tools only represent two methods of measuring source code similarity. MOSS, JPlag, Plaggie and Sim-Grune utilise token-based similarity (or variants thereof). Sherlock-Sydney utilises text-based similarity. While Sherlock-Warwick implements both text and token-based similarity. No SCPDTs that implement metric-based, tree-based, graph-based, or behavioural methods of SCPD are known to be available for reuse. 

The second deficiency of data set availability directly impacts upon the reproducibility and reliability of SCPDT evaluations. In the evaluation of SCPDTs, the utilised evaluation data sets are commonly either not provided, or not comprehensive enough for adequate evaluation. For example, \cite{cheers2020} identified that from a total of 17 SCPDT evaluations:
\begin{itemize}
    \item 3 provided the data set that it was evaluated on \citep{wise1996, chae2013, zhao2015}; although these data sets were not student assignments, or of significant size.
    \item 4 described freely available base programs that were manually/automatically plagiarised for evaluation \citep{liu2006, jhi2011, zhang2014, luo2017}; but the plagiarised variants were subsequently not provided.
    \item 10 did not provide the utilised data set (typically as they contain student assignment submissions) \citep{pikeunknown, grune1989, prechelt2002, schleimer2003, chen2004, jadalla2008, kustanto2009, cosma2012, anjali2015, allyson2019}.
\end{itemize}

It is difficult to evaluate a SCPDT without a data set representative of undergraduate assignment submissions. The most appropriate data sets are collections undergraduate assignment submissions. This is as they represent the intended data to be evaluated by a SCPDT. However, as such data sets contain the works of students, they typically cannot be shared. This is due to privacy concerns (in sharing data sets with known cases of plagiarism) and issues with the ownership of the assignments. For example, at the author's institution, students retain ownership of their assignments. Hence, they cannot be freely shared. 

It must also be considered the quality of any real data set of undergraduate assignment submission for use in bench marking SCPDTs. In any real data set, there is no guarantee that plagiarism exists (at least in a form that is readily detectable for ground-truth comparisons); let alone plagiarism with a diverse range of plagiarism-hiding modifications, that is sourced from students with diverse skill sets. Hence, evaluations of SCPDTs cannot be guaranteed to evaluate a SCPDT against a wide range of plagiarism-hiding transformations, or even identify pervasively transformed cases of plagiarism.

This section will present software tooling used in this evaluation designed to overcome and address these deficiencies. Section \ref{s:naivetools} presents the design of 6 \textit{naive SCPDTs}. This is to bring greater depth into the evaluation of SCPDTs by introducing SCPDTs that implement otherwise unavailable similarity measurement techniques. Section \ref{s:simplag} presents the design of a data set generation tool, \textit{SimPlag}, that can be used for a reproducible method of source code plagiarism detection data set generation. These tools are subsequently combined in Section \ref{s:eval_pipeline} as an automated SCPDT evaluation pipeline, \textit{PrEP}, that is used to facilitate the evaluations performed in this work. 

\subsection{Naive Source Code Plagiarism Detection Tools}\label{s:naivetools}

In order to address the issue of tool availability, the design of six \textit{naive} SCPDTs is presented. These tools are referred to as \textit{naive} as they are simple implementations of source code similarity measurement techniques applied to SCPD. The naive tools do not implement efficiency optimisations, or optimisations to gain greater accuracy or robustness when analysing programs. Of the 6 naive tools, 4 implement similarity measurement techniques of existing SCPDTs (as string and token-based tools), while 2 implement similarity measurement techniques of otherwise unavailable SCPDTs (as tree and graph-based tools). In combination the naive SCPDTs will be used to measure a baseline robustness of the implemented techniques for comparative purposes; and to add greater depth to the performed evaluations (as the number of currently available SCPDTs is limited). However, these tools do not solve the problem of tool availability, nor can their performance be considered to represent that of similar techniques. The naive tools are simply used as an indication of how robust and effective these techniques may be at measuring similarity in the presence of plagiarism-hiding modifications. 

\begin{figure}[htbp]
    \includegraphics[width=\linewidth]{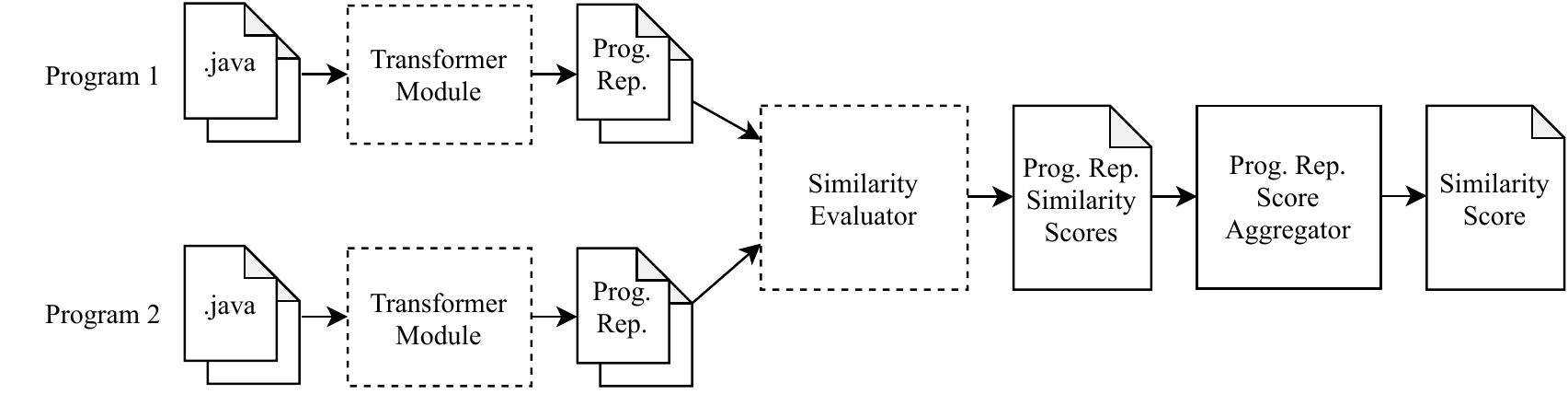}
    \caption{Common naive SCPDT pipeline. Implementations of the transformer and similarity evaluator modules are dependent on specific source code similarity measurement techniques.}
    \label{fig:naivetooldesign}
\end{figure}

The implementation of all naive tools share a common abstract pipeline, presented in Fig. \ref{fig:naivetooldesign}. The tools firstly accept as input a set of assignment submissions as .java source files. Secondly, the source files in each assignment submission are transformed into a specific program representation using a \textit{Transformer Module}. Thirdly, the similarity of all source file pairs between the two submissions are measured by the similarity of the derived program representations using a \textit{Similarity Evaluator}. Finally, the similarity score of the two assignment submissions is then calculated by aggregating the file-wise similarity scores as the average of the best-mapped file-wise similarity scores between the programs, shown in eqn. \ref{e:avgsim}. 

\begin{equation}\label{e:avgsim}
Sim(A, B) = \frac{\begin{aligned}
      \sum_{n=1}^{|A|} max(\{FSim(a_n, b): b \in B\}) + \\ \sum_{m=1}^{|B|} max(\{FSim(b_m, a): a \in A\})
      \end{aligned}}%
  {|A| + |B|}
\end{equation}\\
Where:
\begin{itemize}
    \item[] $A$, $B$ are assignment submissions as .java source files
    \item[] $a_n$, $b_m$ are .java source files $a \in A$, $b \in B$
    \item[] $|A|$, $|B|$ is the number of source files in A, B
    \item[] $FSim(a, b)$ is the file-wise similarity of file pairs $a$, $b$
    \item[] $max(X)$ is the maximum value in set X
\end{itemize}

\subsubsection{Utilised Program Representations}

\input{Fig2}

The naive tools are constructed with four common program representations: 
\begin{itemize}
    \item Text (string)
    \item Token (string)
    \item Abstract Syntax Tree (AST)
    \item Program Dependence Graph (PDG)
\end{itemize}

The text representation interprets a source file as a character string. This is derived simply by reading each source file line-by-line and appending it to a string in memory. The token representation interprets a source file as a string of lexical tokens. Each token represents a lexically-important term in a source file. Tokens are extracted with a tokeniser provided as part of the JavaParser framework\footnote{https://github.com/javaparser/javaparser, last accessed May 1 2020.}. The tree representation interprets a source file as an AST. Each AST represents the source file as the syntactic structure of the Java programming language. ASTs are constructed using the JavaParser framework. The graph representation interprets a program as a set of PDGs \citep{ferrante1987}. A PDG is constructed for each method declared in the source code by analysing the relations between statements. Statements and data are represented by nodes in the PDG. Control edges are created between nodes to indicate dependencies on the execution of statements. Data edges are placed between statements and data nodes to identify common referenced values. These four representations are exemplified in Fig. \ref{fig:representation_example}.

\subsubsection{Similarity Measurement Techniques}

The program representations derived from each file are compared using a \textit{Similarity Evaluator}. Two basic type of similarity evaluation modules are implemented:
\begin{itemize}
    \item Edit distance
    \item Greedy string tiling
\end{itemize}

All edit distance algorithms are conceptually the same irrespective of representation they operate upon. They simply evaluate the number of edits required to turn one program representation into another. For text and tokens, the Apache Commons Text\footnote{http://commons.apache.org/proper/commons-text, last accessed May 1 2020.} implementation of Levenshtein string edit distance is applied. Levenshtein string edit distance evaluates edit distance as the number of character/token deletion, insertion or substitution to turn one sequence into another. Tree edit distance is evaluated with the Java library APTED\footnote{https://github.com/DatabaseGroup/apted, last accessed May 1 2020.}. APTED implements a robust and memory efficient tree edit distance algorithm \citep{pawlik2015, pawlik2016}, that is suitable for comparing many large ASTs. Graph edit distance is evaluated using a recursive greedy edit distance algorithm to efficiently approximate the number of required edits to transform one PDG into another. With all four representations, the edit-distance-derived similarity is calculated with eqn. \ref{e:edsim}.

\begin{equation}\label{e:edsim}
    FSim_{ed}(a, b, d) = 1 - \frac{d}{max(|a|, |b|)}
\end{equation}\\
Where:
\begin{itemize}
    \item[] $a$, $b$ are derived file representations
    \item[] $d$ is the edit distance between $a$ and $b$
    \item[] $|a|$, $|b|$ is the size of each file representation
    \item[] $max(|a|, |b|)$ is the maximum size of program representations $a$ and $b$ 
\end{itemize}

Greedy string tiling attempts to tile subsets of String A over String B to identify the total coverage of A over B. A naive implementation of greedy tiling is approximated by identifying all non-overlapping sub-strings between two documents of length greater than $n$ to measure approximate program coverage. The similarity of two strings (representing source files) with the approximated greedy string tiling is calculated with eqn. \ref{e:gstsim}

\begin{equation}\label{e:gstsim}
    FSim_{gst}(a,b,c) = 2 \times \frac{c}{|a| + |b|}
\end{equation}\\
Where:
\begin{itemize}
    \item[] $a$, $b$ are derived string-based file representations
    \item[] $c$ is number of covered string elements between $a$ and $b$
    \item[] $|a|$, $|b|$ is the size of each file representation
\end{itemize}

\subsubsection{Naive Tool Combinations}

\input{Tab1}

A total of six naive SCPDTs were composed from the listed program representations and similarity evaluators, listed in Table \ref{tab:naivesimilaritytools}. In the performed evaluations, the naive tools will be used as a baseline for robustness for their respective similar approaches without optimisation, in comparison to the compared academic SCPDTs. The implementations of the naive tools are MIT licensed, and can be found at \url{https://github.com/hjc851/NaiveSCPDTools}.

\subsection{Data Set Generation}\label{s:simplag}

The most appropriate method of rectifying the issue of data set availability would be to release a ground truth data set of undergraduate assignment submissions for reuse. This would require identifying and labeling suspicious assignment pairs, and subsequently identifying how they are modified to enable correlation of scores. Such a data set would be similar to the code clone detection bench marking data set BigCloneBench \citep{svajlenko2015}, that is a curated ground-truth data set of code clones. However, a curated data set of undergraduate assignment submissions for plagiarism detection has many associated issues, notably regarding ownership and quality of the data. 

It is common that a student will retain ownership of their assignment submissions. Subsequently, it would border on theft to use these data sets and share them amongst peers for use in evaluating SCPDTs. There are also legal obligations guaranteeing confidentiality in student cases of plagiarism (applied to the author's institution). Therefore, any data set would need to be expertly de-identified with no indication of the student, or even the source of the plagiarised works. It is also ethically ambiguous if student works are used for the evaluation of SCPDTs for research purposes. Furthermore, there are issues of data set quality in real data sets sourced from undergraduate assignment submission. There are no guarantees that in any arbitrary set of undergraduate assignment submission that there exist real cases of plagiarism, nor that they contain diverse examples of plagiarism-hiding modifications.

In the absence of a ground truth data set for SCPD, the generation of test data is an alternative option. This has been utilised in recent works for the evaluation of code similarity tools \citep{svajlenko2013, ko2017, ragkhitwetsagul2018}. The basic idea of these tools is to apply source code modifications to a base program, and use it in the generation of test programs. This idea can be reapplied for the generation of source code plagiarism detection data sets. However, the generated test data must be representative of undergraduate source code plagiarism. Prior works have identified commonly applied modifications used to hide source code plagiarism (for example, \cite{faidhi1987, joy1999, jones2001, mozgovoy2006, freire2007, allyson2019}). Many of these modifications have in common that that they are applied to change the structure and appearance of the source code, while retaining the original semantics and behaviour. Such modifications can be automatically applied, and integrated into a tool that affords the generation of simulated plagiarism representative undergraduate plagiarisers. 

\subsubsection{SimPlag}

\begin{figure}[htbp]
    \includegraphics[width=\linewidth]{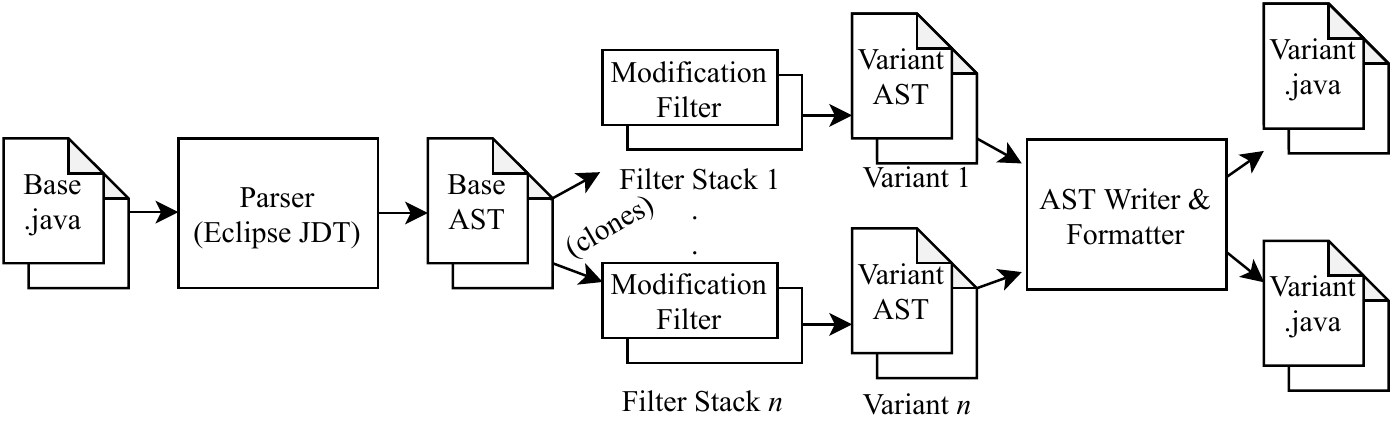}
    \caption{The SimPlag simulated plagiarism generator pipeline.}
    \label{fig:simplagpipeline}
\end{figure}

An automated tool implementing a source code modification framework is developed to generate simulated cases of source code plagiarism representative of undergraduate programmers. This tool is titled \textit{SimPlag}: Simple Plagiarism Generator, and is available at \url{https://github.com/hjc851/SimPlag}. SimPlag does not solve the issue of data set availability, however, it aids in providing a reproducible method of generating test data for comparing SCPDTs. Hence, this tool represents a step towards a solution to this issue.

Fig. \ref{fig:simplagpipeline} presents the architecture of SimPlag. SimPlag is implemented with a pipe-and-filter design pattern to afford a simple but extensible implementation. SimPlag can be scripted to automatically produce simulated plagiarised test programs, that are generated with a configurable selection of source code modifications, that can be applied with a weighted random chance to affect how pervasively modified the simulated plagiarised programs are. The operation of SimPlag is a three step process:
\begin{enumerate}
    \item Pre-processing
    \item Modification
    \item Saving
\end{enumerate}

Firstly, SimPlag accepts as input a single assignment submission. This submission is referred to as a \textit{`base'} program, that is used for the generation of multiple simulated plagiarised \textit{`variant'} programs (referred to as \textit{variants} hereafter). SimPlag will parse the source files of the base program into ASTs using the eclipse Java Development Tools\footnote{https://www.eclipse.org/jdt/, last accessed June 30 2020.}. The AST of the base program is subsequently cloned once per configured number of variants to be produced, resulting in a set of variant ASTs in preparation for the application of source code modifications. Secondly, SimPlag will modify each clone of the base program to simulate plagiarism. The tool will apply a stack of user-configurable \textit{Modification Filters} to the variant ASTs. Each modification filter applies an individual source code modification to the AST. The location and count of all applied source code modifications are recorded for analytical purposes. Finally, the variant ASTs are saved, and written to disk as simulated plagiarised variants of the original base program. As part of this process, a standardised formatting is applied to each program. This is afforded with the Google Java code formatting tool\footnote{https://github.com/google/google-java-format, last accessed June 30 2020}. By formatting the source code with this tool, it is enforced that all produced variant programs are parsable, and can be subsequently analysed by all known available SCPDTs. Furthermore, the source code will appear to be neatly formatted in an effort to make the code appear semi-realistic.

\subsubsection{Source Code Modifications}

A generated data set for source code plagiarism detection needs to simulate as closely as possible a natural data set in terms of the types of modification applied, and the presentation of the program in order to hide the plagiarism. It is logical to conclude that applied plagiarism-hiding modifications must not be so advanced that they are impossible for a novice programmer to apply, but not so simple that the programs remain effectively unchanged. From reviewed literature, two distinct types of source code modifications were identified: source code transformations, and source code injection. Many works reference examples of source code transformations as being largely cosmetic and structural \citep{faidhi1987, joy1999, jones2001}. While examples of injected fragments of source code are typically small, self-contained, and non-functional; but also may result in code being mixed with self-written and plagiarised code \citep{freire2007, granzer2013}. This will be used as a guideline for the implementation of distinct source code transformation and source code injection modification filters.

\paragraph{Transformation Filters.}

\input{Tab2}

SimPlag implements a total of 14 source code transformations, listed in table \ref{tab:simplagmutationoperators}. Each source code transformation is implemented as an individual `transformation modification filter' (referred to as a transformation filter). The source code transformations are applied using a simple tree-walk operation to the filter's nodes of interest (i.e. a node that it can be applied at). However, the transformations are not applied globally to each AST, but applied to randomly selected nodes of interest, determined by a configurable \textit{transformation chance} parameter. The transformation chance indicates how likely each transformation filter will be applied to a node of interest. For example, if a transformation filter is applied with a 10\% transformation chance, there is an approximate 10\% chance the transformation filter will be applied at any node of interest in the AST. Hence, this parameter is used as a configurable method of changing how pervasively each source code transformation is applied. This value is also used to indirectly represent the `effort' a plagiariser may apply to modify a program and hide their plagiarism. Higher modification chances indicate the plagiariser has taken more time and effort to hide their plagiarism, while lower modification changes imply lesser time and effort spent.

SimPlag only implements source code transformations that are characteristic of being applied by undergraduate programmers, and have been listed in prior works \citep{jones2001, mozgovoy2006, allyson2019, freire2007, granzer2013, joy1999, faidhi1987, karnalim2016}. SimPlag does not implement transformations that are considered too difficult or un-representative of a novice programmer with little program understanding to apply. As such, the majority of implemented transformations are simple `swap' or `map' type operations, as well as adding or removing non-functional code to existing code. There are undoubtedly countless many source code transformations that could be implemented by SimPlag. For example, implementing swapping \textit{switch} statements to \textit{if} statements or in-lining method calls. However, the 14 implemented transformations were selected as they are simple to implement, and hence they are considered to be characteristic of novice programmers. Furthermore, many of the implemented transformations are provided as source code refactoring operations by Integrated Development Environments (IDE) such as Eclipse\footnote{ https://www.eclipse.org, last accessed January 15 2021.} or IntelliJ IDEA\footnote{ https://www.jetbrains.com/idea/, last accessed January 15 2021.}. As such, it is conceivable a plagiariser with basic programming skills could use an IDE to facilitate the application of such source code transformations to hide their plagiarism, with minimal effort on their behalf. 

\paragraph{Injection Filters.}

SimPlag implements injection modification filters (referred to as injection filters) to simulate cases where a student has copied code from a peer (e.g. through collaboration), or taken source code verbatim from a website. The tool supports the injection of source code fragments into a program within 4 fragment scopes:
\begin{itemize}
    \item Whole file
    \item Whole class
    \item Whole method
    \item Whole statement
\end{itemize}

The different scopes of source code injection represent different severity of verbatim source code copying. The severity is considered in terms of how hard the plagiarised code is to detect, and for the plagiariser to integrate it into their own work. In its simplest form, a plagiariser may copy an entire source file and submit it in their own work. This is the easiest form of plagiarism to detect, while also requiring the least effort from the plagiariser. This is opposed to plagiarising individual or sequential statements of code. When statements are plagiarised and placed within the plagiariser's own source code, this becomes a much more difficult task to detect, but also much more difficult to integrate into their own work.

Each injected fragment type of source code is implemented as its own distinct injection filter. Like the transformation filters, injection filters have a configurable \textit{injection chance} parameter. This parameter determines how likely each filter is to inject fragments of source code into the variant AST. However, injection modification filters also require a secondary configuration value, limiting the number of times it may inject code. This is largely as a quality concern to stop large quantities of source code fragments being injected into a comparatively smaller program. A limit is defined for all 4 types of injected source code fragments, limiting at most: $f$ files can be injected into any variant, $c$ classes can be injected into any file, $m$ methods injected into any class, and $s$ statements injected into any method. The statement injection filter has a variable upper limit on the number of statements that can be injected, prohibiting it from doubling the size of any existing method. 

All injection filters share access to a user configurable `seed pool' of source code files. The injection filters will use this pool for injecting source code into the variant ASTs. On startup, each injection filter indexes the seed pool in terms of its own required seeded fragments (i.e. the file injection filter will index files from the seed pool, etc) to allow for the later selection of fragments to be injected.

The operation of the injection filters varies slightly to transformation filters. Class, method and statement injection filters operate in a similar manner to transformation filters. They are applied with a simple tree-walk, with the potential to inject a source code fragment at a node of interest. The class injection filter will inject a type declaration into any AST compilation unit (that represents an individual source file); the method injection filter will inject a method declaration into any AST class declaration; and the statement injection filter will inject a randomly selected source code statement into an AST method declaration. At each node of interest, a random boolean value will be rolled (weighted by the filter's injection chance). If this boolean is true and the filter has not exceeded its pre-configured injection limit, a fragment of source code will be randomly selected from the seed pool, and subsequently injected into the AST. However, the file injection filter has a considerably different operation. The file injection filter does not modify the variant AST directly, but injects new ASTs into the variant. The file injection filter will roll a random boolean if it should be applied, and subsequently if true, inject up to $f$ randomly selected files (as ASTs) into the variant.

\subsubsection{Authenticity \& Correctness of Generated Variants}

The source code modifications implemented by SimPlag seek to generate data sets of simulated plagiarism that are characteristic of three scenarios:
\begin{enumerate}
    \item A student has mis-appropriated another's source code in whole, and disguised plagiarism with source code transformations.
    \item A student has mis-appropriated fragments of another's source code and injected them into their own work.
    \item Combinations of the above where fragments are injected into their own work, and subsequently transformed.
\end{enumerate}

The first scenario is accommodated for using the source code transformation capabilities of SimPlag. The second scenario is accommodated for with the source code injection capabilities of SimPlag. While the third scenario is accommodated for using both the source code transformation and injection capabilities of SimPlag. It is argued that by only implementing source code modifications that are known to be representative of undergraduate plagiarisers \citep{faidhi1987, joy1999, mozgovoy2006}, the generated test data is semi-authentic and can be used to represent similar cases of undergraduate source code plagiarism. However, the generated simulated plagiarised variant programs are synthetic test data. While the implemented modifications applied are representative of undergraduate plagiarisers (as referenced from literature), and the base programs are intended to be sourced from real undergraduate assignments, the generated variants are not real cases of plagiarism; and hence, they may be readily identifiable as synthetic to a human reviewer. This is first and foremost the biggest limitation of SimPlag, and its use in the performed evaluations. Hence, SimPlag can only be used for evaluating SCPDTs against the implemented plagiarism-hiding modifications (the purpose of this work), and not for their use in the detection of \textit{real} cases of plagiarism. However, this limitation is balanced by the ability to generate large quantities of simulated plagiarised test data with diverse plagiarism-hiding modifications, as opposed to the collection of real cases of plagiarism that may or may not contain substantial plagiarism-hiding modifications. 

There are also important limitations to the functional correctness of the generated test data by SimPlag. All test data produced by SimPlag is guaranteed to be parsable. Being parsable means the source code can be represented as an abstract syntax tree, and therefore the source code is grammatically correct. However, SimPlag does not enforce that the generated simulated plagiarism is semantically or functionally correct. Developing a tool that can guarantee the functional correctness of code is difficult and requires in-depth analysis of the source code and the impact of modifications. As a result of this, the simulated plagiarised programs produced by SimPlag may not compile, or have strange behaviour at runtime. This limitation is largely caused by the implementation of 2 source code transformations: tRI, and tRS; as well as the source code injection modifications. 

The implementations of tRI and tRS are not guaranteed to invalidate the variants. tRI will globally change user-defined identifiers in the program (limited to class, field, method, parameter and local variable names), and avoids renaming type names and variables declared in system libraries. However, full semantic analysis is not performed, and as such, there is the potential for errors to occur. tRS does not analyse the dependencies between statements when reordering as it is implemented as simple shuffle of statements. This was an intentional design decision as analysing statement dependencies would severely limit the number of statements that could be shuffled, and hence, impact on the number of times tRS could be applied. Hence, to simulate more invasive shuffling of statements, dependency analysis was omitted. Furthermore, having invalid simulated plagiarised programs can add to the realism of the applied plagiarism-hiding modifications. For example, consider by a novice programmer who applies the modifications without the skills to validate the correctness of the program, and consequentially invalidates the program. Such cases still need to be detected by SCPDTs.

The injection filters are likely to invalidate the correctness of the variants to varying extents. The file and class injection modifications should in theory have no or minimal effect on the validity on the generated variants, assuming there are no naming conflicts caused by the injected files or class fragments. However, the method and statement injection filters are expected to potentially invalidate the correctness of the variants, subject to the quality of the seed data. Both the method and statement injection filters do not validate that the injected fragments have external dependencies (i.e. on the declaring class for methods, or within the declaring method for statements). These filters simply inject fragments of source code into the variants, and hence, will invalidate the variants if the seed data has external dependencies. This is again an intentional design decision. If all fragments of source code were required to be self-contained, it would risk impacting upon the quality and complexity of injected source code fragments. This is being able to inject complex source code fragments (that the plagiariser themself may not understand), as opposed to simple one-line fragments of source code with no real meaning.

However, the limits on the functional correctness of the simulated plagiarised programs are not expected to have a profound impact upon the evaluated SCPDTs. Of the 6 known available SCPDTs, none require the source code to parsable, let alone compiled and/or executed. Hence, it will not affect the operation of the SCPDTs. Similarly, for the 6 naive SCPDTs, only Tree ED and Graph ED require the test programs to be parsable. However, Graph ED in particular is expected to be impacted upon by modifications that affect the semantics of the source code as Graph ED measures the semantic similarity of the source code (as the relations between terms) and not the structure of the source code. As tRI, tRS, and the method and statement injection modifications have the potential to invalidate the functional correctness of the program and by extension change the semantic relations, it is expected that these modifications in particular will have a significant impact upon Graph ED's ability to evaluate similarity. However, these limitations are mitigated by the need for confidence in SCPDTs in that they can detect indications of plagiarism, even if the plagiarised work is not functionally correct. 

\subsection{Evaluation Pipeline}\label{s:eval_pipeline}

To facilitate the evaluations of SCPDT robustness performed in this work, a reusable automated evaluation pipeline was developed. This pipeline is titled \textbf{PrEP:} the \textbf{Pr}ogram \textbf{E}valuation \textbf{P}ipeline. PrEP was developed to automate the batch evaluation of SCPDTs. The framework is implemented in Kotlin\footnote{https://kotlinlang.org, last accessed May 1 2020.} and runs on the Java Virtual Machine. It utilises multi-processing to enable the structured evaluation of tools in a manner that is easy to monitor and fault tolerant, but scalable based on computing resources. It exposes SCPDTs through Java bindings, enabling both Java and non-Java SCPDTs to be integrated into a single pipeline. An important feature of this framework is the optional seeding of simulated plagiarised submissions through the integration of SimPlag. This integrates both test data generation and evaluation into a single pipeline approach. The implementation of PrEP is available at \url{https://github.com/hjc851/SCPDT-PrEP}.

\begin{figure}[htbp]
    \includegraphics[scale=0.8]{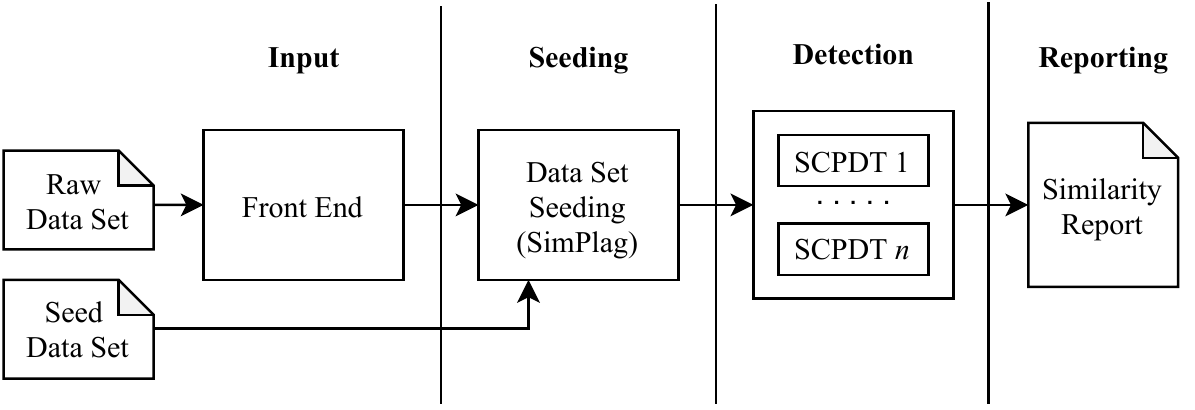}
    \caption{The PrEP tool evaluation pipeline framework.}
    \label{fig:prepframework}
\end{figure}

The PrEP framework is structured as a modular pipeline to enable future extension. Fig. \ref{fig:prepframework} provides an overview of this pipeline. It is divided into four phases:
\begin{enumerate}
    \item Input
    \item Seeding (Optional)
    \item Detection
    \item Reporting
\end{enumerate}

\textit{Input} is the provisioning of test data sets, tool configuration (for both SimPlag and the utilised SCPDTs) and optionally seed data sets to the pipeline. \textit{Seeding} is the addition of simulated plagiarised submissions into the test data set. The simulated plagiarism is optionally sourced from a seed data set. \textit{Detection} is the process of invoking the integrated SCPDTs to evaluate the submissions for similarity. These are invoked on batch for the entire data set, evaluating both submission-wise and file-wise similarities. \textit{Reporting} aggregates the scores from each tool for further comparison and evaluation. In addition to this, a list of simulated plagiarised submissions and their applied source code modifications is generated to enable the accurate analysis of results when used to evaluate SCPDTs.

\subsubsection{Data Set Seeding}

By integrating data set seeding into the framework, an evaluation data set can be generated automatically as part of the evaluation process. SimPlag is used to generate multiple variants of a single assignment submission that can enable benchmarking against any selection of supported plagiarism-hiding modifications. Furthermore, the framework is able to offer ground-truth evaluations. By generating test data, it is already known what submissions are representative of plagiarism, and how they are modified. This is something that cannot be achieved with standard academic data sets without manual review. As a result of this, the framework aids in the accurate evaluation of tools, and the ability to correlate the impact of each modification on the evaluation of similarity.

\subsubsection{Plagiarism Detection Tools}

A total of 11 SCPDTs are embedded in PrEP. Each tool is exposed through a Java binding (i.e. as a Java class), enabling for procedural invocation of the tool from Java code. The binding acts as an adapter, transforming the raw console output of a tool into structured object values. The embedded tools are:
\begin{itemize}
    \item JPlag
    \item Plaggie
    \item Sim-Grune
    \item Sherlock-Sydney
    \item Sherlock-Warwick
    \item The six naive tools (Table \ref{tab:naivesimilaritytools})
\end{itemize}

All tools are exposed through three modes of comparison: file-wise, submission-wise, and batch. File-wise comparison lists the pairwise similarity scores of all files in a pair of submissions. Submission-wise comparison provides the similarity score of two individual assignment submissions. Batch comparison lists the pairwise similarity scores of a set of submissions. Not all tools support all three modes of comparison natively. For example, Sim-Grune and Sherlock-Sydney only support file-wise comparison. The tool bindings accommodate for this where required by aggregating file-wise scores into submission-wise scores (as per eqn. \ref{e:avgsim}). 

While MOSS is known to be an available SCPDT, PrEP omits embedding MOSS into the pipeline. In pilot experiments, it was found that MOSS is unreliable in use for very large data sets. MOSS would often hang for long periods of time while processing the data sets (with each base program and variants submitted as one request), and often fail to provide a response. It could only be used reliably on a small number of comparisons. This is consistent with the observations of MOSS in prior work by \cite{cheers2020}. Hence, for reliability MOSS is omitted.

\subsubsection{Operation of Academic SCPDTs}

JPlag \citep{prechelt2002} operates by applying a token tiling algorithm to cover one source code file with tokens extracted from another. If two source files have a large degree of coverage, they can be considered similar and hence a candidate for plagiarism. First, source code files are converted into a stream of tokens. JPlag uses its own set of tokens which abstract standard language tokens to avoid matching the same token with different meanings. Second, extracted tokens are compared between files to determine similarity by the Running-Karp-Rabin Greedy-String-Tiling algorithm where tokens from one file are covered over another within a tolerance of mis-match. Program similarity is evaluated as the percentage of tokens from one program which can be tiled over another program.

Plaggie \citep{ahtiainen2006} is a tool that is claimed to operate similarly to JPlag. However, it is an entirely local application, compared to JPlag which was originally provided as a web service. No known publication describes the operation of Plaggie; however from examining it's implementation, it operates upon tokenised representations of the source code evaluating similarity by token tiling. Hence, it can be assumed it has similar performance to JPlag.

Sim-Grune \citep{grune1989} analyses programs for structural similarity through the use of string alignment. For two programs, Sim will first parse the source code creating a parse tree. The tool will then represent the parse trees as strings and align them by inserting spaces to obtain a maximal common sub sequence of their contained tokens. The similarity of programs is then evaluated as the quantity of matches.

Sherlock-Warwick \citep{joy1999} implements both text and tokenised comparison methods. In the tool, a pair of programs are compared for similarity 5 times: in their original form, with whitespace removed, with comments removed, with whitespace and comments removed, and as a tokenised file. In all cases, the comparisons measure similarity through the identification of \textit{runs}. A \textit{run} is a sequence of lines common to two files which may be interrupted by anomalies (e.g. extra lines).

Sherlock-Sydney \citep{pikeunknown} analyses programs for lexical similarity. Digital signatures of source code are generated by hashing string token sequences (not lexical tokens) extracted from a text files. The digital signatures are then compared, with the similarity of files being evaluated as the number of digital signatures in common. 

\section{Evaluation Design}

Two distinct evaluations are performed in order to address the research questions guiding this work. Firstly, the available SCPDTs are evaluated for robustness against source code transformations in Section \ref{s:ev1}. Secondly, the available SCPDTs are evaluated for robustness against source code injection in Section \ref{s:ev2}. The two types of plagiarism-hiding modifications are evaluated separately as they have different impacts on the source code. For example, transformations do not necessarily add code, but modifies existing code; while injection adds new code not present in the original program. 

Both evaluations are broken down into distinct experimental cases, allowing for the comparison of the SCPDT robustness against different selections of source code modifications. Different intensities of modification are used to evaluate the SCPDTs against progressively more pervasively modified cases of simulated plagiarism. The intensity of applied plagiarism-hiding modifications is expressed through the transformation chance and injection chance parameters exposed by SimPlag. Test data generated with higher transformation/injection chances are expected to have more modifications applied to the source code itself (i.e. more nodes of interest transformed, and larger quantities of source code injected), and hence such variants are considered to be more pervasively modified. This will allow for the evaluation of the SCPDTs on a sliding scale of pervasiveness of modification.

This section will provide an overview of the common design and setup of the performed evaluations. Firstly, the scope of the evaluations is defined. Secondly, the measures used to compare the robustness of the evaluated SCPDTs is presented. Thirdly, the data sets used in the evaluation, along with the utilised test data generation process is described. Fourthly, the configuration of the utilised SCPDTs is defined and justified.

\subsection{Scope of Evaluations}

The purpose of the performed evaluations are to compare the robustness of SCPDTs to plagiarism-hiding modifications. The evaluations are explicitly not designed to evaluate the accuracy of the compared SCPDTs in detecting instances of plagiarism. An evaluation of accuracy typically compares tools by the number of suspected cases of plagiarism correctly identified. However, a limitation of SCPDTs is that they do not specifically detect plagiarism. Instead, they detect indications of plagiarism, with the identification of plagiarism subject to human review \citep{joy1999, cosma2012}. Hence, as the purpose of this work is to evaluate robustness (as a function of evaluated similarity) and not correct detections of simulated plagiarised works, no conclusion can be made from the results of the experiments in regards to the accuracy of the evaluated SCPDTs. Evaluating the accuracy of SCPDTs is subject to future work.

The performed evaluations do not compare tools from similar domains such as code clone detection (CCD). CCD and SCPD have much in common. For example, they both utilise similar techniques in evaluating source code similarity \citep{roy2007,ragkhitwetsagul2018}, and use similar taxonomies of source code transformations (see the 6-level taxonomy of \cite{faidhi1987} and the common 3-type code clone taxonomy \citep{bellon2007}). As a result of this, it is common to see SCPDTs used in CCD evaluations (e.g. the evaluations performed by \cite{burd2002,schulze2013,ragkhitwetsagul2018}). However, in this article, evaluating Code Clone Detection Tools (CCDT) is considered out of scope. This is fundamentally due to the performed experiments being designed for the evaluation of SCPDTs and not CCDTs.

The purpose of a CCDT is to detect similar fragments of source code. A CCDT is typically evaluated in its ability to detect smaller fragments of source code injected into another larger body of source code (e.g. as evaluated by \cite{bellon2007}). The accuracy of the tool can then be measured in terms of how many injected fragments of source code are correctly identified. If a CCDT was to be evaluated for robustness to modification, it is logical that source code modifications could be applied to the injected code fragments themself. However, the performed experiments are effectively the opposite of this. Whole programs are cloned, and have plagiarism-hiding modifications applied. There is no concept of a traditional code clone to be detected. Instead, these experiments are interested in the evaluation of overall program similarity, and the effect source code modifications have upon it. Hence, it would not be fair to evaluate CCDTs as there are no code clones to detect in these experiments.

In addition to this, there are no directly comparable metrics reported between a SCPDT and CCDT for use in these experiments. In general, a CCDT does not report the similarity of two programs, but the quantity of code clones in common between them. In order to effectively compare CCDTs as SCPDTs, a bridging mechanism would have to be used to convert CCDT results into similarity scores. A simple method for this conversion would be to identify the coverage of identified code clones over the size of the programs (e.g. as used by \cite{ragkhitwetsagul2018}). However, this introduces a new dependent variable in the evaluations: how to calculate the size of a program? Identifying the optimal method for calculating program size is itself a significant undertaking. To focus the performed evaluations, it is considered outside of the scope of this work, and hence, only SCPDTs will be evaluated. It is not to say CCDTs cannot or should not be evaluated in the detection of plagiarism. As future work it would be interesting to compare SCPDTs and CCDTs in the presence of plagiarism-hiding modifications using a modified experimental method that is fair to both tools. However, such an experiment would most likely be focused on the measurement of tool accuracy in the presence of plagiarism-hiding modifications. This is again, not a focus here.

\subsection{Evaluation Data Sets}\label{s:ed-ds}

\input{Tab3}

Both evaluations utilise a large data set of undergraduate assignment submissions as \textit{base} programs, and use these submissions to generate an even larger pool of plagiarised variants as test data. The base evaluation data set is comprised of 6 sets of assignment submissions. These data sets are referred to as assignment sets 1 through 6 (AS1 to AS6). Each assignment set contains undergraduate assignment submissions of varying size and complexity, representing a total of 3 years of undergraduate study, from students with varying skill levels, and assignments implemented with varying technical complexity. Table \ref{table:assignmentstats} presents the average size and metrics of each assignment set, and the whole data set in total. For the purpose of expressing the size of each individual submission, the logical lines of code (LLOC) are used. This formulated from the distinct non-block statement count in each program.

Each individual experiment generates a distinct test data set of simulated plagiarised variant programs with SimPlag. This allows for the generation of a large number of variants, thereby allowing for the identification of the impacts of source code modifications on average to a large sample of data. Each experiment's test data set is generated using a distinct selection of plagiarism-hiding modifications, inline with the goal of the experiment. 6 distinct test data sets are generated in total. Evaluation 1 (section \ref{s:ev1}) applies combinations of source code transformations across 3 experiments, as well as one extended case. Evaluation 2 (section \ref{s:ev2}) applies combinations of source code injection operations across 2 experiments. In each experiment, 5 variants of each base program from the source data set are created, repeated using 6 incremental transformation chance and injection change probabilities: 10\%, 20\%, 40\%, 60\%, 80\%, and 100\%. The lower probabilities will cause SimPlag to apply fewer source code modifications when generating test data, while the higher probabilities will cause more modifications to be applied. Using this test data generation method, between 30 (5 variants at 6 chances of modification) to 420 (5 variants at 6 chances of modification for each of the 14 transformations) variants of each base program are created in each experiment. 

\subsection{Robustness Metrics}\label{s:ed-met}

Robustness is considered to be the ability of a SCPDT to withstand source code modification without decrease in the measurement of similarity. In order to compare the robustness of the evaluated SCPDTs, two comparison metrics are used. 

Firstly, the robustness of the SCPDTs will be compared using a quantitative metric, measuring the average similarity of all variants compared to their respective base programs. The quantitative metric is used to demonstrate the impact of individual source code modifications upon the generated test data sets. It is calculated with eqn. \ref{e:quant}. 

\begin{equation}\label{e:quant}
AvgSim(S) = \frac{\sum_{i=1}^{|S|} S_i}{|S|}
\end{equation}\\
Where:
\begin{itemize}
    \item[] $S$ is the set of similarity scores between each variant and it's base program for a single SCPDT
    \item[] $|S|$ is the number of similarity scores in $S$
\end{itemize}

Using this quantitative metric, SCPDTs that measure a higher average similarity will be considered to be more robust to applied plagiarism-hiding modifications, while SCPDTs that measure a lower average similarity will be considered to be less robust to applied plagiarism-hiding modifications. This value will always be bound between 0 and 100, assuming all scores in $S$ are also in this range. Hence, it will show on average the similarity decrease of the SCPDT as a result of applied plagiarism-hiding modifications.

Secondly, a comparative robustness metric is used to compare the robustness of each SCPDT when evaluating similarity on each test data set. This will compare the SCPDTs by the ratio of applied source code modifications to the total decrease in similarity occurred as a result of the plagiarism-hiding modifications. As the evaluations apply modifications in two forms (transformations and injection), two variations of the comparative metric is introduced to compare SCPDT robustness:
\begin{itemize}
    \item \textit{Robustness to Code Transformation} (RCT), and
    \item \textit{Robustness to Code Injection} (RCI).
\end{itemize}

RCT measures the ratio of source code transformations applied to a program compared to the decrease in measured similarity. This is expressed through eqn. \ref{e:rct} as the number of transformations required to decrease the evaluated similarity by 1\%. 

\begin{equation}\label{e:rct}
    RCT(B,V,n) = \frac{n}{100 - Sim(B,V)}
\end{equation}\\
Where:
\begin{itemize}
    \item[] $V$ is a variant of base program $B$ 
    \item[] $n$ is the greater than zero number of times source code transformations are applied to $B$, transforming it into $V$
    \item[] $Sim(B,V)$ is the similarity of $B$ and $V$ evaluated with a SCPDT
\end{itemize}

RCI measures the ratio of inserted lines of code (LOC) compared to the decrease in measured similarity. This is expressed through eqn. \ref{e:rci} as the number of lines injected in order to decrease the evaluated similarity of a program by 1\%. 

\begin{equation}\label{e:rci}
    RCI(B,V,l) = \frac{l}{100 - Sim(B,V)}
\end{equation}\\
Where:
\begin{itemize}
    \item[] $V$ is a variant of base program $B$ 
    \item[] $l$ is the greater than zero LOC count injected into $B$, transforming it into $V$
    \item[] $Sim(B,V)$ is the similarity of $B$ and $V$ evaluated with a SCPDT 
\end{itemize}

The comparative metrics are strictly for comparing the robustness of SCPDTs on the same data set generated with the same method of applying plagiarism-hiding modifications. Both metrics are similar, however are required as the two types of modification have different impacts upon a body of source code (i.e. transformation changes existing source code, injection adds new fragments of source code). Hence, this requires expressing the impacts of the modifications in terms of the aspects of source code modified. A higher RCT/RCI value will imply a SCPDT is more robust than a SCPDT with a lower RCT/RCI value, and vice-versa. This is reflected by more modifications needing to be applied in order to reduce the evaluated similarity by 1\%. While a low RCT or RCI reflects that few modifications are required to reduce the evaluation similarity by 1\%. However, they are not normalised measures, and cannot be used as a universal determination of robustness when comparing SCPDTs between different data sets. Comparing the RCT and RCI scores for SCPDTs on different data sets is not meaningful. Furthermore, these metrics only account for measuring robustness where there is at least one source code modification applied, that results in a decrease in similarity. Similarly, the two robustness equations should not be considered to imply that the number of modifications needed to reduce the similarity by 1\% is a constant. It is understandable that when the similarity between the two programs is low, then the number of modifications needed to reduce the similarity by 1\% is higher than the case then the two programs share greater commonality. The measured RCT/RCI is simply a sample of the robustness of a SCPDT with a specific selection of modifications applied. The performed experiments will evaluate the similarity of programs that are generated from the same base. In this case they share great commonality, thus, roughly speaking, the score provides a lower bound on the number of modifications needed to reduce similarity by 1\%. 

\subsection{Utilised SCPDTs \& Configurations}

The performed evaluations compare the 11 SCPDTs integrated in PrEP. These consist of the 5 \textit{academic} SCPDTs: JPlag, Plaggie, Sim, Sherlock-Warwick and Sherlock-Sydney; as well as the 6 naive SCPDTs: String Tile, String ED, Token Tile, Token ED, Tree ED and Graph ED. In the performed experiments, all SCPDTs are executed with their default configuration parameters. While prior works have indicated that code similarity tools in general can gain greater performance by the selection of optimal configuration parameters \citep{ragkhitwetsagul2018, ahadi2019}; it is argued that this is not representative of a real-world use of SCPDTs. An academic using a SCPDT will not know in advance the optimal configuration values for any given tool when assessing any arbitrary data set. Furthermore, time will not typically permit an academic to evaluate a data set for plagiarism using multiple tool configurations to find the best possible result, as again, the best possible result is not known in advance. Hence, this work assumes that the original authors of the evaluated SCPDTs have selected default configuration parameters that produce on average acceptable results. 

The naive String Tile and Token Tile tools both require specifying minimum match lengths. String Tile utilises $n = 20$ as this is the approximate average length of expressions in the base data set (being approximately 16, with a buffer of 2 characters each side to avoid false positives). This will cause the tool to match sub-strings of at least this length. Token Tile utilises $n = 12$. This is justified as prior works have suggested values similar to this for token-based tools (e.g. \cite{ragkhitwetsagul2018, ahadi2019}), and that JPlag utilises this same default value \citep{prechelt2002}. The remaining naive SCPDTs (all being edit distance based) do not support configuration, as all edits have a hard-coded weight of 1.



\section{Evaluation 1: Source Code Transformation}\label{s:ev1}

This evaluation will compare the robustness of the SCPDTs to source code transformations. Due to the methods of measuring similarity, some tools are by design more robust to certain transformations. For example, token-based tools ignore commenting and general formatting, hence they are not impacted by such transformations. However, there are numerous source code transformations that can be applied to hide plagiarism, each of which may have a substantial impact on the measurement of similarity.

Three experiments are performed to compare the robustness of the selected tools to specific selections of source code transformations. Firstly, the tools are evaluated using data sets created with each transformation applied in isolation. Secondly, the tools are evaluated using random selections of transformations, over multiple iterations. Thirdly, they are evaluated by applying all transformations in unison. Ideally, each individual selection of transformations would be evaluated in isolation. However, with 14 transformations this leads to 16,383 unique selections, each requiring evaluation separately on each program in the data set. This is computationally infeasible for this evaluation, as it would require generating 282,606,750 variants (from the 575 base programs, create 5 variants at each of the 6 transformation chances for each selection of transformations), requiring comparison with each of the 11 SCPDTs. As such, random selections of transformations are used to gain coverage in the evaluated selections of transformations.

The three experiments are designed to progressively test robustness to source code transformation by combining different types of transformations. The first experiment accommodates for the simplest case with minimal transformations, restricted to only one type of source code transformation applied to each variant. The second experiment simulates more realistic cases of plagiarism, where a plagiariser transforms a program with various types of transformations. Thirdly, all transformations are applied to simulate an extreme case of plagiarism with all supported source code transformations. These three experiments provide coverage to identify what transformations are most effective at reducing the similarity of each tool. Doing so will allow for identifying what transformations each tool is most robust against, and subsequently what transformations they are most vulnerable to; and hence explore RQ1.

The 14 source code transformations are always applied in the same order as listed in Table \ref{tab:simplagmutationoperators}. This decision was made to eliminate the complexity introduced through the number of permutations of transformations if the ordering of application was configurable. This is deemed acceptable as there are no direct interactions between the transformations, except in two cases:
\begin{enumerate}
    \item Add comment, Remove comment, \& Modify comment
    \item Assign default value to variable, \& Split variable declaration and assignment
\end{enumerate}

These interactions will have the greatest impact at the 100\% transformation chance. The first case will cause all comments to be removed. This in theory will only affect the string-based SCPDTs, and in a worst case increase the average similarity scores reported by such tools, as a point of variation is removed from the generated variants. In the second case, it will result in all variable declarations having no assigned default value, with all variables being the target of a assignment expression after declaration. However, it is not expected that these interactions will have a profound impact on results even in the most extreme case.

\subsection{Individual Transformations}\label{s:ev1a}

The purpose of this experiment is to identify the impact of each transformation in isolation on the evaluation of source code similarity. In order to do so, variants of each base program are generated using a single transformation with the 6 transformation chance probabilities. This will allow for the decrease in similarity caused by each transformation to be traced as the probability the transformation is applied at any node of interest is also increased. The generated test data set for this experiment contains 241,500 variant programs. This is broken down into 5 variants for each base program, with each of the 14 transformations applied in isolation, generated with the 6 transformation chances. 

\input{Fig5_ev1a_heatmap}

Fig. \ref{fig:eb2aheatmaps} presents the average similarity of the variants created with each individual transformation at the 6 transformation chances. Six heat maps are presented, one for each transformation chance. Darker colours indicate a higher average level of similarity, while lighter colours indicate lower evaluated average similarity. From this figure, it is clear that the string-based tools (Sherlock-Sydney, String Tile, String ED) are not robust to the application of any source code transformations. This is indicated by the consistent light-coloured rows for these tools. Sherlock-Sydney demonstrates the largest vulnerability to all transformations across all transformation chances. Furthermore, the String Tile, and String ED tools show similarly consistent deceases in similarity. This implies the string-based tools are all impacted upon by any applied transformation. Sherlock-Warwick also suffers a noticeable decrease in similarity across all transformations as it does rely upon string-based metrics to evaluate similarity. However, it's drop in similarity is not as prevalent as the string-only tools, presumably as it integrates token-based similarity measurement.

The results for the non-string-based tools provide insight into the vulnerabilities of token, tree and graph-based techniques. Starting at the 60\% transformation chance and continuing until 100\%, a trend of common vulnerabilities are demonstrated by the lighter columns in the heat maps. While the lighter columns are not consistent amongst the non-string-based tools (implying certain tools show greater robustness to certain transformations), a common trend can be seen by the analysis of the scores with the greatest decrease in similarity.

\input{Tab4_ev1a_trn_ranking}

Table \ref{tab:ev2atrnrankings} presents the rankings of each transformation by greatest impact upon each non-string-based SCPDT's similarity scores at the 100\% transformation chance. The horizontal bar (i.e. `|') delimits transformations that incur a decrease in similarity (on left) from those that incur a negligible decrease in similarity. A common trend of the scores on the right of this delimiter is that they fall within 96\% $\pm$ 1\% similarity. All scores to the left fall below this point, and in some cases, this is by a large margin. Initially it would be expected that transformations with no impact would evaluate a 100\% similarity score, however due to the operation of each tool and potential noise in the data sets, transformations with no impact typically have an average similarity of approximately 96\% over the generated test data set.

There is no consistent ordering of transformations that the tools are vulnerable to (listed left of the bar delimiter). Hence, it can be implied that each tool is more robust against certain transformations. However, for the token and tree-based tools, the 5 most common transformations ranked first are:
\begin{itemize}
    \item tRS (Reorder statements (within methods))
    \item tRM (Reorder class member declarations)
    \item tSO (Swap expression operands)
    \item tFW (Swap \textit{for} statement to \textit{while} statement)
    \item tSD (Split variable declaration and initial assignment)
\end{itemize}

This decrease for the 5 transformations for the token and tree-based tools, can be attributed to fine-grained re-orderings of token sequences. The operation of the token-based tools varies in terms of how they match token sequences. However, in general, to match token sequences, a minimum number of tokens must be matched. If a source file becomes segmented with mis-matched token sequences smaller than this minimum number, it can stop these tools from accurately identifying common token sequences. For example, the Token Tile tool specifies a minimum match length of 12 tokens. For any modifications less than 12 tokens apart, the tool will no longer be able to match the affected sub-sequence between these modifications. Hence, the small re-orderings of tokens applied in these 5 transformations can have a cumulatively large impact on token-based tools. 

The Tree ED tool sees a similar decrease in similarity to these 5 transformations, that is explained by a similar cause to the token-based tools vulnerability. The fine-grained reordering of token sequences is analogous to fine-grained restructuring of an AST. When making large changes to the structure of the tree, the edit distance of the tree also becomes larger, resulting in a lower evaluated similarity. However, the tree edit distance algorithm is comparatively more proficient at handling the lesser more fine-grained statement reordering with tRS; while it is also more vulnerable to the large changes introduced with tRM when comparing against the token-based tools. This greater vulnerability is affected by the naive implementation of the approach. While in an optimal implementation of an AST edit distance tool this may not be an issue; the utilised implementation utilises a Greedy edit distance algorithm. Hence it is subject to errors in the calculation of the optimal edit distance and there are corner cases when a disproportional decrease in similarity can be experienced.
 
The Graph ED tool does not suffer from a large decrease in similarity from all 5 transformations. This is as it focuses on the semantics of the source code, and not the structure. However, it does suffer from a large decrease in similarity when evaluated against tRI (rename identifiers), tFW and tSD; and to a lesser degree tRS. In the case of tRI, this is an unexpected result. A PDG-based tool should not see a large decrease from identifier renaming, as it does not change the semantics of the program. Likewise, tFW and tSD should not impact upon similarity as these are semantics-preserving operations. Upon investigation, this decrease in similarity is caused by the implemented edit distance algorithm. Being a Greedy implementation, it does suffer from a decrease in accuracy. Furthermore, it compares graph nodes by the type of statement it contains. This will therefore see a reduction in similarity caused by modifications that transform statements. In the case of tRS, the implementation of this transformation can modify the semantics of the source code unintentionally. tRS is implemented as a literal shuffling of the statements in code. Hence, it can change the control dependencies in the constructed PDG. Due to these factors, stemming from its naive implementation, the Graph ED tool does suffer from a decrease in similarity.

Other notable modifications that are ranked to the left of the bar (or on the border) include tEA (Expand compound assignment) and tEU (Expand unary expression). However, they do not consistently appear to the left of the bar delimiter for all tools. Hence, they are omitted from this list to focus on the transformations with the greatest average impact. However, it should be noted that these transformations apply similar fine-grained reordering of the source code token sequences. Hence, they should be considered to also be a potential vulnerability of the token and tree-based tools as they share similar characteristics. 

In all cases, this is a clear result. The most effective method of impacting upon a tool is to change the representation of the program in which the tool measures similarity upon. However, it does emphasise that these tools are profoundly vulnerable to such simple transformations. The five identified transformations are not technically complex to implement - and in many cases can be automated by source code editors. Furthermore, when applying all of these transformations in unison it is conceivable that there would be a profound impact upon the token and tree-based tools; and potentially even the graph-based tool.

Overall, the results of this experiment are positive. Firstly, it has reinforced that string-based tools are not robust to any source code transformations. Secondly, it has reinforced that token and tree-based tools are more robust to certain source code transformations, except for those which apply fine-grained modifications to the structure of the source code. This is the most significant impact, as it shows that while token-based tools are more robust to string-based tools, they can still be fooled with simple re-orderings of source code. While the impact of the transformations varies across tools; not being robust to fine-grained reordering of source code is a significant deficiency as this is a common method students use to hide plagiarism. Finally, this experiment has shown that the graph-based tool is more robust to certain source code transformations. However, as discussed, the implementation is not optimal and suffers from a decrease in similarity. This is a major deficiency of the approach; however, this can be attributed to a naive implementation of a PDG-based tool, and as such these results cannot be claimed to be representative of all other PDG-based approaches. 

\subsubsection{Applying Five Transformations}\label{s:ev1ax}

A large impact on the evaluated similarity was found when applying the five individual transformations: tRS, tRM, tSO, tFW, tSD. This subsection extends the isolated transformation experiment to apply these 5 specific transformations in unison to a generated test data set. This is to compare the robustness of the tools to the application of these five transformations, that each apply fine-grained sub-sequence reordering. A second test data set is generated containing 17,250 variants, broken down into 5 variants for each base program (with the 5 transformations applied), generated with each of the 6 transformation chances.

\input{Fig6_ev1ax_avgsimilarity}

Fig. \ref{fig:ev2axavgsimilarity} presents the average similarity of the generated variants as the chance of transformation increases. This figure reinforces the results from Fig. \ref{fig:eb2aheatmaps} in that these five transformations have a significant impact upon the evaluation of source code similarity. Initially, there is a small decrease in average similarity for the token, tree and graph-based tools. This decrease becomes progressively larger for the token-based tools as the chance of transformation increases. This can be explained by the token sequences of the source files containing more and more fine-grained variations as compared to the base program. As a result of this, the token-based tools can no longer match as many token sub-sequences and therefore report lower similarity scores. The tree-based tool also incurs a consistent decrease in similarity, and provides similar scores. However, the graph-based tool does not suffer as much from these transformations, showing the lowest decrease in similarity at all chances of transformation. This is as the dependencies between the statements in source code are not modified to the same degree as the token sequences. Notably, the range of these scores is quite large. However, the standard deviation is generally tight around the average at the lower transformation chances. While it does slowly increase as the chance of transformation also increases, this is seen consistently across most tools. The consistent standard deviation implies that the transformations are having consistent impacts upon the evaluation of similarity by the tools.

\input{Tab5}

\input{Fig7_ev1ax_resilience}

Table \ref{tab:ev2axavgtransformations} presents the average number of transformations applied to each variant at each transformation chance. This represents the average the number of AST nodes of interest that are transformed within each variant, by all 5 transformation filters. The average number of transformations are used to calculate the RCT score for each tool at each chance of transformation. The RCT scores are compared in Fig. \ref{fig:ev2axrctranking}. Higher values indicate a greater robustness to transformation. Comparing the performance of these tools, it is clear that the Graph ED tool is more robust with a large number of applied transformations. This is due to the transformations having a lesser impact upon the representation of the programs through PDGs. However, when considering the scores at the 100\% transformation chance, the average similarity is only approximately 10 points higher than JPlag and the Naive Token Tiling tool. Furthermore, at the lower chances of transformation, JPlag out-performed the Naive PDG-based tool. As the lower transformation chances are more representative of common undergraduate plagiarism-hiding modifications, JPlag can be seen to be more robust to common cases, while in pervasive cases the Graph ED tool appears superior.


Overall, the results of this experiment imply that the Graph ED tool is most robust to the applied transformations. At the 100\% transformation chance, it has the highest RCT score, along with the highest average similarity of variants by approximately 10\%. Hence, the structural string, token and tree-based tools appear to be most vulnerable to such structural transformations; compared to the semantics-based tool that while it does measure a decrease in similarity, is largely due to the previously mentioned deficiencies of its naive implementation. Subsequently, transformations that apply fine-grained modifications to the structure of source code can have a cumulative impact on the effectiveness of the structural tools. Hence, from this it can also be implied that tools which measure the structural similarity of source code are not robust to pervasive applications of source code transformations. This is an interesting point, as many plagiarism-hiding modifications are structural in nature \citep{faidhi1987, joy1999, jones2001, mozgovoy2006, freire2007, allyson2019}, with all known available SCPDTs measuring structural similarity. Hence, evaluating similarity through semantics and not source code structure shows potential in robustness to pervasively applied source code transformations, on the assumption that they are semantics-preserving. 

\subsection{Random Transformations}\label{s:ev1b}

The purpose of this experiment is to compare the robustness of the SCPDTs to random selections of source code transformations. The use of random selections of transformations is arguably more realistic in the generation of test data. A plagiarising student will not restrict them self to using only 1 transformation, instead they would most likely use a diverse selection of transformations. Hence, to add a degree of realism, random selections of transformations are utilised in this experiment. 

The test data set for this experiment is generated in a similar manner to the previous 2 experimental cases. This is with 5 variants generated per base program, for each of the 6 transformation chances, but with a random selection of transformations applied per base program. The selection of applied transformations is restricted in that between 2 and 13 transformations must be selected, and they must not be the same 5 as identified in the previous section. This allows for 17,250 variants to be generated. Using random selections of transformations has the added benefit of gaining partial coverage over the total of number combinations of source code transformations provided by SimPlag. However, in the best-case scenario, the maximum number of selections of transformations used in this experiment is only 575 with one unique selection for each base program. This does not afford producing a statistically significant sample of results, as specific selection of transformations may have a dis-proportional effect on certain base programs.

In order to gain a statistical confidence of 95\%, this experiment would need to be repeated a total of 385 times, with the results then aggregated. This is infeasible for the equipment used in this experiment as each set of 17,250 variants requires approximately, 12 hrs processing; with 385 repetitions this would require approximately 7 months processing time. Hence, as an alternative, the experiment is repeated 10 times, generating 10 test data set and allowing for at most 5,750 selections of transformations to analysed. This is to provide a much more comprehensive analysis of results, that will at least allow for the demonstration of a localised trend within the 10 evaluated repetitions of this experiment.

\input{Fig8_ev1b_iterations}

Fig. \ref{fig:ev1bsample} presents the average similarity of the generated variants over the 10 repetitions using JPlag and Sim. The results for these tools exemplify the results of all 11 SCPDTs. All tools demonstrate a consistent trend of average similarities for the variants generated with the same transformation chance being within a tight range with very little variance. Hence, while the 10 sets of variants generated from each base program are constructed with different selections of transformations over the 10 repetitions, on average, the transformations have a consistent impact over the entire data set, irrespective of the base program they are applied to. Furthermore, an interesting observation over the 10 repetitions is that the standard deviations measured for the SCPDTs scores are remarkably consistent. The maximum standard deviation of all measured standard deviations is at most 0.46. Hence, the distribution of scores measured by each SCPDT on each repetition remain similar, overall indicating a consistent spread of scores. These observations would imply a homogeneous base data set, where the transformations applied have on average a relatively consistent impact upon the generated variants. 

\input{Fig9_ev1b_avgsimilarity}

Fig. \ref{fig:eb2bacgsimilarity} presents the average similarity for all 10 repetitions of the generated variants as the chance of transformation increases. Overall, there is not a profound decrease in similarity as the chance of transformation increases (for non-string-based tools). This is easily explainable as most of the selections of transformations do not have a large impact on the evaluation of similarity. However, there is still a large range of scores, indicating that while the majority of transformation selections do not have a large impact on scores, there are certain selections that the tools are vulnerable to. 

Fig. \ref{fig:eb2bacgsimilarity} also demonstrates as that chance of transformations increases, the average similarity decreases. This is an expected result, but it does imply the average similarity decreases proportionally the the increase in transformations. Furthermore, it again reinforces that the string-based tools are not robust to transformation. This is as all string-based tools at all chances of transformation have relatively poor results. Conversely, the token, tree and graph-based tools do show higher robustness, however they do suffer from a large increase in standard deviation; this indicating a much larger distribution of scores as the chance of transformation increases. This again implies that all string, token and tree-based tools are vulnerable to specific types of transformations.

\input{Fig10_ev1b_resilience}

\input{Tab6}

Table \ref{tab:ev2bavgtransformations} presents the average number of transformations applied to each program variant at each chance of transformation. This is used to calculate the RCT score for each tool at each chance of transformation, which is compared in Fig. \ref{fig:ev2brctranking}. These results show similar rankings to that of the previous experiment, however with one notable difference. The Graph ED tool performs considerably poorer compared to the other tools at all chances of transformation. It no longer has a greater robustness at the high chances of transformation, instead performing approximately on par with other token-based tools. In this case, JPlag and the Token ED tool show a higher robustness to transformation.


\input{Tab7}

In order to gain a better understanding of what selections are most effective at reducing similarity, Table \ref{tab:ev2brankings} ranks the top three selections of transformations for each tool that incurs the greatest decreases in similarity. There is a large degree of overlap between the top three selections. Furthermore, there is correlation between these selections, and the transformations identified from Fig. \ref{fig:eb2aheatmaps} (i.e. tRS, tRM, tSO, tFW, tSD). These selections always include at least one transformation from the five identified in \ref{s:ev1a}. This again strongly implies that selections of these five transformations have a considerable impact upon the tools, and hence the SCPDTs are not robust to such transformations. Table \ref{tab:ev2brankings} also aids in explaining the comparatively poorer results of the Graph ED tool. tRI and tSD are applied in two out of three cases, causing the Graph ED tool to report a lower similarity. Amongst most of applied selections, comment-changing transformations are also included. Realistically, these transformations only affect the string-based tools, as all others simply remove the commenting. Their inclusion in this table is simply due to random chance, and should not be considered to be a contributing factor to being ranked in the top 3 selections for non-string tools.

From these results where randomised transformations are applied; JPlag and the token-based tools in general are the most robust to transformation. However, it also shows that when fine-grained modifications to token sequences are applied, the tools become less robust to transformation. However, a high level of similarity is reported for all non-string-based tools, implying such tools are suitable for detecting real cases source code plagiarism with plagiarism-hiding source code transformations.

\subsection{All Transformations}\label{s:ev1c}

The purpose of this experiment is to identify the impact of all source code transformations being applied in unison to the generated variants. This is to simulate an extreme case of plagiarism, where a plagiariser applies a diverse range of source code transformations pervasively to hide plagiarism. The generated data set for this experiment contains 17,250 variant programs. This is broken down into 5 variants for each of the 575 base programs, generated with the 14 transformations at each of the 6 transformation chances. 

\input{Fig11_ev1c_avgsimilarity}

Fig. \ref{fig:eb2cacgsimilarity} presents the average similarity of variants evaluated with each SCPDT. In addition to the average similarity, the range and standard deviation of scores is also presented to indicate the distribution of variant similarities and where the similarity of most variants lies. Initially at the 10\% transformation chance, the average similarity is high for all non-string-based tools. The similarity then progressively decreases up until the 100\% transformation chance where all averages lie at or below the 60\% similarity mark. Furthermore, the standard deviations do increase slightly as the chance of transformation increases. This indicates that the distribution of scores is increasing with the chance of transformation. 

Most interestingly, all tools are severely impacted at the 80\% and 100\% transformation chances. For example, at 80\% transformation chance, JPlag drops approximately 30\% in its measured average similarity. Furthermore, at the 60\% chance, there is a noticeable decrease in similarity with all but two tools dropping below 70\% similarity. In such cases, it can be conceived that plagiarised programs to this severity could go unnoticed. Another interesting observation is the general plotting of the average scores and standard deviations remains relatively consistent between all chances of transformations. This again implies that the number of transformations applied to the source codes has a proportional impact on the evaluated similarity; and that the applied transformations have consistent impacts across all tools. 

\input{Tab8}

\input{Fig12_ev1c_resilience}

Table \ref{tab:ev2cavgtransformations} identifies the average number of transformations applied to the variants as the chance of transformation increases. This is used to calculate the RCT score for each tool at each chance of transformation, that is compared in Fig. \ref{fig:ev2crctranking}. From this result, initially JPlag is shown to be the most robust tool, however at the 100\% transformation chance Token ED tool shows slightly greater robustness. This is interesting as a naive tool with a similar approach (differing in algorithms to compare similarity) has out-performed a mature academic tool in the most pervasive case of simulated plagiarism. 



The results of this experiment indicates that JPlag and the Token ED tool are the most robust to the pervasive application of all transformations. While their evaluated similarities have the potential to not raise suspicion and evade detection at high chances of transformation, they do show the least decrease in average evaluated similarity. Interestingly, while both of these tools are token-based, they utilise different methods to measure similarity. JPlag uses greedy token tiling, that is approximated in the Token Tile tool. However, the Token Tile shows much poorer performance than JPlag. This can be attributed to the naive implementation of the tool, that approximates greedy token tiling by repetitively identifying the longest common sub-string. While JPlag implements a tolerance for mis-matched token sub-strings, the naive implementation does not. Hence, it is much less robust than JPlag. The Graph ED tool shows a close ranking in third place at the 100\% transformation chance. However, it initially demonstrated a relatively lower RCT score at lower chances of transformation. However, as the number of transformations increase it shows a higher robustness to transformation. This would imply that the PDG-based tool is more robust in the more pervasive cases of applied plagiarism-hiding transformations.

Overall, this is a positive result indicating token-based approaches have a good robustness to transformation. However, there is also a large emphasis on tools implementing a tolerance to error in mis-matched portions of a program in order to gain greater robustness to transformation. The Token ED tool can accommodate for small errors simply through the deletion/swapping of tokens. However, the Token Tile tool implements no such feature and hence suffers from a poor evaluated similarity.

\section{Evaluation 2: Code Injection}\label{s:ev2}

This evaluation will compare the robustness of SCPDTs to source code injection. Code injection poses an issue in the detection of plagiarism as by design, most SCPDTs present program similarity as the aggregation of similarity scores from all source code files. However, by aggregating the scores it has the potential to hide the fact that code has been injected. Furthermore, code injection can be seen to `increase' the size of the injected source code. As a result of this, it adds code that cannot be matched against the source of the plagiarised work.

Two experiments are performed to evaluate the robustness of tools. The first experiment will evaluate the robustness of each SCPDT to data sets generated by injecting one type of source code fragment into each variant (i.e. file, class, methods and statements, individually). The second experiment will evaluate SCPDT robustness to a data set generated by injecting all source code fragments in unison. The source code fragments will be injected with injection chances 10\%, 20\%, 40\%, 60\%, 80\% and 100\%, as with the previous evaluation. However, SimPlag also exposes four injection-specific configuration parameters to limit the quantity of source code fragments injected. Without specifying limits on the quantity of source code fragments injected, it can result in SimPlag generating variants that are substantially larger in size than the original base programs. For the purpose of this experiment, the quantities of injected source code fragments are set to be limited with relative values to the average composition of the base data set (see Table \ref{table:assignmentstats}). This means, the quantity of injected fragments are limited to at most: 
\begin{itemize}
    \item 12 statements into each method (but no more than the method's original statement count),
    \item 9 methods into each class,
    \item 1 class into each file, and
    \item 4 files into each submission.
\end{itemize}

These limits are intended to introduce an element of realism into the generated test data set. This is to avoid injecting an unreasonable quantity of source code fragments into the generated variants. However, this itself is a difficult task to achieve due to the randomised nature of source code injection with SimPlag. In a worst-case scenario, using this setup it can be expected that the generated variants could quadruple in size (by doubling the number of files, classes, methods and statement counts). This should be considered to represent a worst-case-scenario, where for example, a plagiariser has appropriated a large quantity of code, and used it in the implementation of their own assignment submission thereby mixing both plagiarised and original code. However, this is expected to vary by the type of source code fragment injected, and will be discussed as appropriate. For this evaluation, a data set of assignment submissions from unrelated programming courses is used as the seed for injected source code fragments. This is to avoid any unintentional cases where a variant is injected with source code originating from its base program, or from an otherwise similar program. 

When using SimPlag to generate data sets for evaluating the effect of source code injection, there is one notable problem that may cause unreliability in the evaluated results. SimPlag will apply a consistent formatting to all generated variants as to make them appear more realistic, and hence applies by default a simple (but pervasive) source code transformation. This will have no substantial effect on the non-string-based tools as they are immune to such a transformation. However, this has the potential to impact upon the results of all string-based tools, as formatting is not ignored with such SCPTs. Hence, in order to minimise the effect of the consistent formatting applied by SimPlag, the base programs are normalised with the same consistent formatting applied. This is to improve upon the overall reliability of results.

\subsection{Individual Fragment Types Injected}\label{s:ev2a}

This experiment will compare the impact of injecting each of the 4 types of source code fragments upon each SCPDT in isolation. The generated data set for this experiment contains 69,000 variant programs. This is broken down into 5 variants for each of the 575 base programs, generated with the 4 fragments of source code injected individually, at each of the 6 injection chances. 

\input{Tab9}

Table \ref{tab:ev3aavginjectloc} identifies the average lines of code injected into each variant at each chance of injection. At each chance of injection, the quantity of LLOC injected varies significantly. The file and statement data set have similar LLOC injected, but the class and method data sets have comparatively much larger quantities of source code injected. This is a result of the utilised configuration values, and the nature of how fragment injection is implemented in SimPlag. SimPlag uses a shared java.util.Random object (that produces a uniform distribution of random values) to determine if a fragment of source code should be injected into a variant. This is determined at the nodes of interest, that each have different occurrences in a single variant. For example, the file fragment injector has 1 'node' of interest in each variant (being the variant itself), while the class injector has an average of 4.44 nodes of interest in each variant. Using the selected configuration values, SimPlag has a single weighted chance to inject up to 4 files into each variant, while it has on average 4.44 weighted chances to inject 1 class into a file. This slight difference in semantics has a noticeable impact upon the generated variants, with approximately 2-2.5 times the quantity of LLOC being injected into the variants. However, this will not affect the results of this experiment. The RCI score can only be used to compare SCPDTs on the same data set, and not on different data sets. Hence, the generated test data sets will allow for comparison of the SCPDTs against the injection of each of the 4 types of source code fragment in isolation.

\input{Fig13_ev2a_heatmap}

Fig. \ref{fig:ev3aheatmaps} presents the average similarity of the variants created with each of the four individual source code injection operations. Initially, this appears similar to the heat maps in Fig. \ref{fig:eb2aheatmaps}. The similarity at the 10\% injection chance is typically high (larger than 90\%) for all non-string-based tools. However, it then proceeds to drop as the injection chance increases. Comparing Figs. \ref{fig:eb2aheatmaps} and \ref{fig:ev3aheatmaps}, the decrease in similarity is much more profound in the presence of source code injection, as compared to the application of source code transformations. This is expected as new code that cannot be matched to the base program is introduced into the variants, and the relative quantity of injected LLOC.

The string-based tools again produce consistently poor results. However, they are slightly improved compared to the results in Fig. \ref{fig:eb2aheatmaps}, yet still produce considerably low similarity scores. The non-string-based tools initially evaluate high similarity scores for all types of injected source code fragments. However, this quickly begins to drop as the injection chance increases, showing a consistent progression of decreasing similarity. Notably, the similarity of the variants generated with class and method fragment injection drops much faster than the file and statement injection. However, in interpreting this result, it must be considered the greater quantity of LLOC injected for classes and methods, compared to files and statements; and hence, this result can be expected.

For the file injection results, it is interesting to observe that the non-string-based naive SCPDTs demonstrate a higher average similarity than the academic SCPDTs. Comparing the token-based tools, Sim has the greatest decrease in file similarity. This is caused by Sim not providing a submission-wise similarity score (only file-wise). To derive an aggregated submission score, the best similarities for all files are averaged during comparison. Hence, in this case no files are ignored. All naive tools use a similar method of score aggregation, scoring similarity as the intersection of the two programs; i.e. the coverage of matching sections divided by the total size of the compared programs. However, in this case the naive SCPDTs are capable of evaluating a higher similarity than the other SCPDTs. Amongst all other fragment injections, the non-naive SCPDTs demonstrate consistently higher similarity scores. In particular, Plaggie and Sim demonstrate the highest average similarity against class and method injection, while JPlag and Sim evaluate the higher average similarities against statement injection.

\input{Fig14_ev2a_resilience}

In order to rank the tools, the average RCI is calculated for all variants created with each mode of source code injection, at each chance of injection. These scores are compared in Fig. \ref{fig:ev3aTransformationResilience}. The results vary between the tools for different modes of injection. The string-based tools (excluding Sherlock-Warwick) are consistently ranked lower; implying string-based tools are not robust to any form of source code injection. The string-based tools are consistently followed by Sherlock-Warwick, that combines token and string-based methods. Due to this combination of methods, Sherlock-Warwick shows greater robustness than the string-only tools, however it also results in the tool performing worse than the dedicated token-based tools; while the token, tree and graph-based tools consistently demonstrate greater robustness. 

Across file, class and method injection, most tools show a slight trend of increasing in robustness as the chance of transformation also increases. This does not imply the tools are less susceptible to greater quantities of injected source code, simply that the rate of similarity change is decreasing. However, this is not the case for statement injection. Comparatively, much fewer statements need to be injected to reduce the RCI score. As a result of this, statement injection has the greatest impact on similarity, and hence the tools demonstrate less robustness to it. This is related to the fine-grained transformations of token sequences identified in Evaluation 1. Injecting whole statements amongst the existing statements has the potential to interrupt these sequences with new unrelated statements. Hence, there are cases where statement injection lowers the overall score by both increasing the size of the variant programs, and prohibiting the tools from matching existing code. This results in a lower similarity and RCI score.

From these results, it can be generalised that when each mode of source code injection is applied in isolation:
\begin{itemize}
    \item The naive Graph ED and Token ED tools are most robust to file injection, while Plaggie is the most robust non-naive SCPDT;
    \item Plaggie \& Sim show the greatest robustness to class \& method injection; and
    \item JPlag \& Sim are most robust to statement injection.
\end{itemize}

\subsection{All Fragment Types Injected}\label{s:ev2b}

This experiment evaluates the robustness of each SCPDT to the 4 types of source code fragments being injected in unison. This is to compare the robustness of the SCPDTs when large quantities of source code fragments are injected into a program. The generated test data set for this experiment contains 17,250 variant programs. This is broken down into 5 variants for each of the 575 base programs, generated with the 4 types of source code fragments injected at each of the 6 transformation chances. 

\input{Tab10}

Table \ref{tab:ev3bavglocinjected} presents the average LLOC injected into the generated variants. This table shows a consistent increase in lines injected, averaging approximately 170 lines into each variant per 10\% chance of injection. It should be noted that in this case, the number of LLOC injected into the variants is not expected to be representative of simple undergraduate plagiarism. The extreme cases are more representative of students collaborating, or cases where fragments of source code have been appropriated and integrated into an other's own work. However, this does serve to demonstrate the impact on similarity when large quantities of source code are injected.

\input{Fig15_ev2b_avgsimilarity}

Fig. \ref{fig:ev3bavgsimilarity} presents the average similarities of the generated data set for this experiment. Immediately there is a noticeable decrease in similarity for all tools. This is much more profound than the results of the source code fragments injected in isolation. However, one tool consistently ranks with the highest average similarity: Plaggie; being in a consistent range of at most approximately 10 percentage points higher than all other tools. Initially, the results of all non-string-based tools are consistently within a range of 10\%. However, as the number of LLOC increases, this range of scores begins to increase. This implies with large quantities of source code injected, certain tools begin to become less robust to injection. Notably, the String ED tool begins to perform on par with the token, tree and graph-based tools with large quantities of source code injected. This is not implying that String ED is more robust in the presence of large quantities of source code injected. Simply that all other tools are significantly impaired in this scenario.


The large decrease in similarity for all tools is explained by the significantly larger LLOC injected compared to the test data set in Section \ref{s:ev2a}. More fragments of source code being injected implies less source code that can be matched between the variants and their base programs. Hence, the results from this experiment do indicate that in extreme cases where large quantities of source code are injected into a program, the evaluated similarity will drop significantly. With these results it also needs to be considered the dramatic increase in size of the injected LLOC to reduce the similarity of the variants. A `smaller' quantity of injected code does not see a dramatic decrease in similarity for non-string based tools. For example, JPlag requires 189 LLOC to be injected (equivalent to a 56\% increase in LLOC in this data set) to demonstrate a decrease in 20\% similarity. Even at the 20\% injection chance where an average of 362 LOC are injected, the similarity scores typically show a decrease of approximately 30\% similarity. Such scores would most likely raise suspicion to a reviewer, and likewise would require considerable effort on the behalf of a plagiariser to commit. Hence, the tools are generally robust to source code injection within reasonable quantities of injected code. It is only with extremely large amounts of code injected that the tools begin to see a profound decrease in evaluated similarity scores.

\input{Fig16_ev2b_resilience}

In order to rank the tools for robustness against code injection, the average RCI score is evaluated for all tools at each chance of injection. These scores are compared in Fig. \ref{fig:ev3brciranking}. All tools show a common trend of an increasing RCI as the LOC injected is increased. This indicates that smaller quantities of injected code fragments have a comparatively greater impact on the evaluation of similarity. This shows some consistency with the results of Section \ref{s:ev2a}. However, this increase in robustness is considerably larger. The results in Figs. \ref{fig:ev3bavgsimilarity} \& \ref{fig:ev3brciranking} show a clear ranking of tools for this experiment. String-based tools consistently perform poorly and are typically ranked last. While the token, tree and graph-based tools show a greater, and consistently higher robustness to injection. Across all chances of injection, the top 3 rankings of tools from this experiment are:
\begin{enumerate}
    \item Plaggie
    \item Token ED \& Graph ED 
    \item JPlag \& Tree ED.
\end{enumerate}

\section{Discussion of Results}

The evaluations performed in this article were designed explore the robustness of SCPDTs to plagiarism-hiding modifications in order to answer the three research questions:

\textbf{RQ1:} \textit{What are the impacts of source code transformations on SCPDTs?}

\textbf{RQ2:} \textit{What is the impact of source code injection on SCPDTs?}

\textbf{RQ3:} \textit{What SCPDT is most robust to plagiarism-hiding modifications?}

\noindent Each question will be discussed and answered in the following sub-sections.

\subsection{RQ1: Impact of Source Code Transformations}

Evaluation 1 was designed to answer \textbf{RQ1}. This question was explored through 3 experiments (plus an extended case) that evaluated the 11 SCPDTs against different selections of source code transformations, allowing for a comparison of SCPDT robustness. The results of the 3 experiments demonstrated four consistent observations regarding the impacts of source code transformations upon the SCPDTs:
\begin{enumerate}
    \item All string-based tools show poor robustness to all source code transformations, when compared to the non-string-based tools.
    \item The token and tree-based tools demonstrated little impact by cosmetic source code transformations.
    \item The token and tree-based tools demonstrated vulnerability against fine-grained structural source code transformations.
    \item The graph-based tool demonstrated greater robustness against fine-grained structural source code transformations than all other tools, however was vulnerable to transformations that were lexical or modified the program semantics.
\end{enumerate}

These results can be largely explained by considering the SCPDTs in terms of how they represent source code, and how this representation is used to measure source code similarity.  The string, token and tree-based tools represent the structure of source code. The string-based tools represent structure as a literal character string, the token-based tools represent structure as sequences of lexical tokens, and the tree-based tool represent structure as an AST. These three representations of structure are then compared for similarity with two basic techniques: coverage (e.g. through tiling) or edit distance. Any transformation to this structural representation will either impact upon the evaluation of coverage, or increase the edit distance. That is, any transformation that changes: the character string will impact upon string-based tools, the lexical token sequence will impact upon token-based tools, the AST will impact upon tree-based tools. Hence, in all cases, the measurement of similarity was impacted upon.

From the performed experiments it was demonstrated that of the 14 implemented source code transformations:
\begin{itemize}
    \item All 14 modified character strings, and hence impacted upon the string-based tools.
    \item 5 transformations (tRS, tRM, tSO, tFW, tSD) had an observable impact upon token sequences, and hence impacted upon the token-based tools. However, the remaining 9 transformations had no or negligible impact upon the token-based tools.
    \item The same 5 transformations (tRS, tRM, tSO, tFW, tSD) have a similar, but largely less pronounced impact upon the tree-based tool, with the exception of tRM having a extremely pronounced impact upon the tree-based tool.
\end{itemize}

Hence, from these results, it can be summarised that source code transformations impact upon SCPDTs when the representation of source code a SCPDT uses for comparing similarity is modified. This was specifically found with the 5 fine-grained structural transformations: tRS, tRM, tSO, tFW \& tSD in Section \ref{s:ev1ax}; and further emphasised in Sections \ref{s:ev1b} and \ref{s:ev1c}. This is a notable vulnerability with currently available SCPDTs as all implement structure-based measures of source code similarity. Hence, they are all vulnerable to such transformations. This becomes especially pronounced when the transformations are applied pervasively. If such transformations are applied pervasively throughout a body of source code, a large decrease in the evaluated similarity can be observed. Furthermore, this result is consistent with the naive string, token and tree-based tools, indicating such techniques in general are vulnerable to the same structural transformations.

In order to avoid the impact of plagiarism-hiding source code transformations that modify the structural representation of the source code, the simplest method is to ignore the structural aspects of source code that change due to the transformations. This can be seen with the Graph ED tool that represents the program as a set of PDGs. This allows Graph ED to measure the semantic similarity of two programs, through the occurrence of similar relations between statements in each procedure. Assuming that source code transformations do not modify the semantics of the source code - an assumption that, as previously identified, is commonly true in committed source code plagiarism; such an approach should be in theory immune to such changes. This was largely demonstrated for the Graph ED tool in Section \ref{s:ev1a} where it was shown to be impacted upon by transformations that modified the semantics of the program; and in Section \ref{s:ev1ax}, where it demonstrated the greatest robustness to the 5 fine-grained structural transformations. Hence, the theoretical foundation of a PDG-based tool does show merit in the detection of source code plagiarism. 

With this result, it must also be considered how easy the utilised transformations are to apply by a plagiariser. The 14 source code transformations are not technically complex, and with enough time a novice plagiariser could apply them manually. If considering a skilled plagiariser (i.e. somebody proficient at programming) it is conceivable that they would have the necessary skills to apply these transformations, as well as more complex transformations such as obfuscating the control flow of the program. Furthermore, it must be considered that all of the applied transformations are automated, and are in fact features of many source code editors and integrated development environment. Hence, applying source code transformations even in a pervasive manner is a trivial task that needs to be accounted for by SCPDTs, and hence this is a threat to currently available SCPDTs.

\subsection{RQ2: Impact of Source Code Injection}


Evaluation 2 was designed to answer \textbf{RQ2}. This was explored through two experiments comparing the affect of injecting fragments of source code upon the 11 SCPDTs. The impact of source code injection can be generalised from the perspective of by injecting source code, new code is inserted into a program that cannot be matched between it and it's source. However, the results of the two experiments evaluating this impact vary considerably. Similar to evaluation 1, the string-based tools were demonstrated to show great vulnerability to the injection of all fragments of source code; while in comparison the token, tree and graph-based tools demonstrated considerably greater, but varied, robustness against the injection of source code fragments. Overall, the results varied between the individual tools, and the types of source code fragments injected. This would indicate that by design, different tools are more robust to the injection of certain fragments of source code. Hence, interesting observations regarding the impact of source code injection, and how robustness can subsequently be gained. 

Against file injection, the non-string-based naive SCPDTs were demonstrated to have greater robustness than all other tools. This is presumably a result of how the naive SCPDTs aggregate the similarity scores of file pairs into a submission-wise similarity scores. This is through the averaging of similarity scores from the best mapping of file pairs between two submissions. In such a case, it can be assumed that in an ideal case where the naive tools evaluate 100\% similarity between the average 4.41 identical file pairs between a base and variant program, the injection of 1 to 4 files by SimPlag would result in evaluated similarity scores in a range of approximately 69\% to 90\%. This is consistent with the results of Section \ref{s:ev2a}, where at the 100\% injection chance the non-string naive SCPDTs evaluate similarity scores approximately in this range. The remainder SCPDTs utilise similar methods, but do not always `map' individual source files. For example, based on the implementation Plaggie, the tool aggregate all source files together into a single token stream, and use its relative size for the calculation of similarity. This subtle difference results in the tool being unable to measure any significant similarity between the source code in the injected source files. And hence, it produces similar, but in this case lower results.

The injection of class and method fragments effectively adds `junk' to the variant programs, increasing the overall size of the variant sources. Injecting such fragments should not affect matching any existing code (assuming sufficient size in relation to SCPDT minimum match lengths), but merely increase the size of the programs being compared. The theoretical decrease in similarity can again be assumed to be proportional to the size increase. In the case of method injection at the 100\% injection chance, the variants are on average approximately 140\% larger compared to the base programs, equating to approximately 472 LLOC being injected. This is approximately 40\% of the total LLOC being compared between any base program and its variants (the total LLOC being 2 times the average base size of 336 LLOC, plus the injected LLOC). Hence, in a best case scenario, 60\% of the LLOC between the base and variant can be matched, that would result in a similarity scores of approximately 60\%. A similar relation can be assumed for class injection at the 100\% injection chance, where approximately 50\% of the LLOC being compared is injected, resulting in a similarity score of approximately 50\%. This assumption largely holds for the non-string SCPDTs, where against class injection most tools see a 50\% average similarity, while against method injection most tools see a 60\% average similarity. However, Plaggie and Sim are clear outliers in this case, where the tools demonstrate greater robustness. Hence, in this case, Plaggie and Sim are least susceptible to the injection of this junk. 

The injection of statement fragments has a similar impact by again adding `junk' to the variants. If the size increase is again used to predict the similarity of the variant, at the 100\% injection chance it should be assumed that the variants would have an average similarity of approximately 80\%. However, this is largely not the case. Most SCPDTs evaluate the average similarity in a range of 40\% to 60\%, while JPlag and Sim evaluate similarity scores of approximately 75\%. In this case, the injection of individual statements is somewhat analogous to the fine-grained structural transformations identified from evaluation 1. By mixing small fragments of junk with existing code, it can be assumed that this prohibits the matching the base code in the variant. Hence, there is a less than expected evaluation of similarity by the SCPDTs. A similar occurrence is seen when injecting all 4 types of source code fragments. At the 100\% injection chance, with a 485\% increase in variant size, it can be expected approximately 70\% of the analysed LLOC will not be found in the base program, and hence will result in an approximate average similarity of 30\%. This observation holds for most tools. However, Plaggie, Token ED and Graph ED demonstrate higher average similarities at approximately 40\%. Hence, these tools demonstrated greater robustness when all 4 types of source code fragments are injected in unison.

However, in consideration of these results, it must be acknowledged that the quantity of injected fragments of source code border into a `worst-case scenario'. As discussed, at the higher injection chances, the generated data sets would be more representative of a plagiarising student integrating an other's work into their own. Hence, if only the results for non-string-based tools are considered up to the 10-20\% injection chance, all tools were demonstrated to evaluate a high average similarity. However, in this worst case scenario, different SCPDTs demonstrate greater robustness to the injection of different types of source code fragments. This result also emphasises the methods of comparing files for similarity to ignore injected code, and aggregating scores for similarity in providing robustness to source code injection. Hence this should be considered in detection of pervasively modified plagiarised source code.

\subsection{RQ3: Most Robust Tool}

The results of evaluations 1 and 2 address \textbf{RQ3}. From the 5 performed experiments, it is clear that all string-based tools are not robust to any form of plagiarism-hiding modifications. This is demonstrated through the consistently low similarity scores evaluated by the tools; and hence string-based techniques should be avoided for use in SCPD. The token, tree and graph-based tools show much greater robustness to source code modifications. However, identifying what tool is the most robust to plagiarism-hiding modifications is a matter of perspective.

Strictly from the results of evaluation 1, JPlag demonstrated the greatest robustness to source code transformations. This was shown by JPlag most consistently evaluating the highest similarity scores, and hence RCT scores, in the three experiments of evaluation 1 (Sections, \ref{s:ev1a}, \ref{s:ev1b} and \ref{s:ev1c}). However, JPlag is out-performed by the naive Graph ED tool in Section \ref{s:ev1ax} when compared for robustness against the 5 identified fine-grained structural transformations. This is a result of the Graph ED tool being more robust against such transformations when pervasively applied than JPlag, and all other of the evaluated SCPDTs. Hence, overall, from the results of this evaluation, JPlag should be considered to be the most robust SCPDT on average to the evaluated plagiarism-hiding transformations. However, when considering the results against source injection in evaluation 2, Plaggie was shown to be the most robust in 4 of the 5 generated data sets. This is followed by Sim, showing a similar robustness in 3 of the 5 generated data sets. While a notable mention is the Graph ED tool, being consistently ranked high in 4 of the 5 generated data sets. Hence, in considering these results, it is a matter of perspective in determining what tool is most robust. 

To generalise the answer to \textbf{RQ3}, JPlag is the most robust tool to the evaluated plagiarism-hiding transformations, while Plaggie is the most robust tool against the injection of the evaluated source code fragments. But, under certain conditions with pervasively transformed variants, the Graph ED tool does show potential. Hence, utilising such a tool, or potentially one that combines both structural and semantic measures, shows benefit as a future direction of work in SCPD. In Section \ref{s:ev1ax}, this tool demonstrated a greater robustness to pervasively applied transformations which apply fine-grained transformations to the source code structure, while it also ranks consistently high against both source code transformation and injection. While in most experiments it performed approximately on par with the other non-string-based tools, this is attributed to the naive implementation of the tool. It is feasible that an optimal implementation of a PDG-based Graph ED tool would out-perform all other SCPDTs in most, if not all experiments.

However, in typical cases of plagiarism, all non-string-based tools show sufficient robustness to source code modification, being capable of evaluating high similarity scores. But, this result does not mean that all other tools are not suitable for detecting plagiarism is the presence of plagiarism-hiding modifications. Considering the similarity scores of the lesser-transformed program variants (assume $\leq$ 40\% transformation chance); in almost all cases with the non-string-based tools, the evaluated similarity scores are generally enough to raise suspicion. Hence, for typical usage, there is no problem with their robustness to plagiarism-hiding modifications. It is only in the extreme evaluated cases with pervasively applied source code modifications, do JPlag and Plaggie demonstrate greater robustness.

\section{Limitations \& Threats to Validity}

In the evaluations performed in this work, there are numerous design decisions that originate from the utilised tooling and evaluation method. In this section, important limitations of the evaluation, and threats to validity of results are identified and discussed; focusing on configuration bias of the SCPDTs, authenticity and correctness of the generated test data, and the measures used for comparing SCPDT robustness.   

\subsection{Configuration Bias of SCPDTs}

In evaluations of code similarity tools, it is common to identify an `optimal' configuration value for a tool on a given data set. This has been shown to provide greater tool performance (e.g. as seen in \cite{ragkhitwetsagul2018, ahadi2019}). The performed evaluations do not attempt to identify an optimal configuration value. This is an intentional design decision of this evaluation to remove tool bias, as arguably, identifying an optimal configuration has the potential to introduce configuration bias in the performed evaluations. This is due to, as discussed, using optimal configurations is not considered representative of a real-world use of SCPDTs. Using an optimal configuration value for a SCPDT requires the foresight of knowing in advance what submissions are plagiarised. This is of course not the case in a real-world use of SCPDTs and may give a false impression of how robust a SCPDT may be in a real-world use. Hence, to remove any configuration bias, all SCPDTs are executed using their default configuration values under the assumption that the original developers of each SCPDT selected appropriate defaults.  

By extension utilising purpose-built naive SCPDTs has the potential to introduce bias into this evaluation. It would be trivial to use configuration values to improve or skew the results of these tools. E.g. the performance of tiling tools can be improved by decreasing the minimum match length. Hence, to reduce bias through the use of naive SCPDTs, they utilise configuration values that can be sourced from similar SCPDTs, or justified based on the utilised data sets. i.e. the naive String Tile tool uses a minimum match length derived from the average expression size in the data set, while the naive Token Tile tool uses the same minimum match length as JPlag. 

Furthermore, using custom-built SCPDTs for this evaluation is also a source of bias. However, the implementation of these tools is as their name suggests, `naive'. There is very little code written for these tools, and very little room to increase their robustness dis-proportionally to the available academic SCPDTs. They were implemented by re-using and wrapping existing libraries and algorithms into command-line applications. This is with the exception of the Graph ED tool, that uses its own implementation of a PDG and associated edit distance algorithm, but both are still very simple and naive re-applications of existing techniques. In general, the only robustness to plagiarism-hiding modification each tool affords is that intrinsically gained through the respective program representations utilised in each naive SCPDT (e.g. token-based tools being robust to cosmetic changes and renaming identifiers). There is little to no intentional optimisation of these tools.


\subsection{Authenticity \& Correctness of Test Data}

The use of synthetic test data is a potential threat to this evaluation. The generated simulated plagiarised variants are not real cases of plagiarism. Hence this evaluation does not conclusively demonstrate the robustness of the evaluated SCPDTs against real cases of undergraduate source code plagiarism, or real examples of plagiarism-hiding source code modifications. Instead, this evaluation simply evaluates the SCPDTs against source code modifications that are representative of undergraduate plagiarisers. The utilised modifications are termed representative as they have been referenced from literature as being observed to be used by undergraduate plagiarism. However, there are always uncertainties in how representative the complexity of the applied source code modifications are in comparison to real cases of undergraduate source code plagiarism. This is in how pervasively modified a plagiarised work is by the plagiariser. In order to accommodate for this issue, the plagiarism-hiding modifications are applied using a sliding scale of transformation and injection chances. This allows for evaluating SCPDT robustness against lesser and more progressively transformed samples of simulated plagiarised works. It must also be acknowledged that this evaluation only utilises 14 source code transformations, and 4 types of source code fragments injected. This is in contrast to the countless many types of source code transformations that may be applied, or combinations of source code fragments that may be injected. Hence, this work is limited to only evaluating robustness against the utilised source code modifications, and therefore, there may be many more complex source code modifications with a more profound impact upon the evaluation of source code similarity that has not been observed here.

The performed evaluations also only evaluate robustness of the SCPDTs to plagiarism-hiding modifications. They do not evaluate the accuracy of tools (in terms of precision and recall) in the presence of plagiarism-hiding modifications. The results of this experiment show that certain tools and approaches are more robust to specific transformations when applied to the utilised data set. However, the results do not show that certain tools and approaches are more accurate. While the generated data and test conditions try to simulate as closely as possible real-world situations where a student may have plagiarised, this is not a substitute for real-world data. However, also as discussed, such data is generally not available in sufficient quantities for a comprehensive experiment. Hence, requiring the use of synthetic data in this experiment.

There are also potential threats to the correctness of the generated data. As discussed, SimPlag will only ensure that the generated simulated plagiarised variant programs can be parsed (i.e. they are syntactically correct), and not that the variants are compilable or functionally correct. However, plagiarism needs to be identified irrespective of if a plagiarised work compiles and is functionally correct. Hence, this is not perceived to be an issue as this work is focused on evaluating the impact of source code modifications on SCPDT robustness in the case they source code modifications \textit{are} applied, and not in the case that source code modifications \textit{can} be applied. 

Across the data sets there is also a much higher chance than certain transformations will be applied as there are more nodes of interest for their application. For example, a tAC (Add Comment) can be applied to any class, field or method declaration. This is compared to tFW (\textit{for} to \textit{while}) that can only be applied to \textit{for} statements. This can lead to an un-even application of source code transformations. Furthermore, the impact of a single application of a transformation does vary in terms of how much code is modified. For example, a single reordering of a whole block statements and single swapping of operations both count as one transformation. Hence, these transformations are not always uniformly applied to each derived variant, due to the randomness of application, and nature of the data sets; and as such this may impact upon the identification of specific transformations that have the potential to have great impact upon the data sets. 

\subsection{Comparison Measures}

Precision and recall are commonly used metrics in the evaluation of SCPDTs, and code similarity tool evaluations in general \citep{whale1990, novak2016, ragkhitwetsagul2018}. Both metrics express the accuracy of a tool in identifying similar bodies of source code. However, as the purpose of this work is to measure robustness and not accuracy; precision and recall have not been used in this evaluation. This work is focused on measuring the decrease in similarity evaluated by SCPDTs as they are exposed to more pervasively applied plagiarism-hiding source code modifications. This is used as a measure of robustness. Using accuracy metrics would not contribute to the goal of this study as accuracy metrics are derived from a binary choice of if, or if not a SCPDT has detected an indication of plagiarism. As a SCPDT does not directly detect plagiarism, but instead detects indications of plagiarism \citep{joy1999}, using such a method for the evaluation of SCPDT robustness would not contribute to this work.

Two metrics are used in the comparison of robustness: the quantitative absolute decrease in similarity, and the comparative measure expressing the decrease in similarity in ratio to the applied source code modifications (as the RCT and RCI scores). The first metric is used to provide an overview of the impact of source code modifications upon the evaluation of similarity. The second metric is used to compare the SCPDTs. However, the interpretation of the second comparative metric poses a threat to the validity of results in this work. As discussed, it is not a normalised measure that can be used to compare SCPDTs on different data sets. Identifying a `universal' normalised comparison metric is outside of the scope of this work. The RCT and RCI scores are simply used to correlate the impact of source code modifications to the evaluation of similarity, and then ranks the tools under specific experimental conditions on a single data set. Hence, it should be considered that any RCT and RCI measurements can be compared between evaluations on different data sets. Their use is restricted to the comparison and ranking of tools on individual data sets only.

When calculating the RCI, the increase in program size is used. This value is measured as the LLOC, and is calculated as the non-block statement count of a source file. LLOC is used as it is a commonly used metric for expressing the size of code, that counts the executable statements in code without considering formatting, declarations or whitespace. For the purpose of measuring robustness, any size metric could be used in this evaluation (e.g. raw lines of code, token counts, AST node counts, etc) as long as it is consistently used in the evaluation of robustness for all SCPDTs. Using a different size measurement will simply scale the RCI scores of each tool, according to the utilised size measurement. This may give the impression of a greater/lesser robustness for a SCPDTs. However, the RCI (and RCT) scores are relative comparison measures, and not absolute comparison measures. Hence, using a different size value will not change the relative robustness ranking of the SCPDTs.





\subsection{Performance of Naive and Academic Tools}

Throughout the performed experiments, the naive SCPDTs often performed on par, if not better than the academic SCPDTs. It then raises the question, what are the benefits of using the mature academic SCPDTs, when in many cases naive re-applications of existing techniques can be applied with similar results.

\input{Tab11}

The most profound difference between these tools are the runtimes, and by extension the complexity of the implementations. Table \ref{tab:ev0normliasedruntimes} presents the average runtimes of each tool when comparing each program pair in each set of assignment submissions. There are clear differences between the academic and naive tools in the average runtime of their approaches. JPlag is by far the fastest on average for non-string-based tools, while the naive token-based tools run approximately 10 times slower. Hence, it is clear that the academic tools contain optimisations to improve efficiency when compared to the naive tools. Such optimisations are not present in the naive tool implementations.

When considering the results of the experiments which implied the Graph ED tool has potential at being more robust than JPlag; it also has to be considered the runtime of the Graph ED tool. Being graph-based, it is expected that it will have a much higher complexity and hence runtime. However, in this case the Graph ED tool is approximately 42 times slower than JPlag. While the Graph ED tool could feasibly gain greater efficiency through optimisation (e.g. pruning unnecessary comparisons), it raises the question if the potential for greater robustness is worth the substantially greater complexity of the approach.

However, when comparing the academic and naive tools, the accuracy of the tools must also be considered. The performed experiments do not consider the accuracy of the evaluated tools. However, it must be noted that in certain circumstances a highly robust SCPDT may have poor accuracy in the detection of plagiarism. For example, consider if a SCPDT is eager in measuring similarity, such that it reports a high similarity between unrelated programs. In such a case, the tool would have a poor false negative rate, and overall a poor accuracy. Hence, the results of these evaluations should not be considered to imply that when a tool is robust, it is accurate; and therefore, a naive tool that is more robust than an academic tool is not necessarily more accurate than an academic tool. Evaluating the accuracy of the utilised SCPDTs is subject to future work.

\section{Related Works}

Source code plagiarism is a well-explored topic in academia. Subsequently, there exist many works which seek to evaluate or compare SCPDTs. For example, \cite{whale1990,verco1996,lancaster2005,flores2014,ahadi2019}. However, a common theme in prior evaluations is that they evaluate tools for accuracy in detecting plagiarised assignment submissions. This is through the measurement and comparison of the precision and recall of the tools. Similarly, there are a number of similar works in the domain of code clone detection which seek to compare tools. For example, \cite{bellon2007,roy2009,svajlenko2015,walker2020}. However, these works are again more focused on the evaluation of the accuracy of tools in identifying similar programs. There is no emphasis on measuring the robustness of tools against specific source code modifications.

In the performed experiments, the utilised SCPDTs are evaluated for robustness to plagiarism-hiding modifications. This is to measure the impact of applying plagiarism-hiding modifications upon the evaluated similarity of SCPDTs. To the authors knowledge, this is the first work that specifically compares SCPDTs by robustness to plagiarism-hiding modifications that are representative of undergraduate programmers. However, there are three other works with similarities in the theme of evaluating source code similarity tools. \cite{ko2017} evaluated the performance of COAT (a code obfuscation tool) at fooling 4 SCPDTs (Moss, JPlag, Sim and Sherlock). However, this evaluation was against only 8 transformations, many of which are not representative of undergraduate programmers. Furthermore, there is no focus on measuring the impact of the transformations upon the SCPDTs, only the measurement of tool accuracy. \cite{ragkhitwetsagul2018} compare 30 different tools and techniques in their accuracy in evaluating similarity in pervasively modified source code. While similar to this work, Ragkhitwetsagul et al. are focused on the accuracy of tools in detecting code cones. Furthermore, their generation of pervasively modified source code is enabled by Java byte-code obfuscators and decompiler tools. Such modifications to source code are not necessarily representative of undergraduate programmers, and there is no measurement of the impact of specific transformations. \cite{schulze2013} evaluate code clone detection tools for robustness against code obfuscations. Schulze and Meyer applied 5 obfuscations to source code, which could be considered to be representative of plagiarism. However, they only evaluated 3 tools (one of which was JPlag), and the focus of this work was for code clone detection. Hence, while this work is similar, it was not performed at the same scale as the evaluations in this article; and is focused on measuring the accuracy of tools, not measuring the robustness of tools to transformation. The evaluations performed here are a partial extension of previous work in \cite{cheers2020}. The prior work was focused on showing that existing SCPDTs can be fooled with pervasive plagiarism-hiding transformations. While the prior work shares a common theme to this work in evaluating SCPDTs, they are distinct.

The performed evaluations are based on test data generation with SimPlag. There are two other works which enable test data generation for code similarity evaluations: COAT \citep{ko2017} and ForkSim \citep{svajlenko2013}. COAT is the most similar to SimPlag as it is intended for use in plagiarism detection. However, it only implements 8 obfuscations, only supports C source code, and does not appear to have been release for reuse. ForkSim implements an injection/mutation framework to simulate software development activities. This is designed for use in code clone detection activities. The injection capabilities of this tool can be used to simulate cases of verbatim source code copying. However, it only implements basic transformations (referred to as mutations in ForkSim) to source code, typically additions, deletions or substitutions. The authors of this work have also presented SPPlagiarise \citep{cheers2019}, a similar tool to SimPlag which enables the generation of semantics-preserving variants of a base program. There is overlap in the applied transformations of SimPlag and SPPlagiarise, however, SPPlagiarise emphasises maintaining the correctness of the base program and as such cannot apply some of the transformations used in this work (e.g. shuffling statements) without significant re-engineering. 

The performed evaluations are facilitated by the PrEP evaluation pipeline presented as part of this work. There exists one similar tool proposed by \cite{cebrian2009}, which seeks to benchmark plagiarism detection tools. This is through the automatic generation of test cases for comparison. However, it was designed for the APL2 programming language, which is not known to be supported by any commonly available SCPDT. A similar evaluation pipeline exists for code clone detection. BigCloneBench \citep{svajlenko2015} provides an evaluation pipeline and ground truth data set for comparing code clone detection tools. However, BigCloneBench is designed for code clone detection tasks, and as previously discussed, in SCPD it is difficult to obtain a ground-truth evaluation data set. Hence PrEP integrates SimPlag for the generation of test data. 

\section{Conclusion \& Future Work}

In this article the robustness of 11 SCPDTs to plagiarism-hiding modifications have been evaluated. This was performed through two evaluations that firstly, evaluated robustness to source code transformations, and secondly, evaluated robustness to source code injection. The results of these evaluations demonstrate that while in many cases the evaluated SCPDTs are robust to plagiarism-hiding modifications, there are specific source code transformations in which the evaluated SCPDTs are vulnerable to. This is specifically to the application of transformations that apply fine-grained modifications to the structure of a program. For example, reordering statements, reordering members, swapping expression operand orders, mapping statements to semantics equivalents, and splitting statements. Applying such transformations change the structure of the source code, and often resulted in a large impact on the evaluation of program similarity. Hence, for these transformations in particular the evaluated tools mostly did not show a high degree of robustness.

Overall, the results of the evaluations imply that all non-string-based SCPDTs show comparatively good robustness to plagiarism-hiding source code modifications, and as such are not greatly impacted upon by such modifications. However, when source code modifications are most pervasively applied, the results of these experiments demonstrate that the tool JPlag is the most robust to the evaluated plagiarism-hiding source code transformations, while Plaggie is most robust against the injection of source code fragments. The results of the performed evaluations also suggest there is benefit in the use of evaluating program similarity with PDGs to provide robustness to pervasive applications of plagiarism-hiding modifications. This is attributed to while tools such as JPlag measure the similarity of the structure of two programs, PDGs measure the semantic similarity of programs; hence they are more robust to the applied structural modifications that are representative of undergraduate source code plagiarism. The results of the performed evaluations can be summarised as: 
\begin{enumerate}
    \item String-based tools show poor robustness to any modifications;
    \item Non-string-based approaches demonstrate satisfactory robustness to modifications in typical cases of source code plagiarism; 
    \item JPlag shows the greatest robustness to the evaluated plagiarism-hiding transformations;
    \item Plaggie shows the greatest robustness to the injection of fragments of source code; and
    \item PDG-based tools provide indications of greater robustness against pervasively modified source code.
\end{enumerate}


Three directions of future work have been identified. Firstly, this work is intentionally limited to the evaluation of SCPDTs. However, there exist many other tools which evaluate source code similarity for other domains, such as code clone detection. It would be interesting to evaluate the robustness of code clone detection tools to common source code modifications, and compare the results with SCPDTs. However, this would have to be with a revised experimental method which is fair on both tool types. Secondly, there are many more source code transformations which could be evaluated than the 14 used in Section \ref{s:ev1}. Specifically, it would be interesting to evaluate the effect of much more invasive source code transformations which change the structure and semantics of a program, but retain the original behaviour. Finally, the evaluations indicated that PDGs show robustness in the presence of pervasively modified source code. It would be interesting to perform a much more in-depth exploration of semantic-based, and potentially behaviour-based methods of evaluating source code similarity in the presence of pervasive modifications. For example, to see if such approaches can provide greater robustness to pervasive plagiarism-hiding modifications.

%% file: Fig2.tex
\begin{figure}
    \centering
\begin{tabular}{@{}p{.5\linewidth}@{}p{.5\linewidth}@{}}
\begin{verbatim}
public class Test {
  public static void main(String args[]) {
    PrintStream out = System.out;
    out.println("Hello, World");
  }
\end{verbatim} \\

\textbf{(a)} Sample `Hello World' style program.\\

\begin{verbatim}
public class Test { public static 
void main(String args[]) { System
.out.println("Hello, World"); } }
\end{verbatim} &

\begin{verbatim}
PUBL CLAS IDNT LBRC PUBL STAT VOID 
IDNT LPAR IDNT IDNT LBRK RBRK RPAR 
LBRC IDNT DOTT IDNT DOTT IDNT LPAR 
STRN RPAR RBRC RBRC
\end{verbatim}
\\
    
\textbf{(b)} As plain text. & 
\textbf{(c)} As lexical tokens. \\

\includegraphics[scale=0.75]{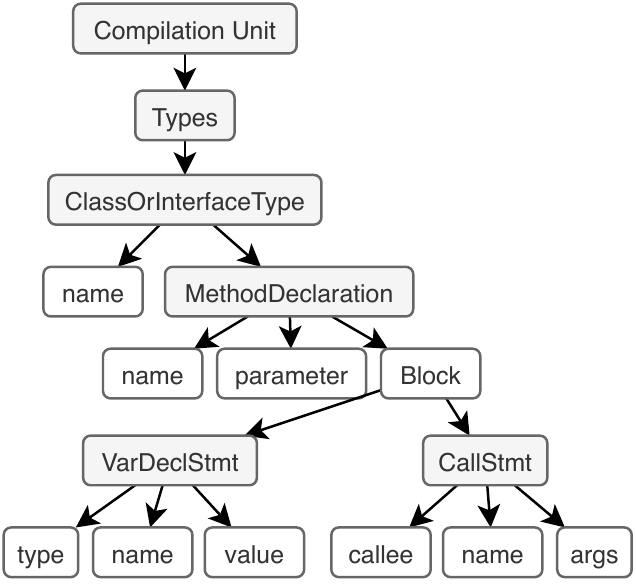} &
\includegraphics[scale=0.75]{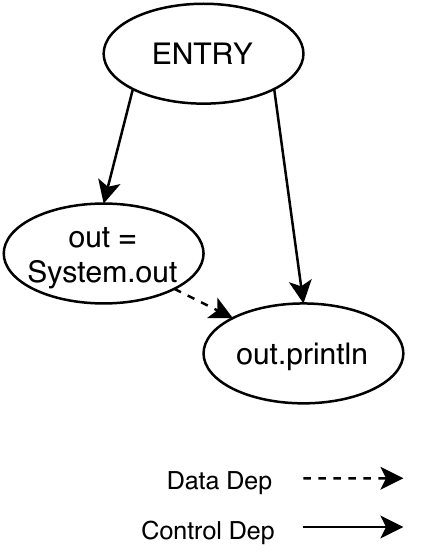} \\

\textbf{(d)} As AST (simplified for brevity) &
\textbf{(e)} As PDG \\

    \end{tabular}
    
    \caption{A Java ``Hello, World'' program representing as text string, token string, abstract syntax tree, and program dependence graph.}
    \label{fig:representation_example}
\end{figure}

%% file: Tab1.tex
\begin{table}[htbp]
    \caption{Naive source code plagiarism detection tools.}
    \label{tab:naivesimilaritytools}
    \begin{tabular}{lll} 
        \hline\noalign{\smallskip}
        {Tool Name} & {Program Representation} & {Similarity Evaluator} \\
        \noalign{\smallskip}\hline\noalign{\smallskip}
        String ED      & Text (String) & Levenshtein Edit Distance \\
        \noalign{\smallskip}
        String Tile    & Text (String) & Greedy String Tiling \\
        \noalign{\smallskip}
        Token ED       & Token         & Levenshtein Edit Distance \\
        \noalign{\smallskip}
        Token Tile     & Token         & Greedy String Tiling \\
        \noalign{\smallskip}
        Tree ED        & AST           & Tree Edit Distance \\
        \noalign{\smallskip}
        Graph ED       & PDG           & Graph Edit Distance \\
        \noalign{\smallskip}\hline
    \end{tabular}
\end{table}

%% file: Tab2.tex
\begin{table}[htbp]
\caption{Source code transformations implemented in SimPlag.}
\label{tab:simplagmutationoperators}
\begin{tabular}{lll} 
    \hline\noalign{\smallskip}
    {Code} & {Source Code Transformation} & {Node(s) of Interest} \\
    \noalign{\smallskip}\hline\noalign{\smallskip}
    tAC & Add comments & Classes, methods, fields, statements \\
    \noalign{\smallskip}
    tRC & Remove comments & Classes, methods, fields, statements \\
    \noalign{\smallskip}
    tMC & Mutate comments & Classes, methods, fields, statements \\
    \noalign{\smallskip}
    tRI & Rename identifiers & Identifiers \\
    \noalign{\smallskip}
    tRS & Reorder statements (within methods) & Block statements \\
    \noalign{\smallskip}
    tRM & Reorder class member declarations & Classes \\
    \noalign{\smallskip}
    rSO & Reorder expression operands & Binary expressions \\
    \noalign{\smallskip}
    tUD & Up-cast primitive types & Primitive type names \\
    \noalign{\smallskip}
    tFW & Swap \textit{for} statement to \textit{while} statement & For statements \\
    \noalign{\smallskip}
    tEA & Expand compound assignment expressions & Compound assignment expressions \\
    \noalign{\smallskip}
    tEU & Expand unary operator expressions & Unary expressions \\
    \noalign{\smallskip}
    tSV & Split group variable declarations & Group variable declarations \\
    \noalign{\smallskip}
    tAD & Assign default value to variable declaration & Variable declarations \\
    \noalign{\smallskip}
    tSD & Split variable declaration and assignment & Variable declarations \\
    \noalign{\smallskip}\hline
\end{tabular}
\end{table}

%% file: Tab3.tex
\begin{table}[htbp]
\caption{Base evaluation data set overview.}
\label{table:assignmentstats}
\begin{tabular}{cccccccc} 
    \hline\noalign{\smallskip}
    \multirow{2}{*}{\shortstack[c]{Assignment\\ Set}} & \multirow{2}{*}{\shortstack[c]{Year\\ Level}} & \multirow{2}{*}{\shortstack[c]{Assignment\\ Count}} & \multicolumn{5}{c}{Average} \\
     & & & {LLOC} & {Files} & {Classes} & {Methods} & {Statements} \\
    \noalign{\smallskip}\hline\noalign{\smallskip}
    AS1 & 1 & 223 & 389.39 & 3.72 & 3.71 & 34.43 & 500.29 \\ 
    \noalign{\smallskip}
    AS2 & 1 & 173 & 396.87 & 3.76 & 3.77 & 46.19 & 525.38 \\   
    \noalign{\smallskip}
    AS3 & 2 & 73 & 225.03 & 5.03 & 5.15 & 29.42 & 294.74 \\  
    \noalign{\smallskip}
    AS4 & 2 & 72 & 227.43 & 5.76 & 5.88 & 39.94 & 312.44 \\   
    \noalign{\smallskip}
    AS5 & 3 & 17 & 46.59 & 2.65 & 2.71 & 9.18 & 165.65 \\ 
    \noalign{\smallskip}
    AS6 & 3 & 17 & 242.82 & 13.47 & 13.29 & 57.53 & 352.06 \\     
    \noalign{\smallskip}\hline\noalign{\smallskip}
    {Total} & - & 575 & 336.03 & 4.41 & 4.44 & 37.96 & 440.99 \\
    \noalign{\smallskip}\hline
\end{tabular}
\end{table}

%% file: Fig5_ev1a_heatmap.tex
\begin{figure}[htbp]
    \begin{tabular}{p{.45\linewidth}p{.45\linewidth}}
         
        \resizebox{.45\textwidth}{!}{
            \includegraphics[width=1.0\linewidth]{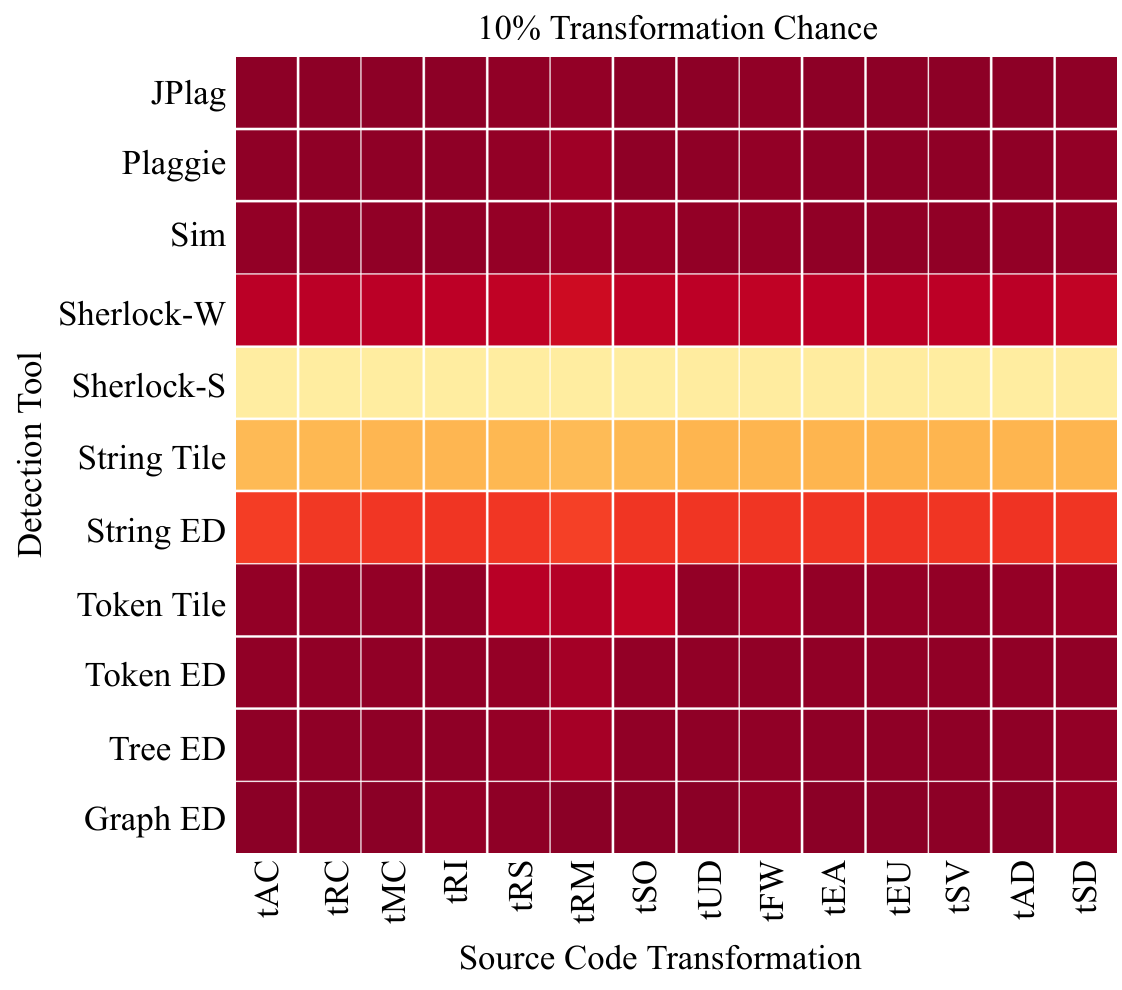}  
        } &
    
        \resizebox{.45\textwidth}{!}{
            \includegraphics[width=1.0\linewidth]{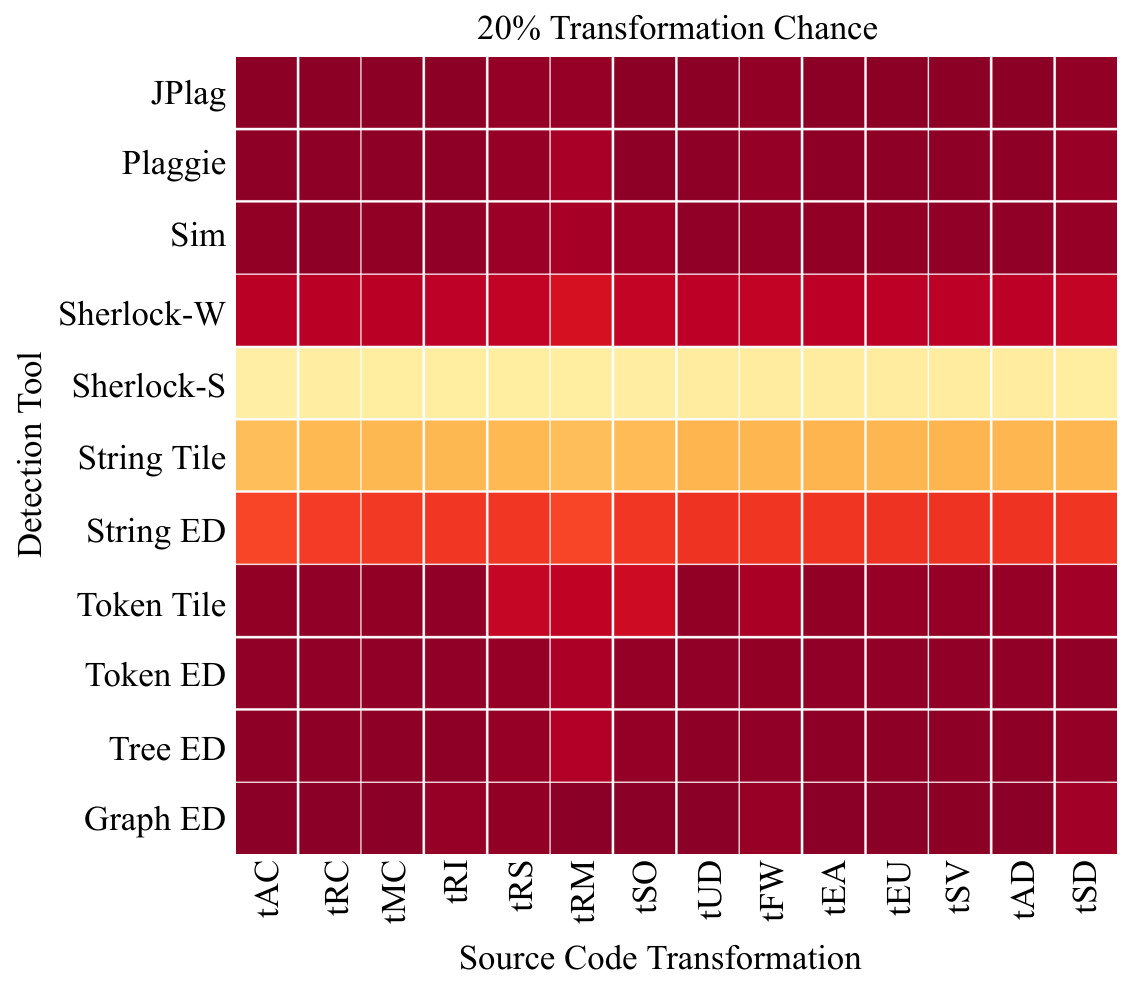}  
        } \\
        
        \resizebox{.45\textwidth}{!}{
            \includegraphics[width=1.0\linewidth]{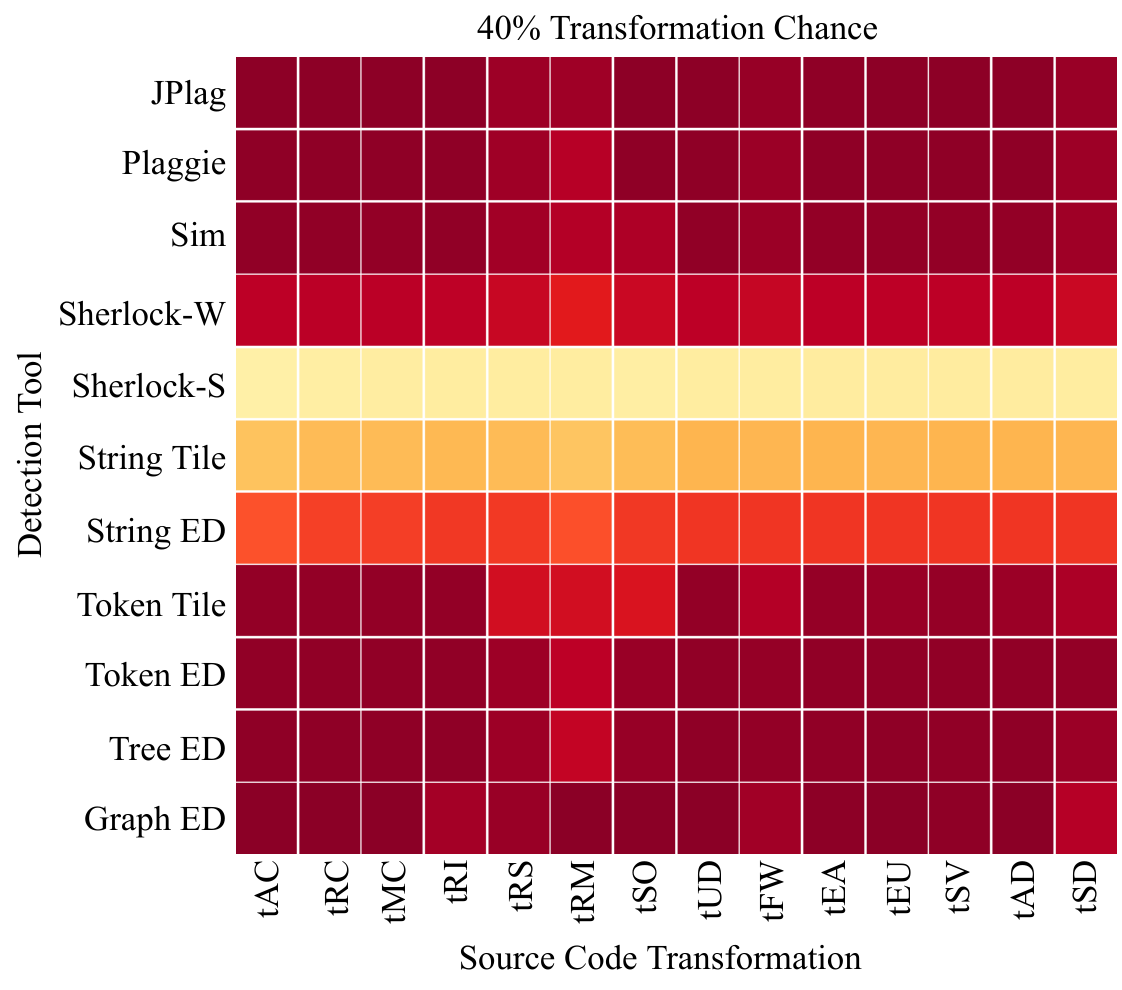}  
        } &
    
        \resizebox{.45\textwidth}{!}{
            \includegraphics[width=1.0\linewidth]{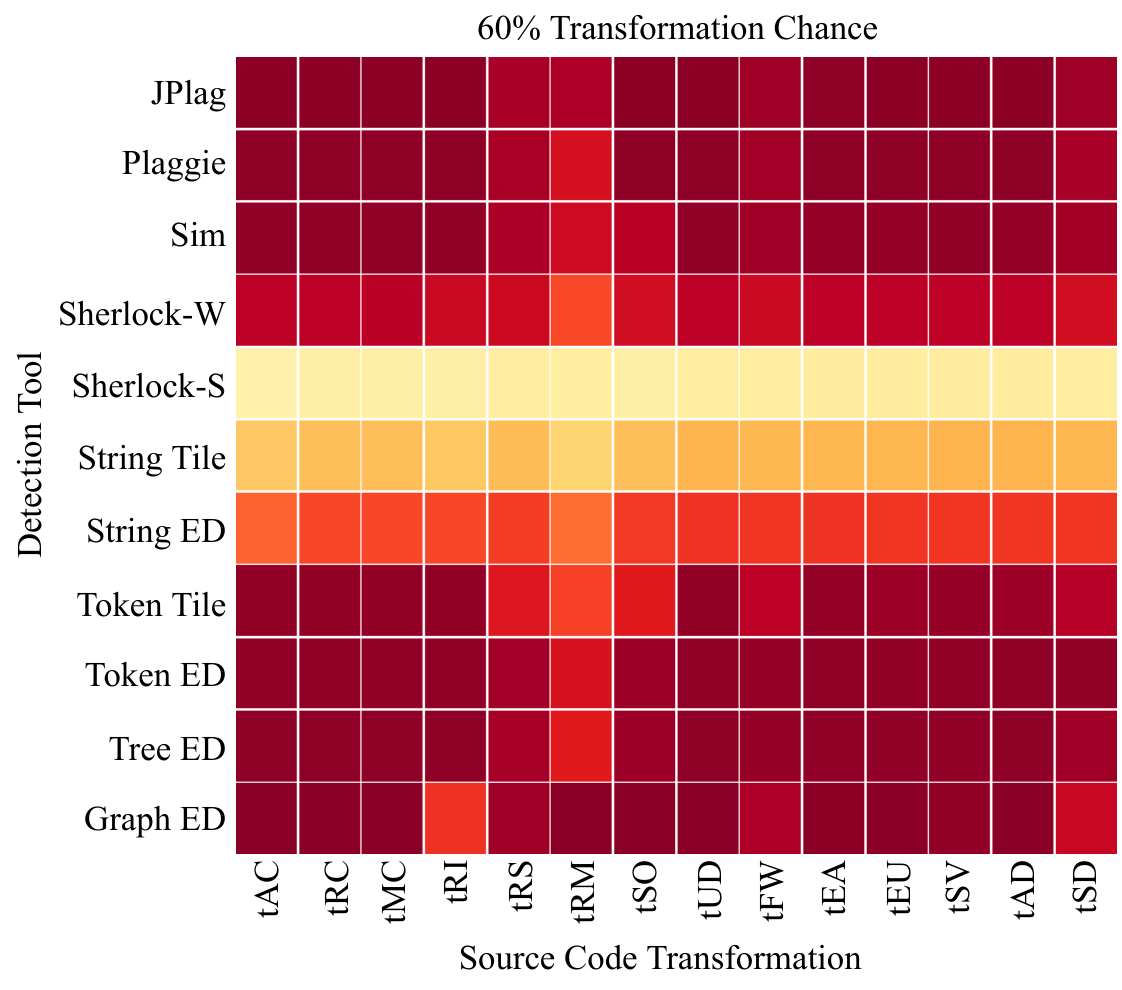}  
        } \\
        
        \resizebox{.45\textwidth}{!}{
            \includegraphics[width=1.0\linewidth]{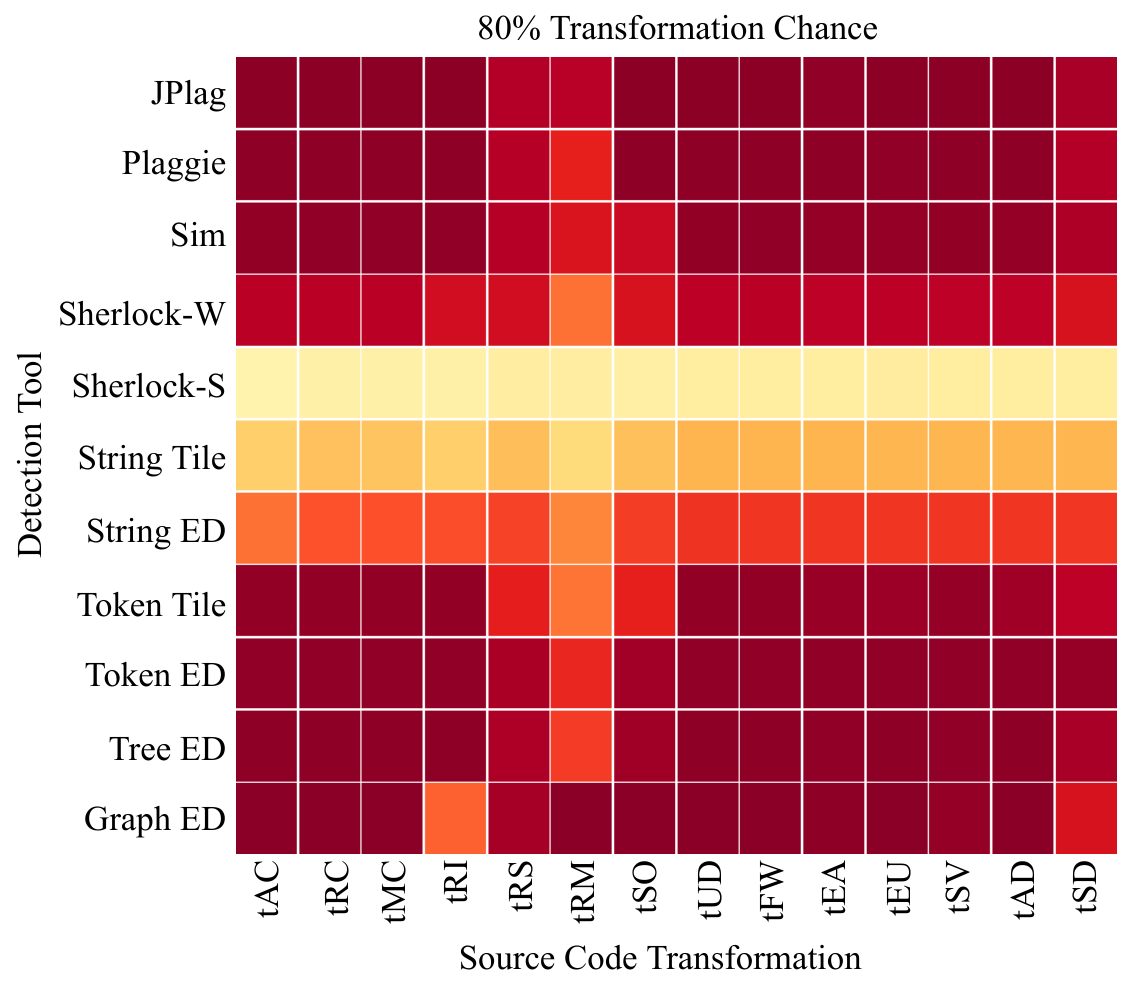}  
        } &
    
        \resizebox{.45\textwidth}{!}{
            \includegraphics[width=1.0\linewidth]{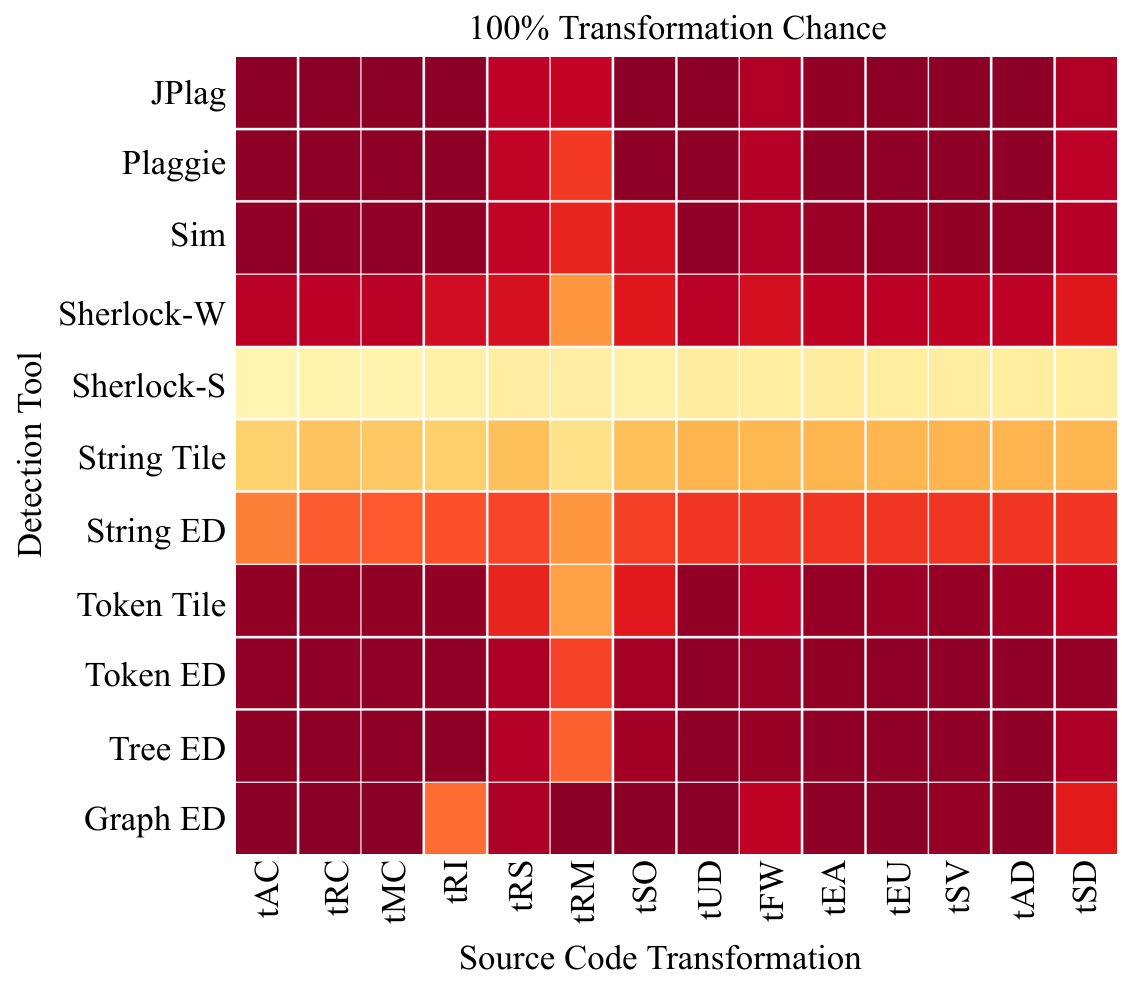}  
        } \\
        
        \multicolumn{2}{c}{
        \resizebox{.45\textwidth}{!}{
            \includegraphics[width=1.0\linewidth]{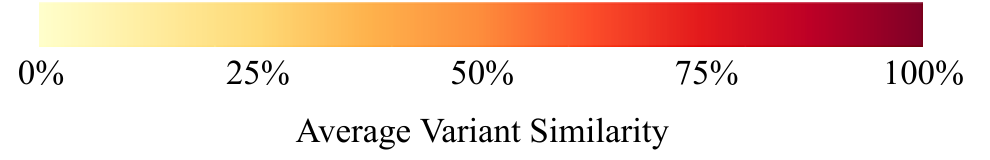} 
        }} \\
        
    \end{tabular}

    \caption{Heat maps representing the average similarity of variants generated with each individual source code source code transformation. Darker colours indicate a higher similarity scores, and hence higher robustness to the applied transformation.}
    \label{fig:eb2aheatmaps}
\end{figure}

%% file: Tab4_ev1a_trn_ranking.tex
\begin{table}[htbp]
    \caption{Ranking of transformations applied to non-string-based SCPDTs at the 100\% transformation chance. Transformations are ranked left (greatest decrease) to right (least decrease). Horizontal bars `|' delimit transformations thath have negligible decrease in similarity. }
    \label{tab:ev2atrnrankings}
    \begin{tabular}{rl}
\hline\noalign{\smallskip}
        {Tool} & {Individual Transformation Rankings} \\
\noalign{\smallskip}\hline\noalign{\smallskip}
        {JPlag}      & {tRM} {tRS} {tFW} {tSD} \textbf{|} tEA tRI tEU tSV {tSO} tAD tRC tUD tMC tAC \\
        \noalign{\smallskip}
        {Plaggie}    & {tRM} {tRS} {tSD} {tFW} \textbf{|} tEU tRC tRI {tSO} tSV tEA tMC tUD tAD tAC \\
        \noalign{\smallskip}
        {Sim}        & {tRM} {tSO} {tRS} {tSD} {tFW} tEA \textbf{|} tEU tAD tRI tSV tRC tMC tAC tUD \\
        \noalign{\smallskip}
        {Token Tile} & {tRM} {tRS} {tSO} {tSD} {tFW} \textbf{|} tAD tEU tEA tSV tUD tAC tMC tRI tRC \\
        \noalign{\smallskip}
        {Token ED}   & {tRM} {tRS} {tSO} {tFW} \textbf{|} {tSD} tEA tAD tSV tEU tRI tMC tUD tRC tAC \\
        \noalign{\smallskip}
        {Tree ED}    & {tRM} {tRS} {tSD} {tSO} {tFW} \textbf{|} tEA tSV tEU tAD tRC tMC tRI tUD tAC \\
        \noalign{\smallskip}
        {Graph ED}   & tRI {tSD} {tFW} {tRS} \textbf{|} tSV tEA {tSO} tEU tUD tAD tRC tAC {tRM} tMC \\
\noalign{\smallskip}\hline
    \end{tabular}
\end{table}

%% file: Fig6_ev1ax_avgsimilarity.tex
\begin{figure}[htbp]

    \begin{tabular}{p{.5\linewidth}p{.5\linewidth}}
         
        \resizebox{.45\textwidth}{!}{
            \begin{tikzpicture}
            \begin{axis}[
                title={10\% Transformation Chance},
                ylabel={Similarity (\%)},
                xmin=-1, xmax=11,
                ymin=0.0, ymax=100,
                xtick={0,1,2,3,4,5,6,7,8,9,10},
                xticklabel style={rotate=90},
                xticklabels={
                    JPlag,Plaggie,Sim,Sherlock-W,Sherlock-S,String Tile,String ED,Token Tile,Token ED,Tree ED,Graph ED
                }
            ]
            
\addplot[mark=*,black] coordinates { (0,93.60) };
\addplot[mark=-,black] coordinates { (0,100.00)(0,17.28) };
\addplot[mark=*,red] coordinates { (0,100.00) };
\addplot[mark=*,red] coordinates { (0,84.07) };

\addplot[mark=*,black] coordinates { (10,92.34) };
\addplot[mark=-,black] coordinates { (10,100.00)(10,0.00) };
\addplot[mark=*,red] coordinates { (10,100.00) };
\addplot[mark=*,red] coordinates { (10,82.10) };

\addplot[mark=*,black] coordinates { (6,65.79) };
\addplot[mark=-,black] coordinates { (6,94.60)(6,29.89) };
\addplot[mark=*,red] coordinates { (6,75.99) };
\addplot[mark=*,red] coordinates { (6,55.59) };

\addplot[mark=*,black] coordinates { (5,33.02) };
\addplot[mark=-,black] coordinates { (5,78.37)(5,5.00) };
\addplot[mark=*,red] coordinates { (5,45.51) };
\addplot[mark=*,red] coordinates { (5,20.53) };

\addplot[mark=*,black] coordinates { (8,91.98) };
\addplot[mark=-,black] coordinates { (8,100.00)(8,0.00) };
\addplot[mark=*,red] coordinates { (8,100.00) };
\addplot[mark=*,red] coordinates { (8,80.12) };

\addplot[mark=*,black] coordinates { (7,76.72) };
\addplot[mark=-,black] coordinates { (7,100.00)(7,0.00) };
\addplot[mark=*,red] coordinates { (7,94.01) };
\addplot[mark=*,red] coordinates { (7,59.42) };

\addplot[mark=*,black] coordinates { (9,90.82) };
\addplot[mark=-,black] coordinates { (9,100.00)(9,30.33) };
\addplot[mark=*,red] coordinates { (9,100.00) };
\addplot[mark=*,red] coordinates { (9,80.67) };

\addplot[mark=*,black] coordinates { (1,91.38) };
\addplot[mark=-,black] coordinates { (1,100.00)(1,0.00) };
\addplot[mark=*,red] coordinates { (1,100.00) };
\addplot[mark=*,red] coordinates { (1,79.30) };

\addplot[mark=*,black] coordinates { (4,12.37) };
\addplot[mark=-,black] coordinates { (4,56.00)(4,0.00) };
\addplot[mark=*,red] coordinates { (4,22.30) };
\addplot[mark=*,red] coordinates { (4,2.44) };

\addplot[mark=*,black] coordinates { (3,79.20) };
\addplot[mark=-,black] coordinates { (3,100.00)(3,23.25) };
\addplot[mark=*,red] coordinates { (3,92.61) };
\addplot[mark=*,red] coordinates { (3,65.79) };

\addplot[mark=*,black] coordinates { (2,90.50) };
\addplot[mark=-,black] coordinates { (2,100.00)(2,20.25) };
\addplot[mark=*,red] coordinates { (2,100.00) };
\addplot[mark=*,red] coordinates { (2,79.75) };
            
            \end{axis}
            \end{tikzpicture}  
        } &
    
    \resizebox{.45\textwidth}{!}{
            \begin{tikzpicture}
            \begin{axis}[
                title={20\% Transformation Chance},
                ylabel={Similarity (\%)},
                xmin=-1, xmax=11,
                ymin=0.0, ymax=100,
                xtick={0,1,2,3,4,5,6,7,8,9,10},
                xticklabel style={rotate=90},
                xticklabels={
                    JPlag,Plaggie,Sim,Sherlock-W,Sherlock-S,String Tile,String ED,Token Tile,Token ED,Tree ED,Graph ED
                }
            ]
            
\addplot[mark=*,black] coordinates { (0,89.74) };
\addplot[mark=-,black] coordinates { (0,100.00)(0,16.76) };
\addplot[mark=*,red] coordinates { (0,99.83) };
\addplot[mark=*,red] coordinates { (0,79.65) };

\addplot[mark=*,black] coordinates { (10,88.24) };
\addplot[mark=-,black] coordinates { (10,100.00)(10,0.00) };
\addplot[mark=*,red] coordinates { (10,100.00) };
\addplot[mark=*,red] coordinates { (10,76.16) };

\addplot[mark=*,black] coordinates { (6,62.77) };
\addplot[mark=-,black] coordinates { (6,94.60)(6,25.75) };
\addplot[mark=*,red] coordinates { (6,73.11) };
\addplot[mark=*,red] coordinates { (6,52.43) };

\addplot[mark=*,black] coordinates { (5,30.43) };
\addplot[mark=-,black] coordinates { (5,76.78)(5,3.53) };
\addplot[mark=*,red] coordinates { (5,43.06) };
\addplot[mark=*,red] coordinates { (5,17.80) };

\addplot[mark=*,black] coordinates { (8,87.50) };
\addplot[mark=-,black] coordinates { (8,100.00)(8,0.00) };
\addplot[mark=*,red] coordinates { (8,99.81) };
\addplot[mark=*,red] coordinates { (8,75.19) };

\addplot[mark=*,black] coordinates { (7,67.07) };
\addplot[mark=-,black] coordinates { (7,100.00)(7,0.00) };
\addplot[mark=*,red] coordinates { (7,85.41) };
\addplot[mark=*,red] coordinates { (7,48.73) };

\addplot[mark=*,black] coordinates { (9,85.06) };
\addplot[mark=-,black] coordinates { (9,100.00)(9,26.88) };
\addplot[mark=*,red] coordinates { (9,96.46) };
\addplot[mark=*,red] coordinates { (9,73.66) };

\addplot[mark=*,black] coordinates { (1,85.57) };
\addplot[mark=-,black] coordinates { (1,100.00)(1,0.00) };
\addplot[mark=*,red] coordinates { (1,99.66) };
\addplot[mark=*,red] coordinates { (1,71.47) };

\addplot[mark=*,black] coordinates { (4,12.01) };
\addplot[mark=-,black] coordinates { (4,55.25)(4,0.00) };
\addplot[mark=*,red] coordinates { (4,21.60) };
\addplot[mark=*,red] coordinates { (4,2.42) };

\addplot[mark=*,black] coordinates { (3,71.32) };
\addplot[mark=-,black] coordinates { (3,100.00)(3,11.75) };
\addplot[mark=*,red] coordinates { (3,85.87) };
\addplot[mark=*,red] coordinates { (3,56.78) };

\addplot[mark=*,black] coordinates { (2,84.14) };
\addplot[mark=-,black] coordinates { (2,100.00)(2,18.50) };
\addplot[mark=*,red] coordinates { (2,96.40) };
\addplot[mark=*,red] coordinates { (2,71.88) };
            
            \end{axis}
            \end{tikzpicture}  
        } \\
        
                \resizebox{.45\textwidth}{!}{
            \begin{tikzpicture}
            \begin{axis}[
                title={40\% Transformation Chance},
                ylabel={Similarity (\%)},
                xmin=-1, xmax=11,
                ymin=0.0, ymax=100,
                xtick={0,1,2,3,4,5,6,7,8,9,10},
                xticklabel style={rotate=90},
                xticklabels={
                    JPlag,Plaggie,Sim,Sherlock-W,Sherlock-S,String Tile,String ED,Token Tile,Token ED,Tree ED,Graph ED
                }
            ]
            
\addplot[mark=*,black] coordinates { (0,82.97) };
\addplot[mark=-,black] coordinates { (0,100.00)(0,15.03) };
\addplot[mark=*,red] coordinates { (0,94.06) };
\addplot[mark=*,red] coordinates { (0,71.87) };

\addplot[mark=*,black] coordinates { (10,82.40) };
\addplot[mark=-,black] coordinates { (10,100.00)(10,0.00) };
\addplot[mark=*,red] coordinates { (10,96.44) };
\addplot[mark=*,red] coordinates { (10,68.35) };

\addplot[mark=*,black] coordinates { (6,57.86) };
\addplot[mark=-,black] coordinates { (6,94.60)(6,25.95) };
\addplot[mark=*,red] coordinates { (6,68.41) };
\addplot[mark=*,red] coordinates { (6,47.32) };

\addplot[mark=*,black] coordinates { (5,27.19) };
\addplot[mark=-,black] coordinates { (5,77.21)(5,3.37) };
\addplot[mark=*,red] coordinates { (5,39.69) };
\addplot[mark=*,red] coordinates { (5,14.69) };

\addplot[mark=*,black] coordinates { (8,80.36) };
\addplot[mark=-,black] coordinates { (8,100.00)(8,0.00) };
\addplot[mark=*,red] coordinates { (8,93.33) };
\addplot[mark=*,red] coordinates { (8,67.38) };

\addplot[mark=*,black] coordinates { (7,56.62) };
\addplot[mark=-,black] coordinates { (7,100.00)(7,0.00) };
\addplot[mark=*,red] coordinates { (7,75.89) };
\addplot[mark=*,red] coordinates { (7,37.36) };

\addplot[mark=*,black] coordinates { (9,75.82) };
\addplot[mark=-,black] coordinates { (9,100.00)(9,18.23) };
\addplot[mark=*,red] coordinates { (9,88.73) };
\addplot[mark=*,red] coordinates { (9,62.90) };

\addplot[mark=*,black] coordinates { (1,75.62) };
\addplot[mark=-,black] coordinates { (1,100.00)(1,0.00) };
\addplot[mark=*,red] coordinates { (1,92.50) };
\addplot[mark=*,red] coordinates { (1,58.74) };

\addplot[mark=*,black] coordinates { (4,11.33) };
\addplot[mark=-,black] coordinates { (4,54.00)(4,0.00) };
\addplot[mark=*,red] coordinates { (4,20.44) };
\addplot[mark=*,red] coordinates { (4,2.22) };

\addplot[mark=*,black] coordinates { (3,59.25) };
\addplot[mark=-,black] coordinates { (3,100.00)(3,11.25) };
\addplot[mark=*,red] coordinates { (3,75.30) };
\addplot[mark=*,red] coordinates { (3,43.19) };

\addplot[mark=*,black] coordinates { (2,73.58) };
\addplot[mark=-,black] coordinates { (2,100.00)(2,18.25) };
\addplot[mark=*,red] coordinates { (2,87.84) };
\addplot[mark=*,red] coordinates { (2,59.33) };
            
            \end{axis}
            \end{tikzpicture}  
        } &
    
    \resizebox{.45\textwidth}{!}{
            \begin{tikzpicture}
            \begin{axis}[
                title={60\% Transformation Chance},
                ylabel={Similarity (\%)},
                xmin=-1, xmax=11,
                ymin=0.0, ymax=100,
                xtick={0,1,2,3,4,5,6,7,8,9,10},
                xticklabel style={rotate=90},
                xticklabels={
                    JPlag,Plaggie,Sim,Sherlock-W,Sherlock-S,String Tile,String ED,Token Tile,Token ED,Tree ED,Graph ED
                }
            ]
            
\addplot[mark=*,black] coordinates { (0,75.33) };
\addplot[mark=-,black] coordinates { (0,100.00)(0,14.51) };
\addplot[mark=*,red] coordinates { (0,87.55) };
\addplot[mark=*,red] coordinates { (0,63.11) };

\addplot[mark=*,black] coordinates { (10,77.75) };
\addplot[mark=-,black] coordinates { (10,100.00)(10,0.00) };
\addplot[mark=*,red] coordinates { (10,92.50) };
\addplot[mark=*,red] coordinates { (10,62.99) };

\addplot[mark=*,black] coordinates { (6,52.84) };
\addplot[mark=-,black] coordinates { (6,92.81)(6,26.57) };
\addplot[mark=*,red] coordinates { (6,63.19) };
\addplot[mark=*,red] coordinates { (6,42.48) };

\addplot[mark=*,black] coordinates { (5,23.80) };
\addplot[mark=-,black] coordinates { (5,74.64)(5,3.60) };
\addplot[mark=*,red] coordinates { (5,35.93) };
\addplot[mark=*,red] coordinates { (5,11.67) };

\addplot[mark=*,black] coordinates { (8,73.21) };
\addplot[mark=-,black] coordinates { (8,100.00)(8,0.00) };
\addplot[mark=*,red] coordinates { (8,86.04) };
\addplot[mark=*,red] coordinates { (8,60.38) };

\addplot[mark=*,black] coordinates { (7,47.46) };
\addplot[mark=-,black] coordinates { (7,100.00)(7,0.00) };
\addplot[mark=*,red] coordinates { (7,65.68) };
\addplot[mark=*,red] coordinates { (7,29.24) };

\addplot[mark=*,black] coordinates { (9,66.61) };
\addplot[mark=-,black] coordinates { (9,100.00)(9,22.31) };
\addplot[mark=*,red] coordinates { (9,79.62) };
\addplot[mark=*,red] coordinates { (9,53.61) };

\addplot[mark=*,black] coordinates { (1,64.23) };
\addplot[mark=-,black] coordinates { (1,100.00)(1,0.00) };
\addplot[mark=*,red] coordinates { (1,82.51) };
\addplot[mark=*,red] coordinates { (1,45.96) };

\addplot[mark=*,black] coordinates { (4,10.70) };
\addplot[mark=-,black] coordinates { (4,45.67)(4,0.00) };
\addplot[mark=*,red] coordinates { (4,19.33) };
\addplot[mark=*,red] coordinates { (4,2.07) };

\addplot[mark=*,black] coordinates { (3,48.44) };
\addplot[mark=-,black] coordinates { (3,95.25)(3,7.50) };
\addplot[mark=*,red] coordinates { (3,64.47) };
\addplot[mark=*,red] coordinates { (3,32.42) };

\addplot[mark=*,black] coordinates { (2,61.47) };
\addplot[mark=-,black] coordinates { (2,100.00)(2,6.50) };
\addplot[mark=*,red] coordinates { (2,76.66) };
\addplot[mark=*,red] coordinates { (2,46.27) };
            
            \end{axis}
            \end{tikzpicture}  
        } \\
        
                \resizebox{.45\textwidth}{!}{
            \begin{tikzpicture}
            \begin{axis}[
                title={80\% Transformation Chance},
                ylabel={Similarity (\%)},
                xmin=-1, xmax=11,
                ymin=0.0, ymax=100,
                xtick={0,1,2,3,4,5,6,7,8,9,10},
                xticklabel style={rotate=90},
                xticklabels={
                    JPlag,Plaggie,Sim,Sherlock-W,Sherlock-S,String Tile,String ED,Token Tile,Token ED,Tree ED,Graph ED
                }
            ]
            
\addplot[mark=*,black] coordinates { (0,67.35) };
\addplot[mark=-,black] coordinates { (0,98.95)(0,12.49) };
\addplot[mark=*,red] coordinates { (0,80.71) };
\addplot[mark=*,red] coordinates { (0,54.00) };

\addplot[mark=*,black] coordinates { (10,74.14) };
\addplot[mark=-,black] coordinates { (10,100.00)(10,0.00) };
\addplot[mark=*,red] coordinates { (10,89.65) };
\addplot[mark=*,red] coordinates { (10,58.62) };

\addplot[mark=*,black] coordinates { (6,47.62) };
\addplot[mark=-,black] coordinates { (6,90.18)(6,22.62) };
\addplot[mark=*,red] coordinates { (6,57.24) };
\addplot[mark=*,red] coordinates { (6,37.99) };

\addplot[mark=*,black] coordinates { (5,20.30) };
\addplot[mark=-,black] coordinates { (5,74.06)(5,2.06) };
\addplot[mark=*,red] coordinates { (5,31.91) };
\addplot[mark=*,red] coordinates { (5,8.70) };

\addplot[mark=*,black] coordinates { (8,66.07) };
\addplot[mark=-,black] coordinates { (8,97.43)(8,0.00) };
\addplot[mark=*,red] coordinates { (8,78.30) };
\addplot[mark=*,red] coordinates { (8,53.85) };

\addplot[mark=*,black] coordinates { (7,38.54) };
\addplot[mark=-,black] coordinates { (7,91.82)(7,0.00) };
\addplot[mark=*,red] coordinates { (7,55.77) };
\addplot[mark=*,red] coordinates { (7,21.30) };

\addplot[mark=*,black] coordinates { (9,57.41) };
\addplot[mark=-,black] coordinates { (9,94.84)(9,25.66) };
\addplot[mark=*,red] coordinates { (9,69.70) };
\addplot[mark=*,red] coordinates { (9,45.13) };

\addplot[mark=*,black] coordinates { (1,52.46) };
\addplot[mark=-,black] coordinates { (1,100.00)(1,0.00) };
\addplot[mark=*,red] coordinates { (1,70.75) };
\addplot[mark=*,red] coordinates { (1,34.17) };

\addplot[mark=*,black] coordinates { (4,10.13) };
\addplot[mark=-,black] coordinates { (4,45.67)(4,0.00) };
\addplot[mark=*,red] coordinates { (4,18.39) };
\addplot[mark=*,red] coordinates { (4,1.87) };

\addplot[mark=*,black] coordinates { (3,38.26) };
\addplot[mark=-,black] coordinates { (3,92.50)(3,0.00) };
\addplot[mark=*,red] coordinates { (3,53.32) };
\addplot[mark=*,red] coordinates { (3,23.19) };

\addplot[mark=*,black] coordinates { (2,49.16) };
\addplot[mark=-,black] coordinates { (2,96.00)(2,6.00) };
\addplot[mark=*,red] coordinates { (2,64.14) };
\addplot[mark=*,red] coordinates { (2,34.17) };

            \end{axis}
            \end{tikzpicture}  
        } &
    
    \resizebox{.45\textwidth}{!}{
            \begin{tikzpicture}
            \begin{axis}[
                title={100\% Transformation Chance},
                ylabel={Similarity (\%)},
                xmin=-1, xmax=11,
                ymin=0.0, ymax=100,
                xtick={0,1,2,3,4,5,6,7,8,9,10},
                xticklabel style={rotate=90},
                xticklabels={
                    JPlag,Plaggie,Sim,Sherlock-W,Sherlock-S,String Tile,String ED,Token Tile,Token ED,Tree ED,Graph ED
                }
            ]
            
\addplot[mark=*,black] coordinates { (0,59.55) };
\addplot[mark=-,black] coordinates { (0,89.99)(0,12.25) };
\addplot[mark=*,red] coordinates { (0,73.61) };
\addplot[mark=*,red] coordinates { (0,45.49) };

\addplot[mark=*,black] coordinates { (10,71.03) };
\addplot[mark=-,black] coordinates { (10,100.00)(10,0.00) };
\addplot[mark=*,red] coordinates { (10,87.07) };
\addplot[mark=*,red] coordinates { (10,54.99) };

\addplot[mark=*,black] coordinates { (6,42.50) };
\addplot[mark=-,black] coordinates { (6,83.91)(6,22.31) };
\addplot[mark=*,red] coordinates { (6,50.67) };
\addplot[mark=*,red] coordinates { (6,34.32) };

\addplot[mark=*,black] coordinates { (5,16.94) };
\addplot[mark=-,black] coordinates { (5,71.31)(5,2.25) };
\addplot[mark=*,red] coordinates { (5,27.60) };
\addplot[mark=*,red] coordinates { (5,6.27) };

\addplot[mark=*,black] coordinates { (8,59.09) };
\addplot[mark=-,black] coordinates { (8,91.97)(8,0.00) };
\addplot[mark=*,red] coordinates { (8,70.04) };
\addplot[mark=*,red] coordinates { (8,48.13) };

\addplot[mark=*,black] coordinates { (7,30.77) };
\addplot[mark=-,black] coordinates { (7,80.47)(7,0.00) };
\addplot[mark=*,red] coordinates { (7,45.99) };
\addplot[mark=*,red] coordinates { (7,15.55) };

\addplot[mark=*,black] coordinates { (9,48.43) };
\addplot[mark=-,black] coordinates { (9,85.80)(9,21.48) };
\addplot[mark=*,red] coordinates { (9,58.88) };
\addplot[mark=*,red] coordinates { (9,37.98) };

\addplot[mark=*,black] coordinates { (1,41.21) };
\addplot[mark=-,black] coordinates { (1,87.63)(1,0.00) };
\addplot[mark=*,red] coordinates { (1,58.64) };
\addplot[mark=*,red] coordinates { (1,23.78) };

\addplot[mark=*,black] coordinates { (4,9.46) };
\addplot[mark=-,black] coordinates { (4,39.75)(4,0.00) };
\addplot[mark=*,red] coordinates { (4,17.25) };
\addplot[mark=*,red] coordinates { (4,1.68) };

\addplot[mark=*,black] coordinates { (3,29.46) };
\addplot[mark=-,black] coordinates { (3,84.50)(3,3.75) };
\addplot[mark=*,red] coordinates { (3,42.12) };
\addplot[mark=*,red] coordinates { (3,16.81) };

\addplot[mark=*,black] coordinates { (2,36.79) };
\addplot[mark=-,black] coordinates { (2,81.75)(2,1.25) };
\addplot[mark=*,red] coordinates { (2,50.28) };
\addplot[mark=*,red] coordinates { (2,23.29) };
            
            \end{axis}
            \end{tikzpicture}  
        } \\
        
    \end{tabular}

    \caption{Average similarity of variants generated with the 5 identified source code transformations, evaluated using all 11 SCPDTs. Bars indicate range of similarity scores. Red marks indicate range of standard deviation around the average.}
    \label{fig:ev2axavgsimilarity}
\end{figure}

%% file: Tab5.tex
\begin{table}[htbp]
    \caption{Average number of applied transformations to each program variant created with the 5 identified source code transformations.}
    \label{tab:ev2axavgtransformations}
    \begin{tabular}{p{4cm}|cccccc}
        \hline\noalign{\smallskip}
        {Transformation Chance} & {10\%} & {20\%} & {40\%} & {60\%} & {80\%} & {100\%} \\
        \noalign{\smallskip}
        \cline{2-7}
        \noalign{\smallskip}
        {Avg. No. of Transformations} & 12.79 & 23.43 & 41.37 & 58.09 & 74.04 & 89.33 \\
        \noalign{\smallskip}\hline
    \end{tabular}
\end{table}

%% file: Fig7_ev1ax_resilience.tex
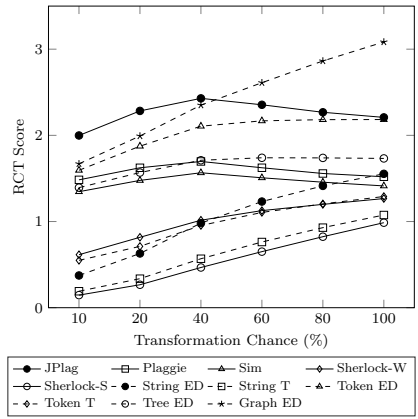
\begin{figure}[htbp]
    \centering
    \resizebox{.45\linewidth}{!}{
     \begin{tikzpicture}
        \begin{axis}[
            xlabel={Transformation Chance (\%)},
            ylabel={RCT Score},
            xmin=0.5, xmax=6.5,
            ymin=0.0, ymax=3.5,
            xtick={0,1,2,3,4,5,6,7},
            xticklabels={
                ,10,20,40,60,80,100
            }
        ]
            
	\addplot[mark=*,black] coordinates {
        (1,1.9984375)
        (2,2.283625731)
        (3,2.429242513)
        (4,2.3546818)
        (5,2.267687596)
        (6,2.208405439)
	};

	\addplot[dashed,color=black,mark=star,mark options={solid}] coordinates {
        (1,1.669712794)
        (2,1.992346939)
        (3,2.350568182)
        (4,2.610786517)
        (5,2.863109049)
        (6,3.083534691)
	};

	\addplot[dashed,color=black,mark=*,mark options={solid}] coordinates {
        (1,0.37386729)
        (2,0.629331185)
        (3,0.981727575)
        (4,1.231764207)
        (5,1.413516609)
        (6,1.553565217)
	};

	\addplot[dashed,color=black,mark=square,mark options={solid}] coordinates {
        (1,0.190952523)
        (2,0.336783096)
        (3,0.568191183)
        (4,0.762335958)
        (5,0.928983689)
        (6,1.075487599)
	};

	\addplot[dashed,color=black,mark=triangle,mark options={solid}] coordinates {
        (1,1.594763092)
        (2,1.8744)
        (3,2.106415479)
        (4,2.168346398)
        (5,2.182139699)
        (6,2.183573698)
	};

	\addplot[dashed,color=black,mark=diamond,mark options={solid}] coordinates {
        (1,0.549398625)
        (2,0.711509262)
        (3,0.953665284)
        (4,1.105633803)
        (5,1.204685975)
        (6,1.290336559)
	};

	\addplot[dashed,color=black,mark=o,mark options={solid}] coordinates {
        (1,1.393246187)
        (2,1.568273092)
        (3,1.710918114)
        (4,1.739742438)
        (5,1.738436253)
        (6,1.732208648)
	};

	\addplot[color=black,mark=square] coordinates {
        (1,1.483758701)
        (2,1.623700624)
        (3,1.696882691)
        (4,1.623986581)
        (5,1.557425326)
        (6,1.519476101)
	};

	\addplot[color=black,mark=o] coordinates {
        (1,0.145954582)
        (2,0.266280259)
        (3,0.466561407)
        (4,0.650503919)
        (5,0.823856682)
        (6,0.986635741)
	};

	\addplot[color=black,mark=diamond] coordinates {
        (1,0.614903846)
        (2,0.816945607)
        (3,1.015214724)
        (4,1.126648565)
        (5,1.199222546)
        (6,1.266373689)
	};

	\addplot[color=black,mark=triangle] coordinates {	
        (1,1.346315789)
        (2,1.477301387)
        (3,1.565859198)
        (4,1.507656372)
        (5,1.456333596)
        (6,1.413225755)
	};
            
        \end{axis}
    \end{tikzpicture}  
    }
    \\
    \resizebox{.45\linewidth}{!}{
    \begin{tikzpicture}
        \begin{axis}[
            hide axis,
            xmin=0,
            xmax=0,
            ymin=0,
            ymax=0,
            legend style={draw=white!15!black,legend cell align=left},
            legend columns = 4
        ]
        
        \addlegendimage{color=black,mark=*}
        \addlegendentry{JPlag}
        
       \addlegendimage{color=black,mark=square}
        \addlegendentry{Plaggie}
        
        \addlegendimage{color=black,mark=triangle}
        \addlegendentry{Sim}
        
        \addlegendimage{color=black,mark=diamond}
        \addlegendentry{Sherlock-W}
        
        \addlegendimage{color=black,mark=o}
        \addlegendentry{Sherlock-S}
        
        \addlegendimage{dashed,color=black,mark=*,mark options={solid}}
        \addlegendentry{String ED}
        
        \addlegendimage{dashed,color=black,mark=square,mark options={solid}}
        \addlegendentry{String T}
        
        \addlegendimage{dashed,color=black,mark=triangle,mark options={solid}}
        \addlegendentry{Token ED}
        
        \addlegendimage{dashed,color=black,mark=diamond,mark options={solid}}
        \addlegendentry{Token T}
        
        \addlegendimage{dashed,color=black,mark=o,mark options={solid}}
        \addlegendentry{Tree ED}
        
        \addlegendimage{dashed,color=black,mark=star,mark options={solid}}
        \addlegendentry{Graph ED}
    
        \end{axis}
    \end{tikzpicture}
    }
    \caption{Average RCT scores for variants generated with the 5 identified transformations.}
    \label{fig:ev2axrctranking}
\end{figure}

%% file: Fig8_ev1b_iterations.tex
\begin{figure}
    \centering
    \begin{tabular}{@{}p{.5\linewidth}@{}p{.5\linewidth}@{}}
        \resizebox{.45\textwidth}{!}{
\begin{tikzpicture}
    \begin{axis}[
        title={JPlag},
        xlabel={Repetition},
        ylabel={Avg Similairty (\%)},
        ymax=100.0,ymin=65.0
    ]
    
\addplot[color=black,mark=*,mark options={solid}] table[x index=0,y index=1] {1b-jplag.txt};
\addplot[color=black,mark=square,mark options={solid}] table[x index=0,y index=2] {1b-jplag.txt};
\addplot[color=black,mark=triangle,mark options={solid}] table[x index=0,y index=3] {1b-jplag.txt};
\addplot[color=black,mark=diamond,mark options={solid}] table[x index=0,y index=4] {1b-jplag.txt};
\addplot[color=black,mark=o,mark options={solid}] table[x index=0,y index=5] {1b-jplag.txt};
\addplot[color=black,mark=star,mark options={solid}] table[x index=0,y index=6] {1b-jplag.txt};
\legend{};

    \end{axis}
\end{tikzpicture}
        } &
        \resizebox{.45\textwidth}{!}{
\begin{tikzpicture}
    \begin{axis}[
        title={Sim},
        xlabel={Repetition},
        ylabel={Avg Similairty (\%)},
        ymax=100.0,ymin=65.0
    ]
    
\addplot[color=black,mark=*,mark options={solid}] table[x index=0,y index=1] {1b-sim.txt};
\addplot[color=black,mark=square,mark options={solid}] table[x index=0,y index=2] {1b-sim.txt};
\addplot[color=black,mark=triangle,mark options={solid}] table[x index=0,y index=3] {1b-sim.txt};
\addplot[color=black,mark=diamond,mark options={solid}] table[x index=0,y index=4] {1b-sim.txt};
\addplot[color=black,mark=o,mark options={solid}] table[x index=0,y index=5] {1b-sim.txt};
\addplot[color=black,mark=star,mark options={solid}] table[x index=0,y index=6] {1b-sim.txt};
\legend{};

    \end{axis}
\end{tikzpicture}
        } \\
        
\multicolumn{2}{c}{
\resizebox{.45\textwidth}{!}{
\begin{tikzpicture}[font=\footnotesize]
    \begin{axis}[
        hide axis,
        xmin=0,
        xmax=0,
        ymin=0,
        ymax=0,
        legend style={draw=white!15!black,legend cell align=left},
        legend columns = 6
    ]
    
\addlegendimage{color=black,mark=*,mark options={solid}}
\addlegendentry{10\%}

\addlegendimage{color=black,mark=square,mark options={solid}}
\addlegendentry{20\%}

\addlegendimage{color=black,mark=triangle,mark options={solid}}
\addlegendentry{40\%}

\addlegendimage{color=black,mark=diamond,mark options={solid}}
\addlegendentry{60\%}

\addlegendimage{color=black,mark=o,mark options={solid}}
\addlegendentry{80\%}

\addlegendimage{color=black,mark=star,mark options={solid}}
\addlegendentry{100\%}

    \end{axis}
\end{tikzpicture}
}
} \\
        
    \end{tabular}
    \caption{Average similarity of variants generated using random selections of transformations at each transformation chance over 10 repetitions for SCPDTs JPlag and Sim.}
    \label{fig:ev1bsample}
\end{figure}
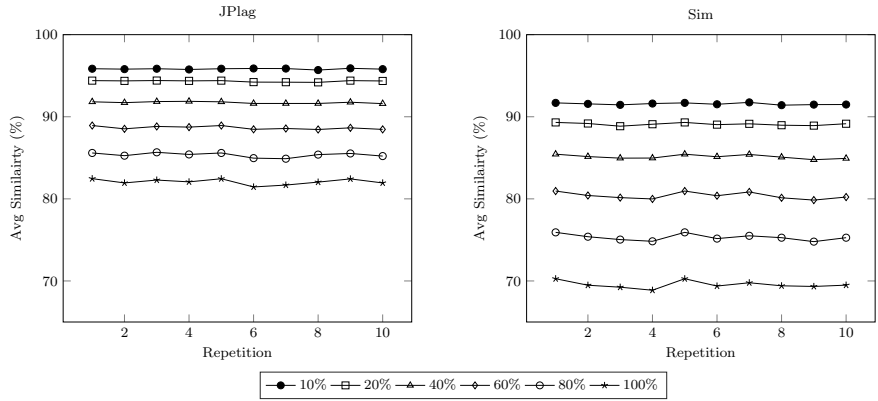

%% file: Fig9_ev1b_avgsimilarity.tex
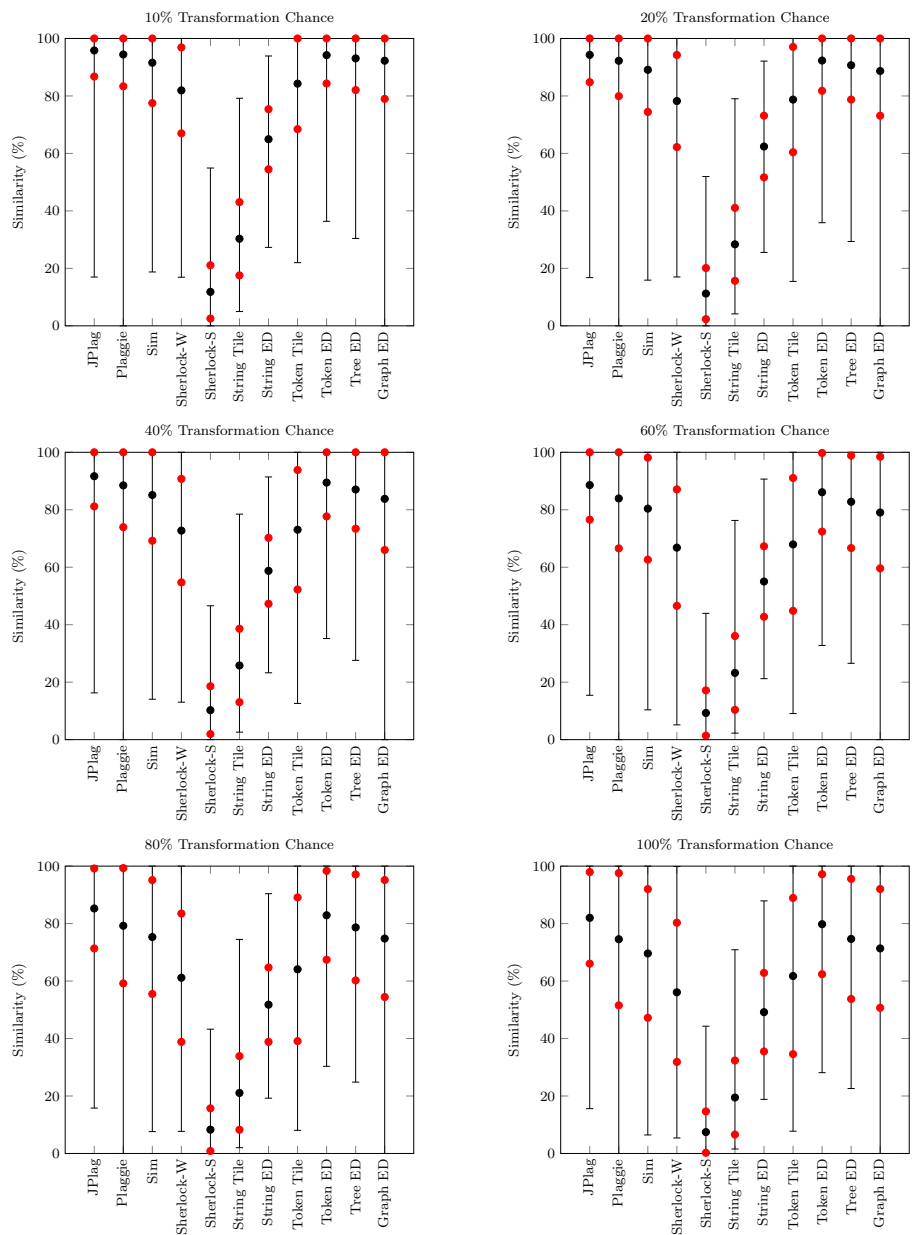
\begin{figure}[htbp]

    \begin{tabular}{p{.5\linewidth}p{.5\linewidth}}
         
        \resizebox{.45\textwidth}{!}{
            \begin{tikzpicture}
            \begin{axis}[
                title={10\% Transformation Chance},
                ylabel={Similarity (\%)},
                xmin=-1, xmax=11,
                ymin=0.0, ymax=100,
                xtick={0,1,2,3,4,5,6,7,8,9,10},
                xticklabel style={rotate=90},
                xticklabels={
                    JPlag,Plaggie,Sim,Sherlock-W,Sherlock-S,String Tile,String ED,Token Tile,Token ED,Tree ED,Graph ED
                }
            ]
            
\addplot[mark=*,black] coordinates {(0, 95.83074977592196)};
\addplot[mark=-,black] coordinates { (0,16.974951833333332)(0,100.0) };
\addplot[mark=*,red] coordinates { (0,86.75463067506631) };
\addplot[mark=*,red] coordinates { (0,100.0) };

\addplot[mark=*,black] coordinates {(10, 92.25488414689953)};
\addplot[mark=-,black] coordinates { (10,0.0)(10,100.0) };
\addplot[mark=*,red] coordinates { (10,78.98115965905777) };
\addplot[mark=*,red] coordinates { (10,100.0) };

\addplot[mark=*,black] coordinates {(6, 64.94173717844915)};
\addplot[mark=-,black] coordinates { (6,27.30438590542875)(6,93.90914813835691) };
\addplot[mark=*,red] coordinates { (6,54.47951833506231) };
\addplot[mark=*,red] coordinates { (6,75.40395602183598) };

\addplot[mark=*,black] coordinates {(5, 30.29942123664773)};
\addplot[mark=-,black] coordinates { (5,4.9729886034989805)(5,79.19515365938209) };
\addplot[mark=*,red] coordinates { (5,17.549789263227325) };
\addplot[mark=*,red] coordinates { (5,43.04905321006814) };

\addplot[mark=*,black] coordinates {(8, 94.193413123298)};
\addplot[mark=-,black] coordinates { (8,36.36908784236977)(8,100.0) };
\addplot[mark=*,red] coordinates { (8,84.33801224589996) };
\addplot[mark=*,red] coordinates { (8,100.0) };

\addplot[mark=*,black] coordinates {(7, 84.28291800786218)};
\addplot[mark=-,black] coordinates { (7,21.982461125851277)(7,100.0) };
\addplot[mark=*,red] coordinates { (7,68.44961762901609) };
\addplot[mark=*,red] coordinates { (7,100.0) };

\addplot[mark=*,black] coordinates {(9, 93.065152026966)};
\addplot[mark=-,black] coordinates { (9,30.41168398508714)(9,100.0) };
\addplot[mark=*,red] coordinates { (9,82.0587790027436) };
\addplot[mark=*,red] coordinates { (9,100.0) };

\addplot[mark=*,black] coordinates {(1, 94.42192877982642)};
\addplot[mark=-,black] coordinates { (1,0.0)(1,100.0) };
\addplot[mark=*,red] coordinates { (1,83.34553069491727) };
\addplot[mark=*,red] coordinates { (1,100.0) };

\addplot[mark=*,black] coordinates {(4, 11.813950050732814)};
\addplot[mark=-,black] coordinates { (4,0.0)(4,54.916666666666664) };
\addplot[mark=*,red] coordinates { (4,2.539105004219703) };
\addplot[mark=*,red] coordinates { (4,21.088795097245924) };

\addplot[mark=*,black] coordinates {(3, 81.93490663661117)};
\addplot[mark=-,black] coordinates { (3,16.916666666666668)(3,100.0) };
\addplot[mark=*,red] coordinates { (3,66.97702974950488) };
\addplot[mark=*,red] coordinates { (3,96.89278352371747) };

\addplot[mark=*,black] coordinates {(2, 91.55726266277964)};
\addplot[mark=-,black] coordinates { (2,18.743055555555557)(2,100.0) };
\addplot[mark=*,red] coordinates { (2,77.52136661595645) };
\addplot[mark=*,red] coordinates { (2,100.0) };

            \end{axis}
            \end{tikzpicture}  
        } &
    
    \resizebox{.45\textwidth}{!}{
            \begin{tikzpicture}
            \begin{axis}[
                title={20\% Transformation Chance},
                ylabel={Similarity (\%)},
                xmin=-1, xmax=11,
                ymin=0.0, ymax=100,
                xtick={0,1,2,3,4,5,6,7,8,9,10},
                xticklabel style={rotate=90},
                xticklabels={
                    JPlag,Plaggie,Sim,Sherlock-W,Sherlock-S,String Tile,String ED,Token Tile,Token ED,Tree ED,Graph ED
                }
            ]
            
\addplot[mark=*,black] coordinates {(0, 94.29601046504793)};
\addplot[mark=-,black] coordinates { (0,16.772639833333333)(0,100.0) };
\addplot[mark=*,red] coordinates { (0,84.79351665279442) };
\addplot[mark=*,red] coordinates { (0,100.0) };

\addplot[mark=*,black] coordinates {(10, 88.6700230093396)};
\addplot[mark=-,black] coordinates { (10,0.0)(10,100.0) };
\addplot[mark=*,red] coordinates { (10,73.12324972057039) };
\addplot[mark=*,red] coordinates { (10,100.0) };

\addplot[mark=*,black] coordinates {(6, 62.394815486587625)};
\addplot[mark=-,black] coordinates { (6,25.551504839396618)(6,92.11850436407695) };
\addplot[mark=*,red] coordinates { (6,51.66628039087082) };
\addplot[mark=*,red] coordinates { (6,73.12335058230444) };

\addplot[mark=*,black] coordinates {(5, 28.354726586365317)};
\addplot[mark=-,black] coordinates { (5,4.1676841739921615)(5,79.02254940641329) };
\addplot[mark=*,red] coordinates { (5,15.668527271168129) };
\addplot[mark=*,red] coordinates { (5,41.0409259015625) };

\addplot[mark=*,black] coordinates {(8, 92.32250385383533)};
\addplot[mark=-,black] coordinates { (8,35.88154374779707)(8,100.0) };
\addplot[mark=*,red] coordinates { (8,81.79150172551728) };
\addplot[mark=*,red] coordinates { (8,100.0) };

\addplot[mark=*,black] coordinates {(7, 78.73249386302957)};
\addplot[mark=-,black] coordinates { (7,15.457765858770301)(7,100.0) };
\addplot[mark=*,red] coordinates { (7,60.40270046013373) };
\addplot[mark=*,red] coordinates { (7,97.06228726592542) };

\addplot[mark=*,black] coordinates {(9, 90.69254621244905)};
\addplot[mark=-,black] coordinates { (9,29.34089872867865)(9,100.0) };
\addplot[mark=*,red] coordinates { (9,78.7437778109523) };
\addplot[mark=*,red] coordinates { (9,100.0) };

\addplot[mark=*,black] coordinates {(1, 92.23803427194603)};
\addplot[mark=-,black] coordinates { (1,0.0)(1,100.0) };
\addplot[mark=*,red] coordinates { (1,79.92200143445812) };
\addplot[mark=*,red] coordinates { (1,100.0) };

\addplot[mark=*,black] coordinates {(4, 11.221522803953574)};
\addplot[mark=-,black] coordinates { (4,0.0)(4,51.958333333333336) };
\addplot[mark=*,red] coordinates { (4,2.3322584069800065) };
\addplot[mark=*,red] coordinates { (4,20.110787200927142) };

\addplot[mark=*,black] coordinates {(3, 78.2318520902191)};
\addplot[mark=-,black] coordinates { (3,17.0)(3,100.0) };
\addplot[mark=*,red] coordinates { (3,62.20138155636273) };
\addplot[mark=*,red] coordinates { (3,94.26232262407548) };

\addplot[mark=*,black] coordinates {(2, 89.08892638991966)};
\addplot[mark=-,black] coordinates { (2,15.895833333333334)(2,100.0) };
\addplot[mark=*,red] coordinates { (2,74.44362176957932) };
\addplot[mark=*,red] coordinates { (2,100.0) };

            \end{axis}
            \end{tikzpicture}  
        } \\
        
                \resizebox{.45\textwidth}{!}{
            \begin{tikzpicture}
            \begin{axis}[
                title={40\% Transformation Chance},
                ylabel={Similarity (\%)},
                xmin=-1, xmax=11,
                ymin=0.0, ymax=100,
                xtick={0,1,2,3,4,5,6,7,8,9,10},
                xticklabel style={rotate=90},
                xticklabels={
                    JPlag,Plaggie,Sim,Sherlock-W,Sherlock-S,String Tile,String ED,Token Tile,Token ED,Tree ED,Graph ED
                }
            ]
            
\addplot[mark=*,black] coordinates {(0, 91.67326795859923)};
\addplot[mark=-,black] coordinates { (0,16.271676166666666)(0,100.0) };
\addplot[mark=*,red] coordinates { (0,81.16501177392337) };
\addplot[mark=*,red] coordinates { (0,100.0) };

\addplot[mark=*,black] coordinates {(10, 83.79422291795373)};
\addplot[mark=-,black] coordinates { (10,0.0)(10,100.0) };
\addplot[mark=*,red] coordinates { (10,65.99869631557125) };
\addplot[mark=*,red] coordinates { (10,100.0) };

\addplot[mark=*,black] coordinates {(6, 58.753956127006084)};
\addplot[mark=-,black] coordinates { (6,23.268315719487987)(6,91.41309996837008) };
\addplot[mark=*,red] coordinates { (6,47.274449199332714) };
\addplot[mark=*,red] coordinates { (6,70.23346305467945) };

\addplot[mark=*,black] coordinates {(5, 25.810834050906177)};
\addplot[mark=-,black] coordinates { (5,2.6107548869221797)(5,78.4716785681211) };
\addplot[mark=*,red] coordinates { (5,13.025594660352292) };
\addplot[mark=*,red] coordinates { (5,38.596073441460064) };

\addplot[mark=*,black] coordinates {(8, 89.46131900279084)};
\addplot[mark=-,black] coordinates { (8,35.191703226562694)(8,100.0) };
\addplot[mark=*,red] coordinates { (8,77.6919264090501) };
\addplot[mark=*,red] coordinates { (8,100.0) };

\addplot[mark=*,black] coordinates {(7, 73.04825848485724)};
\addplot[mark=-,black] coordinates { (7,12.607588657705783)(7,100.0) };
\addplot[mark=*,red] coordinates { (7,52.264574015515166) };
\addplot[mark=*,red] coordinates { (7,93.83194295419932) };

\addplot[mark=*,black] coordinates {(9, 87.05535539714633)};
\addplot[mark=-,black] coordinates { (9,27.59570999341629)(9,100.0) };
\addplot[mark=*,red] coordinates { (9,73.4031560875235) };
\addplot[mark=*,red] coordinates { (9,100.0) };

\addplot[mark=*,black] coordinates {(1, 88.4829454022464)};
\addplot[mark=-,black] coordinates { (1,0.0)(1,100.0) };
\addplot[mark=*,red] coordinates { (1,73.94164976951627) };
\addplot[mark=*,red] coordinates { (1,100.0) };

\addplot[mark=*,black] coordinates {(4, 10.257653996181066)};
\addplot[mark=-,black] coordinates { (4,0.0)(4,46.583333333333336) };
\addplot[mark=*,red] coordinates { (4,1.9341075653049185) };
\addplot[mark=*,red] coordinates { (4,18.58120042705721) };

\addplot[mark=*,black] coordinates {(3, 72.74563661501945)};
\addplot[mark=-,black] coordinates { (3,13.041666666666666)(3,100.0) };
\addplot[mark=*,red] coordinates { (3,54.730930073013106) };
\addplot[mark=*,red] coordinates { (3,90.7603431570258) };

\addplot[mark=*,black] coordinates {(2, 85.12705258490693)};
\addplot[mark=-,black] coordinates { (2,14.0625)(2,100.0) };
\addplot[mark=*,red] coordinates { (2,69.23281117023026) };
\addplot[mark=*,red] coordinates { (2,100.0) };
            
            \end{axis}
            \end{tikzpicture}  
        } &
    
    \resizebox{.45\textwidth}{!}{
            \begin{tikzpicture}
            \begin{axis}[
                title={60\% Transformation Chance},
                ylabel={Similarity (\%)},
                xmin=-1, xmax=11,
                ymin=0.0, ymax=100,
                xtick={0,1,2,3,4,5,6,7,8,9,10},
                xticklabel style={rotate=90},
                xticklabels={
                    JPlag,Plaggie,Sim,Sherlock-W,Sherlock-S,String Tile,String ED,Token Tile,Token ED,Tree ED,Graph ED
                }
            ]
            
\addplot[mark=*,black] coordinates {(0, 88.58844190789)};
\addplot[mark=-,black] coordinates { (0,15.44316)(0,100.0) };
\addplot[mark=*,red] coordinates { (0,76.54695689037158) };
\addplot[mark=*,red] coordinates { (0,100.0) };

\addplot[mark=*,black] coordinates {(10, 79.04004459987942)};
\addplot[mark=-,black] coordinates { (10,0.0)(10,100.0) };
\addplot[mark=*,red] coordinates { (10,59.611754746375325) };
\addplot[mark=*,red] coordinates { (10,98.46833445338352) };

\addplot[mark=*,black] coordinates {(6, 55.02488579941407)};
\addplot[mark=-,black] coordinates { (6,21.22248615496534)(6,90.678509537487) };
\addplot[mark=*,red] coordinates { (6,42.7652072312704) };
\addplot[mark=*,red] coordinates { (6,67.28456436755775) };

\addplot[mark=*,black] coordinates {(5, 23.225296626069156)};
\addplot[mark=-,black] coordinates { (5,2.2662384490370857)(5,76.2884980393876) };
\addplot[mark=*,red] coordinates { (5,10.379331977119865) };
\addplot[mark=*,red] coordinates { (5,36.07126127501844) };

\addplot[mark=*,black] coordinates {(8, 86.08964884136032)};
\addplot[mark=-,black] coordinates { (8,32.78270840212279)(8,100.0) };
\addplot[mark=*,red] coordinates { (8,72.41394477215239) };
\addplot[mark=*,red] coordinates { (8,99.76535291056824) };

\addplot[mark=*,black] coordinates {(7, 67.94074662431716)};
\addplot[mark=-,black] coordinates { (7,9.077412972851542)(7,100.0) };
\addplot[mark=*,red] coordinates { (7,44.83065300270535) };
\addplot[mark=*,red] coordinates { (7,91.05084024592897) };

\addplot[mark=*,black] coordinates {(9, 82.76498567863825)};
\addplot[mark=-,black] coordinates { (9,26.56241429038977)(9,100.0) };
\addplot[mark=*,red] coordinates { (9,66.66691141099274) };
\addplot[mark=*,red] coordinates { (9,98.86305994628377) };

\addplot[mark=*,black] coordinates {(1, 83.9383891209198)};
\addplot[mark=-,black] coordinates { (1,0.0)(1,100.0) };
\addplot[mark=*,red] coordinates { (1,66.56400831673903) };
\addplot[mark=*,red] coordinates { (1,100.0) };

\addplot[mark=*,black] coordinates {(4, 9.277728050930033)};
\addplot[mark=-,black] coordinates { (4,0.0)(4,43.958333333333336) };
\addplot[mark=*,red] coordinates { (4,1.4150447095419558) };
\addplot[mark=*,red] coordinates { (4,17.14041139231811) };

\addplot[mark=*,black] coordinates {(3, 66.82159752669293)};
\addplot[mark=-,black] coordinates { (3,5.145833333333333)(3,100.0) };
\addplot[mark=*,red] coordinates { (3,46.55138068702071) };
\addplot[mark=*,red] coordinates { (3,87.09181436636514) };

\addplot[mark=*,black] coordinates {(2, 80.3973362714399)};
\addplot[mark=-,black] coordinates { (2,10.375)(2,100.0) };
\addplot[mark=*,red] coordinates { (2,62.64714861440452) };
\addplot[mark=*,red] coordinates { (2,98.14752392847527) };
            
            \end{axis}
            \end{tikzpicture}  
        } \\
        
                \resizebox{.45\textwidth}{!}{
            \begin{tikzpicture}
            \begin{axis}[
                title={80\% Transformation Chance},
                ylabel={Similarity (\%)},
                xmin=-1, xmax=11,
                ymin=0.0, ymax=100,
                xtick={0,1,2,3,4,5,6,7,8,9,10},
                xticklabel style={rotate=90},
                xticklabels={
                    JPlag,Plaggie,Sim,Sherlock-W,Sherlock-S,String Tile,String ED,Token Tile,Token ED,Tree ED,Graph ED
                }
            ]
            
\addplot[mark=*,black] coordinates {(0, 85.26319468337282)};
\addplot[mark=-,black] coordinates { (0,15.78998075)(0,100.0) };
\addplot[mark=*,red] coordinates { (0,71.34698486699435) };
\addplot[mark=*,red] coordinates { (0,99.17940449975129) };

\addplot[mark=*,black] coordinates {(10, 74.79628612484932)};
\addplot[mark=-,black] coordinates { (10,0.0)(10,100.0) };
\addplot[mark=*,red] coordinates { (10,54.442055143830075) };
\addplot[mark=*,red] coordinates { (10,95.15051710586856) };

\addplot[mark=*,black] coordinates {(6, 51.77601022198481)};
\addplot[mark=-,black] coordinates { (6,19.237601585058258)(6,90.39876822300363) };
\addplot[mark=*,red] coordinates { (6,38.83703247368899) };
\addplot[mark=*,red] coordinates { (6,64.71498797028063) };

\addplot[mark=*,black] coordinates {(5, 21.062045915402198)};
\addplot[mark=-,black] coordinates { (5,2.00667366259656)(5,74.48841920145388) };
\addplot[mark=*,red] coordinates { (5,8.248938589534463) };
\addplot[mark=*,red] coordinates { (5,33.875153241269935) };

\addplot[mark=*,black] coordinates {(8, 82.88883470747203)};
\addplot[mark=-,black] coordinates { (8,30.327250661963234)(8,100.0) };
\addplot[mark=*,red] coordinates { (8,67.42341403280979) };
\addplot[mark=*,red] coordinates { (8,98.35425538213427) };

\addplot[mark=*,black] coordinates {(7, 64.10070832899854)};
\addplot[mark=-,black] coordinates { (7,8.05074987312795)(7,100.0) };
\addplot[mark=*,red] coordinates { (7,39.10506037087457) };
\addplot[mark=*,red] coordinates { (7,89.0963562871225) };

\addplot[mark=*,black] coordinates {(9, 78.6669475271619)};
\addplot[mark=-,black] coordinates { (9,24.82836510834082)(9,100.0) };
\addplot[mark=*,red] coordinates { (9,60.22727300836726) };
\addplot[mark=*,red] coordinates { (9,97.10662204595653) };

\addplot[mark=*,black] coordinates {(1, 79.2489654984056)};
\addplot[mark=-,black] coordinates { (1,0.0)(1,100.0) };
\addplot[mark=*,red] coordinates { (1,59.1922666380741) };
\addplot[mark=*,red] coordinates { (1,99.30566435873709) };

\addplot[mark=*,black] coordinates {(4, 8.28579991031076)};
\addplot[mark=-,black] coordinates { (4,0.0)(4,43.25) };
\addplot[mark=*,red] coordinates { (4,0.8603387015593702) };
\addplot[mark=*,red] coordinates { (4,15.711261119062147) };

\addplot[mark=*,black] coordinates {(3, 61.16045486749948)};
\addplot[mark=-,black] coordinates { (3,7.694444444444444)(3,100.0) };
\addplot[mark=*,red] coordinates { (3,38.833605418864394) };
\addplot[mark=*,red] coordinates { (3,83.48730431613457) };

\addplot[mark=*,black] coordinates {(2, 75.32114450092007)};
\addplot[mark=-,black] coordinates { (2,7.611111111111111)(2,100.0) };
\addplot[mark=*,red] coordinates { (2,55.501699267664875) };
\addplot[mark=*,red] coordinates { (2,95.14058973417526) };
            
            \end{axis}
            \end{tikzpicture}  
        } &
    
    \resizebox{.45\textwidth}{!}{
            \begin{tikzpicture}
            \begin{axis}[
                title={100\% Transformation Chance},
                ylabel={Similarity (\%)},
                xmin=-1, xmax=11,
                ymin=0.0, ymax=100,
                xtick={0,1,2,3,4,5,6,7,8,9,10},
                xticklabel style={rotate=90},
                xticklabels={
                    JPlag,Plaggie,Sim,Sherlock-W,Sherlock-S,String Tile,String ED,Token Tile,Token ED,Tree ED,Graph ED
                }
            ]
            
\addplot[mark=*,black] coordinates {(0, 81.99637421652885)};
\addplot[mark=-,black] coordinates { (0,15.606936500000002)(0,100.0) };
\addplot[mark=*,red] coordinates { (0,66.05426065761068) };
\addplot[mark=*,red] coordinates { (0,97.93848777544702) };

\addplot[mark=*,black] coordinates {(10, 71.3607666892703)};
\addplot[mark=-,black] coordinates { (10,0.0)(10,100.0) };
\addplot[mark=*,red] coordinates { (10,50.68395713382876) };
\addplot[mark=*,red] coordinates { (10,92.03757624471186) };

\addplot[mark=*,black] coordinates {(6, 49.18020226199372)};
\addplot[mark=-,black] coordinates { (6,18.83556911733861)(6,87.90201214704813) };
\addplot[mark=*,red] coordinates { (6,35.510016086525596) };
\addplot[mark=*,red] coordinates { (6,62.850388437461845) };

\addplot[mark=*,black] coordinates {(5, 19.471095852517948)};
\addplot[mark=-,black] coordinates { (5,1.559712339834541)(5,70.8939113026593) };
\addplot[mark=*,red] coordinates { (5,6.587231496893688) };
\addplot[mark=*,red] coordinates { (5,32.35496020814221) };

\addplot[mark=*,black] coordinates {(8, 79.78331236706931)};
\addplot[mark=-,black] coordinates { (8,28.117963096473503)(8,100.0) };
\addplot[mark=*,red] coordinates { (8,62.38936187758606) };
\addplot[mark=*,red] coordinates { (8,97.17726285655257) };

\addplot[mark=*,black] coordinates {(7, 61.75489013292543)};
\addplot[mark=-,black] coordinates { (7,7.753365378492346)(7,100.0) };
\addplot[mark=*,red] coordinates { (7,34.574559032229885) };
\addplot[mark=*,red] coordinates { (7,88.93522123362098) };

\addplot[mark=*,black] coordinates {(9, 74.66640013507286)};
\addplot[mark=-,black] coordinates { (9,22.596607677730336)(9,100.0) };
\addplot[mark=*,red] coordinates { (9,53.770392274271614) };
\addplot[mark=*,red] coordinates { (9,95.56240799587411) };

\addplot[mark=*,black] coordinates {(1, 74.55868971884503)};
\addplot[mark=-,black] coordinates { (1,0.0)(1,100.0) };
\addplot[mark=*,red] coordinates { (1,51.54541533177685) };
\addplot[mark=*,red] coordinates { (1,97.57196410591321) };

\addplot[mark=*,black] coordinates {(4, 7.436385088927157)};
\addplot[mark=-,black] coordinates { (4,0.0)(4,44.291666666666664) };
\addplot[mark=*,red] coordinates { (4,0.23495747673935075) };
\addplot[mark=*,red] coordinates { (4,14.637812701114964) };

\addplot[mark=*,black] coordinates {(3, 56.07834600558139)};
\addplot[mark=-,black] coordinates { (3,5.395833333333333)(3,99.875) };
\addplot[mark=*,red] coordinates { (3,31.859066346578597) };
\addplot[mark=*,red] coordinates { (3,80.29762566458417) };

\addplot[mark=*,black] coordinates {(2, 69.61040141299917)};
\addplot[mark=-,black] coordinates { (2,6.4444444444444455)(2,100.0) };
\addplot[mark=*,red] coordinates { (2,47.219694024436706) };
\addplot[mark=*,red] coordinates { (2,92.00110880156163) };

            \end{axis}
            \end{tikzpicture}  
        } \\
        
    \end{tabular}

    \caption{Average similarity of variants generated with random source code transformations, evaluated using all 11 SCPDTs. Bars indicate range of similarity scores. Red marks indicate range of standard deviation around the average.}
    \label{fig:eb2bacgsimilarity}
\end{figure}

%% file: Fig10_ev1b_resilience.tex
\begin{figure}[htbp]
    \centering
    \resizebox{.45\linewidth}{!}{
     \begin{tikzpicture}
        \begin{axis}[
            xlabel={Transformation Chance (\%)},
            ylabel={RCT Score},
            xmin=0, xmax=110,
            ymin=0.0, ymax=6.5,
        ]
            
\addplot[mark=*,black] table [x index=0,y index=1] {ev1b-resilience.txt};

\addplot[dashed,color=black,mark=star,mark options={solid}] table [x index=0,y index=2] {ev1b-resilience.txt};

\addplot[dashed,color=black,mark=*,mark options={solid}] table [x index=0,y index=3] {ev1b-resilience.txt};

\addplot[dashed,color=black,mark=square,mark options={solid}] table [x index=0,y index=4] {ev1b-resilience.txt};

\addplot[dashed,color=black,mark=triangle,mark options={solid}] table [x index=0,y index=5] {ev1b-resilience.txt};

\addplot[dashed,color=black,mark=diamond,mark options={solid}] table [x index=0,y index=6] {ev1b-resilience.txt};

\addplot[dashed,color=black,mark=o,mark options={solid}] table [x index=0,y index=7] {ev1b-resilience.txt};

\addplot[color=black,mark=square] table [x index=0,y index=8] {ev1b-resilience.txt};

\addplot[color=black,mark=o] table [x index=0,y index=9] {ev1b-resilience.txt};

\addplot[color=black,mark=diamond] table [x index=0,y index=10] {ev1b-resilience.txt};

\addplot[color=black,mark=triangle] table [x index=0,y index=11] {ev1b-resilience.txt};
            
        \end{axis}
    \end{tikzpicture}
    }
    \\
    \resizebox{.45\linewidth}{!}{
    \begin{tikzpicture}
        \begin{axis}[
            hide axis,
            xmin=0,
            xmax=0,
            ymin=0,
            ymax=0,
            legend style={draw=white!15!black,legend cell align=left},
            legend columns = 4
        ]
        
        \addlegendimage{color=black,mark=*}
        \addlegendentry{JPlag}
        
       \addlegendimage{color=black,mark=square}
        \addlegendentry{Plaggie}
        
        \addlegendimage{color=black,mark=triangle}
        \addlegendentry{Sim}
        
        \addlegendimage{color=black,mark=diamond}
        \addlegendentry{Sherlock-W}
        
        \addlegendimage{color=black,mark=o}
        \addlegendentry{Sherlock-S}
        
        \addlegendimage{dashed,color=black,mark=*,mark options={solid}}
        \addlegendentry{String ED}
        
        \addlegendimage{dashed,color=black,mark=square,mark options={solid}}
        \addlegendentry{String T}
        
        \addlegendimage{dashed,color=black,mark=triangle,mark options={solid}}
        \addlegendentry{Token ED}
        
        \addlegendimage{dashed,color=black,mark=diamond,mark options={solid}}
        \addlegendentry{Token T}
        
        \addlegendimage{dashed,color=black,mark=o,mark options={solid}}
        \addlegendentry{Tree ED}
        
        \addlegendimage{dashed,color=black,mark=star,mark options={solid}}
        \addlegendentry{Graph ED}
    
        \end{axis}
    \end{tikzpicture}
    }
    \caption{Average RCT scores for variants generated with randomly selected transformations.}
    \label{fig:ev2brctranking}
\end{figure}
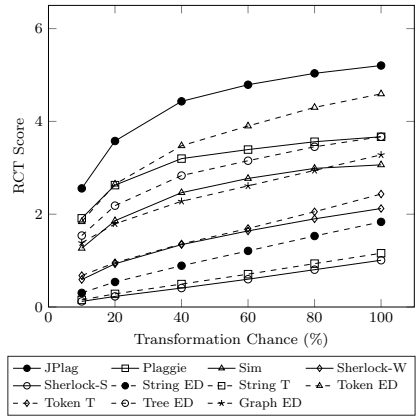

%% file: Tab6.tex
\begin{table}[htbp]
    \caption{Average number of applied transformations to each program variant with randomly selected transformations.}
    \label{tab:ev2bavgtransformations}
    \begin{tabular}{p{4cm}|cccccc}
\hline\noalign{\smallskip}
        {Transformation Chance} & {10\%} & {20\%} & {40\%} & {60\%} & {80\%} & {100\%} \\
        \noalign{\smallskip}
        \cline{2-7}
        \noalign{\smallskip}
        {Avg. No. of Transformations} & 10.67 & 20.28 & 36.65 & 54.35 & 73.77 & 93.29 \\
\noalign{\smallskip}\hline
    \end{tabular}
\end{table}

%% file: Tab7.tex
\begin{table}[htbp]
    \caption{Top 3 selections of transformations each SCPDT is most vulnerable to, identified from the random selections of applied source code transformations. Bold ids are from the 5 identified transformations in Section \ref{s:ev1ax}.}
    \label{tab:ev2brankings}
    \begin{tabular}{ll}
        \hline\noalign{\smallskip}
        {Tool} & {Top 3 Selections of Transformations} \\
        \noalign{\smallskip}\hline\noalign{\smallskip}
        \multirow{3}{*}{{JPlag}}
        & tRC tMC tRI \textbf{tRM} \textbf{tSO} \textbf{tFW} tAD \textbf{tSD} \\
        & tAC tRC tRI \textbf{tRS} \textbf{tSO} tEA \textbf{tSD} \\
        & tMC \textbf{tRS} \textbf{tFW} tSV \\
        \noalign{\smallskip}
        \multirow{3}{*}{{Plaggie}}
        & tAC \textbf{tRS} \textbf{tRM} \textbf{tFW} tEU \\
        & tAC tRC tMC \textbf{tRS} \textbf{tRM} \textbf{tSO} \textbf{tFW} tEA \textbf{tSD} \\
        & tAC tMC tRI \textbf{tRM} \textbf{tSO} tUD \textbf{tFW} tEA tSV tAD \textbf{tSD} \\
        \noalign{\smallskip}
        \multirow{3}{*}{{Sim}}
        & tRC \textbf{tRM} \textbf{tSO} tAD \textbf{tSD} \\
        & tAC tRC tMC \textbf{tRS} \textbf{tRM} \textbf{tSO} \textbf{tFW} tEA \textbf{tSD} \\
        & tRC \textbf{tRM} \textbf{tSO} tAD \textbf{tSD} \\
        \noalign{\smallskip}
        \multirow{3}{*}{{Sherlock-W}}
        & tAC tMC tRI \textbf{tRS} \textbf{tRM} \textbf{tSO} tUD \textbf{tFW} tEU \\
        & tRC tRI \textbf{tRS} \textbf{tRM} \textbf{tSO} \textbf{tFW} tEA tSV tAD \\
        & tAC tRC tMC \textbf{tRS} \textbf{tRM} \textbf{tSO} tUD tEU tSV \textbf{tSD} \\
        \noalign{\smallskip}
        \multirow{3}{*}{{Sherlock-S}}
        & tRC tRI \textbf{tRS} \textbf{tRM} tUD tEU tSV \textbf{tSD} \\
        & tMC tRI \textbf{tSO} tUD tEU tSV tAD \\
        & tMC tRI tEA tEU \textbf{tSD} \\
        \noalign{\smallskip}
        \multirow{3}{*}{{String Tile}}
        & tAC tRI \textbf{tRS} \textbf{tRM} \textbf{tSO} tUD \textbf{tFW} tEA tEU tSV \\
        & tRC tRI \textbf{tRS} \textbf{tRM} \textbf{tSO} tUD tAD \\
        & tMC tRI \textbf{tRM} \textbf{tSO} tUD tEA tEU tSV \textbf{tSD} \\
        \noalign{\smallskip}
        \multirow{3}{*}{{String ED}}
        & tAC tRC tMC \textbf{tRM} \textbf{tFW} tSV \textbf{tSD} \\
        & tAC tRC tMC tRI tAD \\
        & tMC tRI \textbf{tRS} \textbf{tRM} tUD tEA tEU tSV \textbf{tSD} \\
        \noalign{\smallskip}
        \multirow{3}{*}{{Token Tile}}
        & tMC tRI \textbf{tRS} \textbf{tRM} tUD tEU tSV \\
        & tAC tRC tMC \textbf{tRM} \textbf{tSO} tUD tEU tSV tAD \textbf{tSD} \\
        & tRS \textbf{tRM} \textbf{tSO} tEU tSV \\
        \noalign{\smallskip}
        \multirow{3}{*}{{Token ED}}
        & tAC tRC tMC \textbf{tRS} \textbf{tRM} tUD tEA tAD tAD \textbf{tSD} \\
        & tRC \textbf{tRM} \textbf{tSO} tAD \textbf{tSD} \\
        & tAC tRC tMC \textbf{tRS} \textbf{tRM} \textbf{tSO} \textbf{tFW} tEA \textbf{tSD} \\
        \noalign{\smallskip}
        \multirow{3}{*}{{Tree ED}}
        & tAC \textbf{tRS} \textbf{tRM} \textbf{tFW} tEU \\
        & tMC \textbf{tRS} \textbf{tRM} tSV tAD \\
        & \textbf{tRS} \textbf{tRM} \textbf{tSO} tEU tSV \\
        \noalign{\smallskip}
        \multirow{3}{*}{{Graph ED}}
        & tAC tRC tMC tRI \textbf{tRS} \textbf{tRM} \textbf{tSO} \textbf{tFW} tEA \textbf{tSD} \\
        & tRC \textbf{tRS} \textbf{tRM} \textbf{tFW} tSV \textbf{tSD} \\
        & tAC tRC tMC tRI \textbf{tRS} \textbf{tRM} tUD tEA tAD \textbf{tSD} \\
        \noalign{\smallskip}\hline
    \end{tabular}
\end{table}

%% file: Fig11_ev1c_avgsimilarity.tex
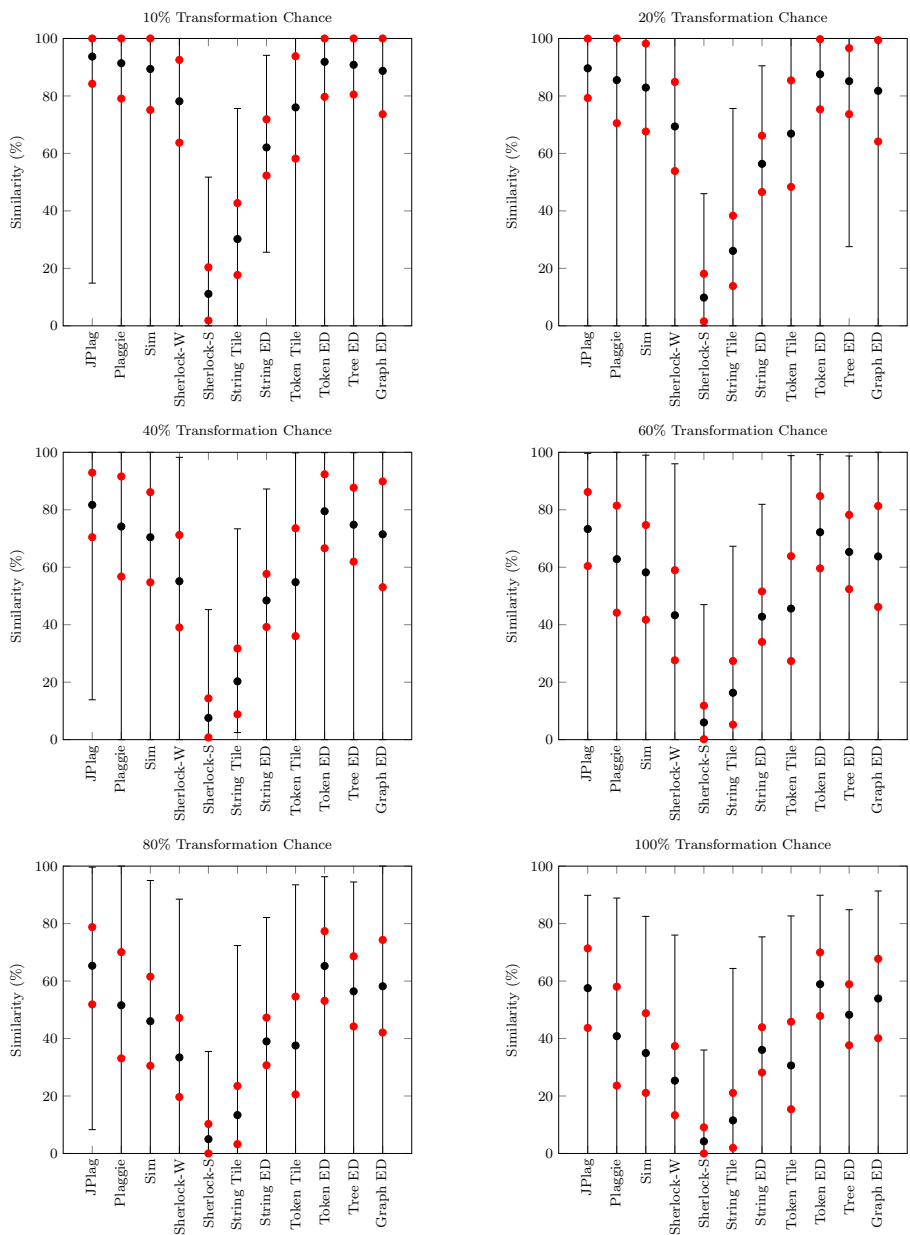
\begin{figure}[htbp]

    \begin{tabular}{p{.5\linewidth}p{.5\linewidth}}
         
        \resizebox{.45\textwidth}{!}{
            \begin{tikzpicture}
            \begin{axis}[
                title={10\% Transformation Chance},
                ylabel={Similarity (\%)},
                xmin=-1, xmax=11,
                ymin=0.0, ymax=100,
                xtick={0,1,2,3,4,5,6,7,8,9,10},
                xticklabel style={rotate=90},
                xticklabels={
                    JPlag,Plaggie,Sim,Sherlock-W,Sherlock-S,String Tile,String ED,Token Tile,Token ED,Tree ED,Graph ED
                }
            ]
            
\addplot[mark=*,black] coordinates { (0,93.70) };
\addplot[mark=-,black] coordinates { (0,100.00)(0,14.86) };
\addplot[mark=*,red] coordinates { (0,100.00) };
\addplot[mark=*,red] coordinates { (0,84.23) };

\addplot[mark=*,black] coordinates { (10,88.73) };
\addplot[mark=-,black] coordinates { (10,100.00)(10,0.00) };
\addplot[mark=*,red] coordinates { (10,100.00) };
\addplot[mark=*,red] coordinates { (10,73.64) };

\addplot[mark=*,black] coordinates { (6,62.09) };
\addplot[mark=-,black] coordinates { (6,94.14)(6,25.61) };
\addplot[mark=*,red] coordinates { (6,71.89) };
\addplot[mark=*,red] coordinates { (6,52.28) };

\addplot[mark=*,black] coordinates { (5,30.20) };
\addplot[mark=-,black] coordinates { (5,75.63)(5,0.00) };
\addplot[mark=*,red] coordinates { (5,42.71) };
\addplot[mark=*,red] coordinates { (5,17.69) };

\addplot[mark=*,black] coordinates { (8,91.88) };
\addplot[mark=-,black] coordinates { (8,100.00)(8,0.00) };
\addplot[mark=*,red] coordinates { (8,100.00) };
\addplot[mark=*,red] coordinates { (8,79.68) };

\addplot[mark=*,black] coordinates { (7,76.02) };
\addplot[mark=-,black] coordinates { (7,100.00)(7,0.00) };
\addplot[mark=*,red] coordinates { (7,93.83) };
\addplot[mark=*,red] coordinates { (7,58.20) };

\addplot[mark=*,black] coordinates { (9,90.84) };
\addplot[mark=-,black] coordinates { (9,100.00)(9,0.00) };
\addplot[mark=*,red] coordinates { (9,100.00) };
\addplot[mark=*,red] coordinates { (9,80.49) };

\addplot[mark=*,black] coordinates { (1,91.39) };
\addplot[mark=-,black] coordinates { (1,100.00)(1,0.00) };
\addplot[mark=*,red] coordinates { (1,100.00) };
\addplot[mark=*,red] coordinates { (1,79.09) };

\addplot[mark=*,black] coordinates { (4,11.11) };
\addplot[mark=-,black] coordinates { (4,51.75)(4,0.00) };
\addplot[mark=*,red] coordinates { (4,20.37) };
\addplot[mark=*,red] coordinates { (4,1.85) };

\addplot[mark=*,black] coordinates { (3,78.14) };
\addplot[mark=-,black] coordinates { (3,100.00)(3,0.00) };
\addplot[mark=*,red] coordinates { (3,92.55) };
\addplot[mark=*,red] coordinates { (3,63.74) };

\addplot[mark=*,black] coordinates { (2,89.39) };
\addplot[mark=-,black] coordinates { (2,100.00)(2,0.00) };
\addplot[mark=*,red] coordinates { (2,100.00) };
\addplot[mark=*,red] coordinates { (2,75.11) };

            \end{axis}
            \end{tikzpicture}  
        } &
    
    \resizebox{.45\textwidth}{!}{
            \begin{tikzpicture}
            \begin{axis}[
                title={20\% Transformation Chance},
                ylabel={Similarity (\%)},
                xmin=-1, xmax=11,
                ymin=0.0, ymax=100,
                xtick={0,1,2,3,4,5,6,7,8,9,10},
                xticklabel style={rotate=90},
                xticklabels={
                    JPlag,Plaggie,Sim,Sherlock-W,Sherlock-S,String Tile,String ED,Token Tile,Token ED,Tree ED,Graph ED
                }
            ]
            
\addplot[mark=*,black] coordinates { (0,89.62) };
\addplot[mark=-,black] coordinates { (0,100.00)(0,0.00) };
\addplot[mark=*,red] coordinates { (0,99.96) };
\addplot[mark=*,red] coordinates { (0,79.29) };

\addplot[mark=*,black] coordinates { (10,81.78) };
\addplot[mark=-,black] coordinates { (10,100.00)(10,0.00) };
\addplot[mark=*,red] coordinates { (10,99.43) };
\addplot[mark=*,red] coordinates { (10,64.13) };

\addplot[mark=*,black] coordinates { (6,56.37) };
\addplot[mark=-,black] coordinates { (6,90.49)(6,0.00) };
\addplot[mark=*,red] coordinates { (6,66.16) };
\addplot[mark=*,red] coordinates { (6,46.57) };

\addplot[mark=*,black] coordinates { (5,26.09) };
\addplot[mark=-,black] coordinates { (5,75.63)(5,0.00) };
\addplot[mark=*,red] coordinates { (5,38.33) };
\addplot[mark=*,red] coordinates { (5,13.85) };

\addplot[mark=*,black] coordinates { (8,87.55) };
\addplot[mark=-,black] coordinates { (8,100.00)(8,0.00) };
\addplot[mark=*,red] coordinates { (8,99.76) };
\addplot[mark=*,red] coordinates { (8,75.34) };

\addplot[mark=*,black] coordinates { (7,66.88) };
\addplot[mark=-,black] coordinates { (7,100.00)(7,0.00) };
\addplot[mark=*,red] coordinates { (7,85.42) };
\addplot[mark=*,red] coordinates { (7,48.35) };

\addplot[mark=*,black] coordinates { (9,85.14) };
\addplot[mark=-,black] coordinates { (9,100.00)(9,27.55) };
\addplot[mark=*,red] coordinates { (9,96.64) };
\addplot[mark=*,red] coordinates { (9,73.65) };

\addplot[mark=*,black] coordinates { (1,85.51) };
\addplot[mark=-,black] coordinates { (1,100.00)(1,0.00) };
\addplot[mark=*,red] coordinates { (1,100.00) };
\addplot[mark=*,red] coordinates { (1,70.53) };

\addplot[mark=*,black] coordinates { (4,9.84) };
\addplot[mark=-,black] coordinates { (4,46.00)(4,0.00) };
\addplot[mark=*,red] coordinates { (4,18.11) };
\addplot[mark=*,red] coordinates { (4,1.56) };

\addplot[mark=*,black] coordinates { (3,69.38) };
\addplot[mark=-,black] coordinates { (3,100.00)(3,0.00) };
\addplot[mark=*,red] coordinates { (3,84.87) };
\addplot[mark=*,red] coordinates { (3,53.88) };

\addplot[mark=*,black] coordinates { (2,82.92) };
\addplot[mark=-,black] coordinates { (2,100.00)(2,0.00) };
\addplot[mark=*,red] coordinates { (2,98.21) };
\addplot[mark=*,red] coordinates { (2,67.62) };

            \end{axis}
            \end{tikzpicture}  
        } \\
        
                \resizebox{.45\textwidth}{!}{
            \begin{tikzpicture}
            \begin{axis}[
                title={40\% Transformation Chance},
                ylabel={Similarity (\%)},
                xmin=-1, xmax=11,
                ymin=0.0, ymax=100,
                xtick={0,1,2,3,4,5,6,7,8,9,10},
                xticklabel style={rotate=90},
                xticklabels={
                    JPlag,Plaggie,Sim,Sherlock-W,Sherlock-S,String Tile,String ED,Token Tile,Token ED,Tree ED,Graph ED
                }
            ]
            
\addplot[mark=*,black] coordinates { (0,81.69) };
\addplot[mark=-,black] coordinates { (0,100.00)(0,13.87) };
\addplot[mark=*,red] coordinates { (0,92.91) };
\addplot[mark=*,red] coordinates { (0,70.46) };

\addplot[mark=*,black] coordinates { (10,71.46) };
\addplot[mark=-,black] coordinates { (10,100.00)(10,0.00) };
\addplot[mark=*,red] coordinates { (10,89.85) };
\addplot[mark=*,red] coordinates { (10,53.07) };

\addplot[mark=*,black] coordinates { (6,48.44) };
\addplot[mark=-,black] coordinates { (6,87.23)(6,0.00) };
\addplot[mark=*,red] coordinates { (6,57.67) };
\addplot[mark=*,red] coordinates { (6,39.21) };

\addplot[mark=*,black] coordinates { (5,20.29) };
\addplot[mark=-,black] coordinates { (5,73.37)(5,2.47) };
\addplot[mark=*,red] coordinates { (5,31.74) };
\addplot[mark=*,red] coordinates { (5,8.84) };

\addplot[mark=*,black] coordinates { (8,79.48) };
\addplot[mark=-,black] coordinates { (8,99.87)(8,0.00) };
\addplot[mark=*,red] coordinates { (8,92.35) };
\addplot[mark=*,red] coordinates { (8,66.61) };

\addplot[mark=*,black] coordinates { (7,54.80) };
\addplot[mark=-,black] coordinates { (7,99.80)(7,0.00) };
\addplot[mark=*,red] coordinates { (7,73.57) };
\addplot[mark=*,red] coordinates { (7,36.03) };

\addplot[mark=*,black] coordinates { (9,74.80) };
\addplot[mark=-,black] coordinates { (9,99.85)(9,0.00) };
\addplot[mark=*,red] coordinates { (9,87.67) };
\addplot[mark=*,red] coordinates { (9,61.93) };

\addplot[mark=*,black] coordinates { (1,74.16) };
\addplot[mark=-,black] coordinates { (1,100.00)(1,0.00) };
\addplot[mark=*,red] coordinates { (1,91.59) };
\addplot[mark=*,red] coordinates { (1,56.74) };

\addplot[mark=*,black] coordinates { (4,7.56) };
\addplot[mark=-,black] coordinates { (4,45.25)(4,0.00) };
\addplot[mark=*,red] coordinates { (4,14.37) };
\addplot[mark=*,red] coordinates { (4,0.75) };

\addplot[mark=*,black] coordinates { (3,55.13) };
\addplot[mark=-,black] coordinates { (3,98.25)(3,0.00) };
\addplot[mark=*,red] coordinates { (3,71.22) };
\addplot[mark=*,red] coordinates { (3,39.05) };

\addplot[mark=*,black] coordinates { (2,70.44) };
\addplot[mark=-,black] coordinates { (2,100.00)(2,0.00) };
\addplot[mark=*,red] coordinates { (2,86.12) };
\addplot[mark=*,red] coordinates { (2,54.75) };
            
            \end{axis}
            \end{tikzpicture}  
        } &
    
    \resizebox{.45\textwidth}{!}{
            \begin{tikzpicture}
            \begin{axis}[
                title={60\% Transformation Chance},
                ylabel={Similarity (\%)},
                xmin=-1, xmax=11,
                ymin=0.0, ymax=100,
                xtick={0,1,2,3,4,5,6,7,8,9,10},
                xticklabel style={rotate=90},
                xticklabels={
                    JPlag,Plaggie,Sim,Sherlock-W,Sherlock-S,String Tile,String ED,Token Tile,Token ED,Tree ED,Graph ED
                }
            ]
            
\addplot[mark=*,black] coordinates { (0,73.30) };
\addplot[mark=-,black] coordinates { (0,99.64)(0,0.00) };
\addplot[mark=*,red] coordinates { (0,86.18) };
\addplot[mark=*,red] coordinates { (0,60.42) };

\addplot[mark=*,black] coordinates { (10,63.75) };
\addplot[mark=-,black] coordinates { (10,100.00)(10,0.00) };
\addplot[mark=*,red] coordinates { (10,81.32) };
\addplot[mark=*,red] coordinates { (10,46.18) };

\addplot[mark=*,black] coordinates { (6,42.80) };
\addplot[mark=-,black] coordinates { (6,81.90)(6,0.00) };
\addplot[mark=*,red] coordinates { (6,51.58) };
\addplot[mark=*,red] coordinates { (6,34.01) };

\addplot[mark=*,black] coordinates { (5,16.31) };
\addplot[mark=-,black] coordinates { (5,67.30)(5,0.00) };
\addplot[mark=*,red] coordinates { (5,27.38) };
\addplot[mark=*,red] coordinates { (5,5.24) };

\addplot[mark=*,black] coordinates { (8,72.19) };
\addplot[mark=-,black] coordinates { (8,99.19)(8,0.00) };
\addplot[mark=*,red] coordinates { (8,84.76) };
\addplot[mark=*,red] coordinates { (8,59.61) };

\addplot[mark=*,black] coordinates { (7,45.61) };
\addplot[mark=-,black] coordinates { (7,98.84)(7,0.00) };
\addplot[mark=*,red] coordinates { (7,63.85) };
\addplot[mark=*,red] coordinates { (7,27.36) };

\addplot[mark=*,black] coordinates { (9,65.31) };
\addplot[mark=-,black] coordinates { (9,98.72)(9,0.00) };
\addplot[mark=*,red] coordinates { (9,78.23) };
\addplot[mark=*,red] coordinates { (9,52.38) };

\addplot[mark=*,black] coordinates { (1,62.81) };
\addplot[mark=-,black] coordinates { (1,100.00)(1,0.00) };
\addplot[mark=*,red] coordinates { (1,81.44) };
\addplot[mark=*,red] coordinates { (1,44.18) };

\addplot[mark=*,black] coordinates { (4,5.99) };
\addplot[mark=-,black] coordinates { (4,47.00)(4,0.00) };
\addplot[mark=*,red] coordinates { (4,11.83) };
\addplot[mark=*,red] coordinates { (4,0.15) };

\addplot[mark=*,black] coordinates { (3,43.29) };
\addplot[mark=-,black] coordinates { (3,96.00)(3,0.00) };
\addplot[mark=*,red] coordinates { (3,58.97) };
\addplot[mark=*,red] coordinates { (3,27.61) };

\addplot[mark=*,black] coordinates { (2,58.21) };
\addplot[mark=-,black] coordinates { (2,99.00)(2,0.00) };
\addplot[mark=*,red] coordinates { (2,74.68) };
\addplot[mark=*,red] coordinates { (2,41.75) };
            
            \end{axis}
            \end{tikzpicture}  
        } \\
        
                \resizebox{.45\textwidth}{!}{
            \begin{tikzpicture}
            \begin{axis}[
                title={80\% Transformation Chance},
                ylabel={Similarity (\%)},
                xmin=-1, xmax=11,
                ymin=0.0, ymax=100,
                xtick={0,1,2,3,4,5,6,7,8,9,10},
                xticklabel style={rotate=90},
                xticklabels={
                    JPlag,Plaggie,Sim,Sherlock-W,Sherlock-S,String Tile,String ED,Token Tile,Token ED,Tree ED,Graph ED
                }
            ]
            
\addplot[mark=*,black] coordinates { (0,65.32) };
\addplot[mark=-,black] coordinates { (0,99.64)(0,8.29) };
\addplot[mark=*,red] coordinates { (0,78.78) };
\addplot[mark=*,red] coordinates { (0,51.87) };

\addplot[mark=*,black] coordinates { (10,58.20) };
\addplot[mark=-,black] coordinates { (10,100.00)(10,0.00) };
\addplot[mark=*,red] coordinates { (10,74.32) };
\addplot[mark=*,red] coordinates { (10,42.09) };

\addplot[mark=*,black] coordinates { (6,39.00) };
\addplot[mark=-,black] coordinates { (6,82.11)(6,0.00) };
\addplot[mark=*,red] coordinates { (6,47.28) };
\addplot[mark=*,red] coordinates { (6,30.71) };

\addplot[mark=*,black] coordinates { (5,13.36) };
\addplot[mark=-,black] coordinates { (5,72.34)(5,0.00) };
\addplot[mark=*,red] coordinates { (5,23.48) };
\addplot[mark=*,red] coordinates { (5,3.24) };

\addplot[mark=*,black] coordinates { (8,65.24) };
\addplot[mark=-,black] coordinates { (8,96.31)(8,0.00) };
\addplot[mark=*,red] coordinates { (8,77.36) };
\addplot[mark=*,red] coordinates { (8,53.11) };

\addplot[mark=*,black] coordinates { (7,37.57) };
\addplot[mark=-,black] coordinates { (7,93.49)(7,0.00) };
\addplot[mark=*,red] coordinates { (7,54.60) };
\addplot[mark=*,red] coordinates { (7,20.54) };

\addplot[mark=*,black] coordinates { (9,56.42) };
\addplot[mark=-,black] coordinates { (9,94.48)(9,0.00) };
\addplot[mark=*,red] coordinates { (9,68.63) };
\addplot[mark=*,red] coordinates { (9,44.20) };

\addplot[mark=*,black] coordinates { (1,51.58) };
\addplot[mark=-,black] coordinates { (1,100.00)(1,0.00) };
\addplot[mark=*,red] coordinates { (1,70.08) };
\addplot[mark=*,red] coordinates { (1,33.09) };

\addplot[mark=*,black] coordinates { (4,4.98) };
\addplot[mark=-,black] coordinates { (4,35.50)(4,0.00) };
\addplot[mark=*,red] coordinates { (4,10.25) };
\addplot[mark=*,red] coordinates { (4,0.00) };

\addplot[mark=*,black] coordinates { (3,33.42) };
\addplot[mark=-,black] coordinates { (3,88.50)(3,0.00) };
\addplot[mark=*,red] coordinates { (3,47.19) };
\addplot[mark=*,red] coordinates { (3,19.65) };

\addplot[mark=*,black] coordinates { (2,46.02) };
\addplot[mark=-,black] coordinates { (2,95.00)(2,0.00) };
\addplot[mark=*,red] coordinates { (2,61.53) };
\addplot[mark=*,red] coordinates { (2,30.51) };
            
            \end{axis}
            \end{tikzpicture}  
        } &
    
    \resizebox{.45\textwidth}{!}{
            \begin{tikzpicture}
            \begin{axis}[
                title={100\% Transformation Chance},
                ylabel={Similarity (\%)},
                xmin=-1, xmax=11,
                ymin=0.0, ymax=100,
                xtick={0,1,2,3,4,5,6,7,8,9,10},
                xticklabel style={rotate=90},
                xticklabels={
                    JPlag,Plaggie,Sim,Sherlock-W,Sherlock-S,String Tile,String ED,Token Tile,Token ED,Tree ED,Graph ED
                }
            ]
            
\addplot[mark=*,black] coordinates { (0,57.54) };
\addplot[mark=-,black] coordinates { (0,89.84)(0,0.00) };
\addplot[mark=*,red] coordinates { (0,71.37) };
\addplot[mark=*,red] coordinates { (0,43.70) };

\addplot[mark=*,black] coordinates { (10,53.92) };
\addplot[mark=-,black] coordinates { (10,91.35)(10,0.00) };
\addplot[mark=*,red] coordinates { (10,67.74) };
\addplot[mark=*,red] coordinates { (10,40.10) };

\addplot[mark=*,black] coordinates { (6,36.05) };
\addplot[mark=-,black] coordinates { (6,75.37)(6,0.00) };
\addplot[mark=*,red] coordinates { (6,43.94) };
\addplot[mark=*,red] coordinates { (6,28.16) };

\addplot[mark=*,black] coordinates { (5,11.52) };
\addplot[mark=-,black] coordinates { (5,64.39)(5,0.00) };
\addplot[mark=*,red] coordinates { (5,21.08) };
\addplot[mark=*,red] coordinates { (5,1.95) };

\addplot[mark=*,black] coordinates { (8,58.91) };
\addplot[mark=-,black] coordinates { (8,89.86)(8,0.00) };
\addplot[mark=*,red] coordinates { (8,69.98) };
\addplot[mark=*,red] coordinates { (8,47.84) };

\addplot[mark=*,black] coordinates { (7,30.62) };
\addplot[mark=-,black] coordinates { (7,82.66)(7,0.00) };
\addplot[mark=*,red] coordinates { (7,45.83) };
\addplot[mark=*,red] coordinates { (7,15.40) };

\addplot[mark=*,black] coordinates { (9,48.27) };
\addplot[mark=-,black] coordinates { (9,84.83)(9,0.00) };
\addplot[mark=*,red] coordinates { (9,58.88) };
\addplot[mark=*,red] coordinates { (9,37.66) };

\addplot[mark=*,black] coordinates { (1,40.83) };
\addplot[mark=-,black] coordinates { (1,88.89)(1,0.00) };
\addplot[mark=*,red] coordinates { (1,58.05) };
\addplot[mark=*,red] coordinates { (1,23.61) };

\addplot[mark=*,black] coordinates { (4,4.21) };
\addplot[mark=-,black] coordinates { (4,36.00)(4,0.00) };
\addplot[mark=*,red] coordinates { (4,9.09) };
\addplot[mark=*,red] coordinates { (4,0.00) };

\addplot[mark=*,black] coordinates { (3,25.34) };
\addplot[mark=-,black] coordinates { (3,76.00)(3,0.00) };
\addplot[mark=*,red] coordinates { (3,37.37) };
\addplot[mark=*,red] coordinates { (3,13.32) };

\addplot[mark=*,black] coordinates { (2,34.95) };
\addplot[mark=-,black] coordinates { (2,82.50)(2,0.00) };
\addplot[mark=*,red] coordinates { (2,48.82) };
\addplot[mark=*,red] coordinates { (2,21.08) };
            
            \end{axis}
            \end{tikzpicture}  
        } \\
        
    \end{tabular}

    \caption{Average similarity of variants generated with all 14 source code transformations, evaluated using all 11 SCPDTs. Bars indicate range of similarity scores. Red marks indicate range of standard deviation around the average.}
    \label{fig:eb2cacgsimilarity}
\end{figure}

%% file: Tab8.tex
\begin{table}[htbp]
    \caption{Average number of applied transformations to each program variant created with all transformations.}
    \label{tab:ev2cavgtransformations}
    \begin{tabular}{p{4cm}|cccccc}
        \hline\noalign{\smallskip}
        {Transformation Chance} & {10\%} & {20\%} & {40\%} & {60\%} & {80\%} & {100\%} \\
        \noalign{\smallskip}
        \cline{2-7}
        \noalign{\smallskip}
        {Avg. No. of Transformations} & 21.63 & 41.76 & 80.03 & 120.71 & 165.4 & 212.3 \\
        \noalign{\smallskip}\hline
    \end{tabular}
\end{table}

%% file: Fig12_ev1c_resilience.tex
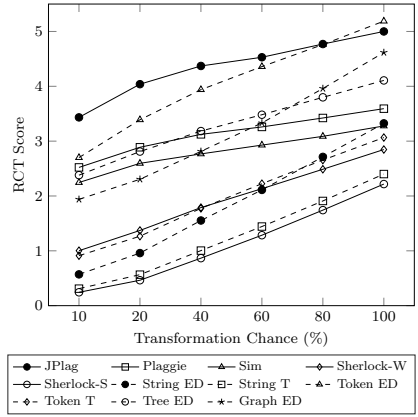
\begin{figure}[htbp]
    \centering
    \resizebox{.45\linewidth}{!}{
     \begin{tikzpicture}
        \begin{axis}[
            xlabel={Transformation Chance (\%)},
            ylabel={RCT Score},
            xmin=0.5, xmax=6.5,
            ymin=0.0, ymax=5.5,
            xtick={0,1,2,3,4,5,6,7},
            xticklabels={
                ,10,20,40,60,80,100
            }
        ]
            
	\addplot[mark=*,black] coordinates {
        (1,3.433333333)
        (2,4.03868472)
        (3,4.370835609)
        (4,4.52945591)
        (5,4.770695125)
        (6,5)
	};

	\addplot[dashed,color=black,mark=star,mark options={solid}] coordinates {
        (1,1.938172043)
        (2,2.304635762)
        (3,2.813005272)
        (4,3.331769252)
        (5,3.959779746)
        (6,4.616220918)
	};

	\addplot[dashed,color=black,mark=*,mark options={solid}] coordinates {
        (1,0.570561857)
        (2,0.958677686)
        (3,1.552774544)
        (4,2.110683686)
        (5,2.712809578)
        (6,3.324459756)
	};

	\addplot[dashed,color=black,mark=square,mark options={solid}] coordinates {
        (1,0.310107527)
        (2,0.565240931)
        (3,1.004014553)
        (4,1.442691526)
        (5,1.909269306)
        (6,2.399683509)
	};

	\addplot[dashed,color=black,mark=triangle,mark options={solid}] coordinates {
        (1,2.700374532)
        (2,3.38961039)
        (3,3.938484252)
        (4,4.3593355)
        (5,4.761082326)
        (6,5.188172043)
	};

	\addplot[dashed,color=black,mark=diamond,mark options={solid}] coordinates {
        (1,0.911120472)
        (2,1.266606005)
        (3,1.774107737)
        (4,2.223839352)
        (5,2.655321882)
        (6,3.065703971)
	};

	\addplot[dashed,color=black,mark=o,mark options={solid}] coordinates {
        (1,2.379537954)
        (2,2.810228802)
        (3,3.183373111)
        (4,3.481684453)
        (5,3.798805696)
        (6,4.107177404)
	};

	\addplot[color=black,mark=square] coordinates {
        (1,2.520979021)
        (2,2.887966805)
        (3,3.12373146)
        (4,3.258909287)
        (5,3.422304987)
        (6,3.592216582)
	};

	\addplot[color=black,mark=o] coordinates {
        (1,0.243608514)
        (2,0.463536464)
        (3,0.866313055)
        (4,1.284695615)
        (5,1.741969458)
        (6,2.217232376)
	};

	\addplot[color=black,mark=diamond] coordinates {
        (1,1.00231696)
        (2,1.371428571)
        (3,1.788379888)
        (4,2.133062379)
        (5,2.487218045)
        (6,2.846989406)
	};

	\addplot[color=black,mark=triangle] coordinates {	
        (1,2.246105919)
        (2,2.59540087)
        (3,2.767289073)
        (4,2.927722532)
        (5,3.087549001)
        (6,3.276740238)
	};
            
        \end{axis}
    \end{tikzpicture}  
    }
    \\
    \resizebox{.45\linewidth}{!}{
    \begin{tikzpicture}
        \begin{axis}[
            hide axis,
            xmin=0,
            xmax=0,
            ymin=0,
            ymax=0,
            legend style={draw=white!15!black,legend cell align=left},
            legend columns = 4
        ]
        
        \addlegendimage{color=black,mark=*}
        \addlegendentry{JPlag}
        
       \addlegendimage{color=black,mark=square}
        \addlegendentry{Plaggie}
        
        \addlegendimage{color=black,mark=triangle}
        \addlegendentry{Sim}
        
        \addlegendimage{color=black,mark=diamond}
        \addlegendentry{Sherlock-W}
        
        \addlegendimage{color=black,mark=o}
        \addlegendentry{Sherlock-S}
        
        \addlegendimage{dashed,color=black,mark=*,mark options={solid}}
        \addlegendentry{String ED}
        
        \addlegendimage{dashed,color=black,mark=square,mark options={solid}}
        \addlegendentry{String T}
        
        \addlegendimage{dashed,color=black,mark=triangle,mark options={solid}}
        \addlegendentry{Token ED}
        
        \addlegendimage{dashed,color=black,mark=diamond,mark options={solid}}
        \addlegendentry{Token T}
        
        \addlegendimage{dashed,color=black,mark=o,mark options={solid}}
        \addlegendentry{Tree ED}
        
        \addlegendimage{dashed,color=black,mark=star,mark options={solid}}
        \addlegendentry{Graph ED}
    
        \end{axis}
    \end{tikzpicture}
    }
    \caption{Average RCT scores for variants generated with all transformations.}
    \label{fig:ev2crctranking}
\end{figure}

%% file: Tab9.tex
\begin{table}[htbp]
    \caption{Average logical lines of code injected and increase in size (\%) for each type of injected source code fragment at each chance of injection.}
    \label{tab:ev3aavginjectloc}
    \begin{tabular}{rcccccc}
        \hline\noalign{\smallskip}
        \multirow{2}{*}{} & \multicolumn{6}{c}{{Injection Chance}} \\
        & {10\%} & {20\%} & {40\%} & {60\%} & {80\%} & {100\%} \\
        \noalign{\smallskip}\hline\noalign{\smallskip}
        \multirow{2}{*}{{File}}     & 38.22 & 71.27 & 115.38 & 187.19 & 282.88 & 322.85 \\
                                    & 11.37\% & 21.21\% & 34.34\% & 55.71\% & 84.19\% & 96.08\% \\
        \noalign{\smallskip}
        \multirow{2}{*}{{Class}}        & 222.01 & 242.71 & 312.69 & 437.72 & 598.56 & 793.96 \\
                                        & 66.07\% & 72.23\% & 93.05\% & 130.26\% & 178.13\% & 236.28\% \\
        \noalign{\smallskip}
        \multirow{2}{*}{{Method}}       & 75.87 & 107.25 & 173.07 & 265.28 & 360.55 & 471.87 \\
                                        & 22.58\% & 31.92\% & 51.50\% & 78.95\% & 107.30\% & 140.42\% \\
        \noalign{\smallskip}
        \multirow{2}{*}{{Statement}}    & 35.86 & 52.64 & 83.35 & 117.09 & 151.54 & 185.60 \\
                                        & 10.67\% & 15.67\% & 24.80\% & 34.85\% & 45.10\% & 55.23\% \\
        \noalign{\smallskip}\hline
    \end{tabular}
\end{table}

%% file: Fig13_ev2a_heatmap.tex
\begin{figure}[htbp]
    \begin{tabular}{p{0.45\linewidth}p{0.45\linewidth}}
         
        \resizebox{0.45\textwidth}{!}{
            
            \includegraphics[width=1.0\linewidth]{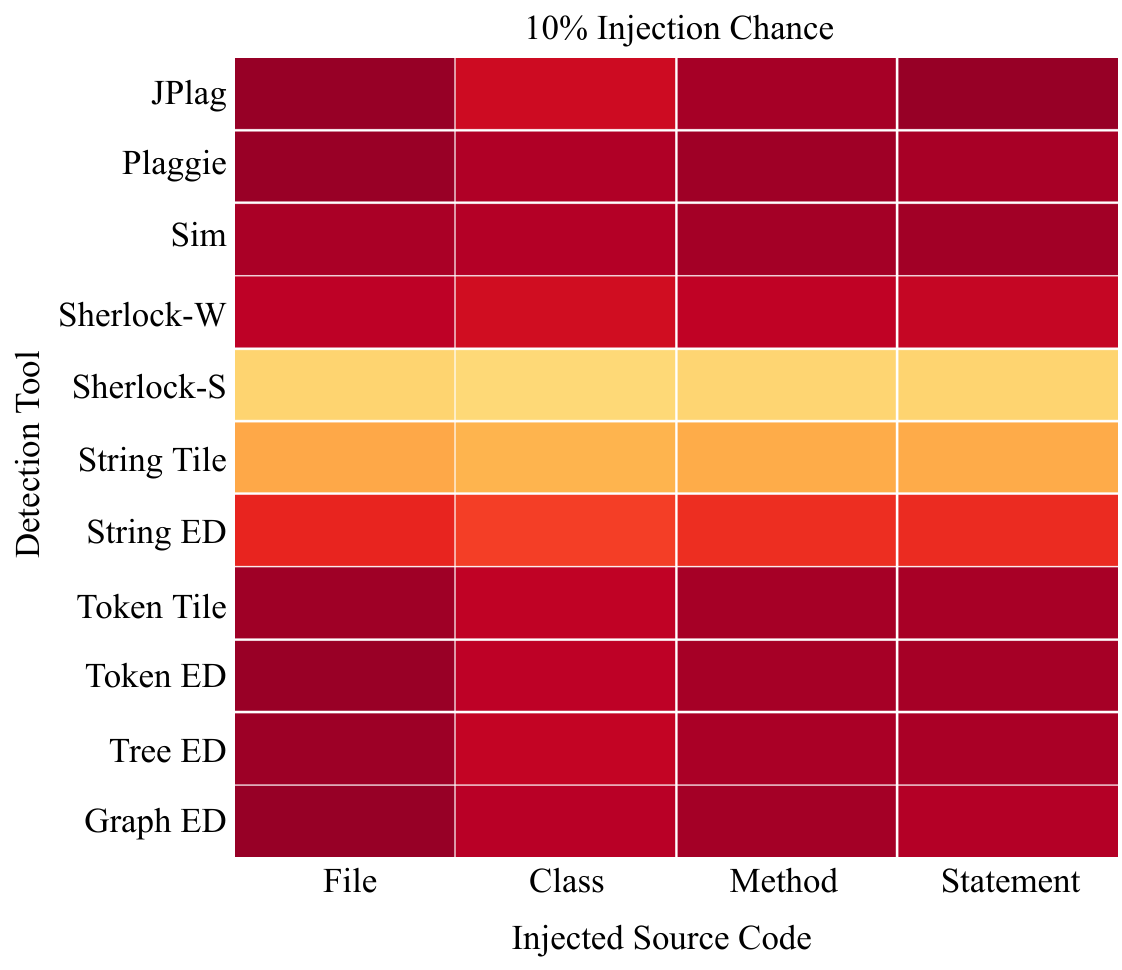}  
        } &
    
        \resizebox{0.45\textwidth}{!}{
            
            \includegraphics[width=1.0\linewidth]{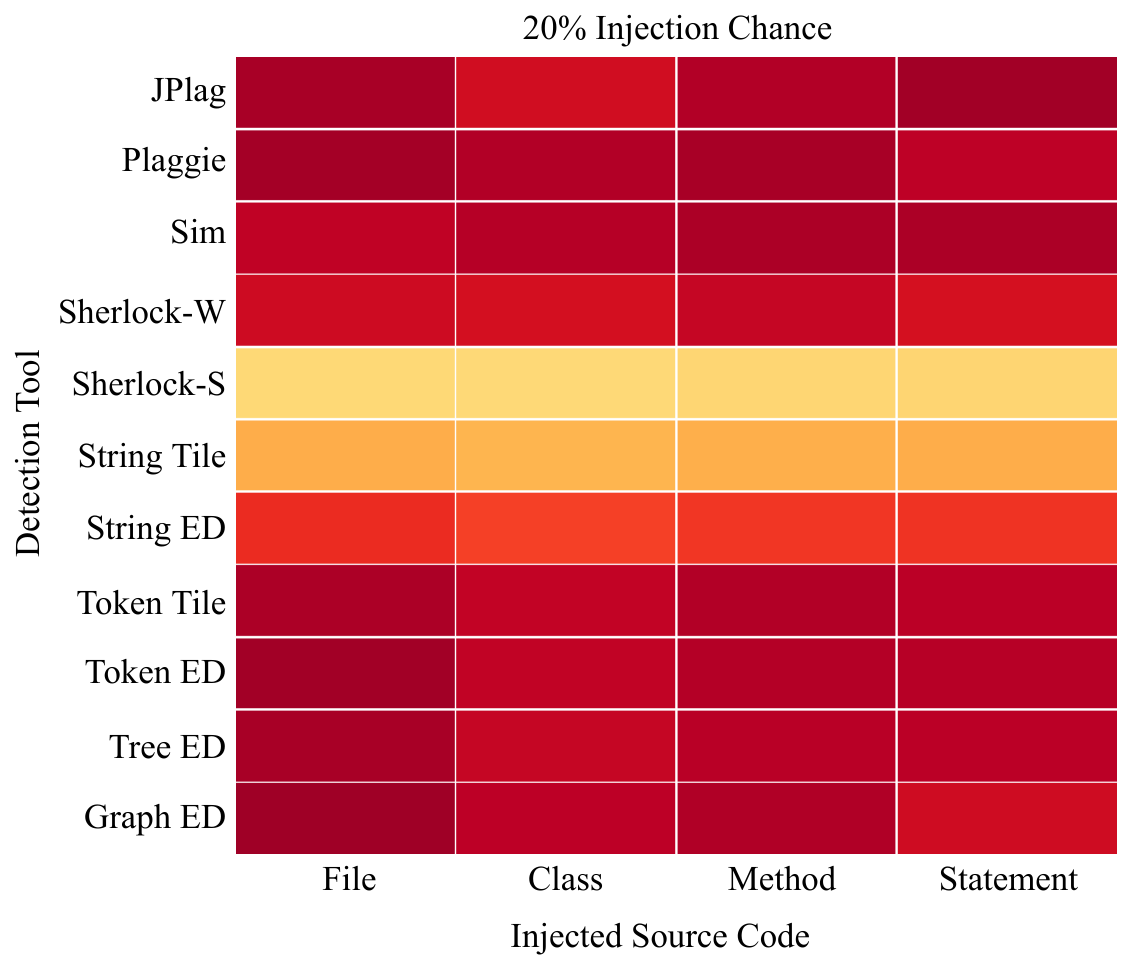}  
        } \\
        
        \resizebox{0.45\textwidth}{!}{
            
            \includegraphics[width=1.0\linewidth]{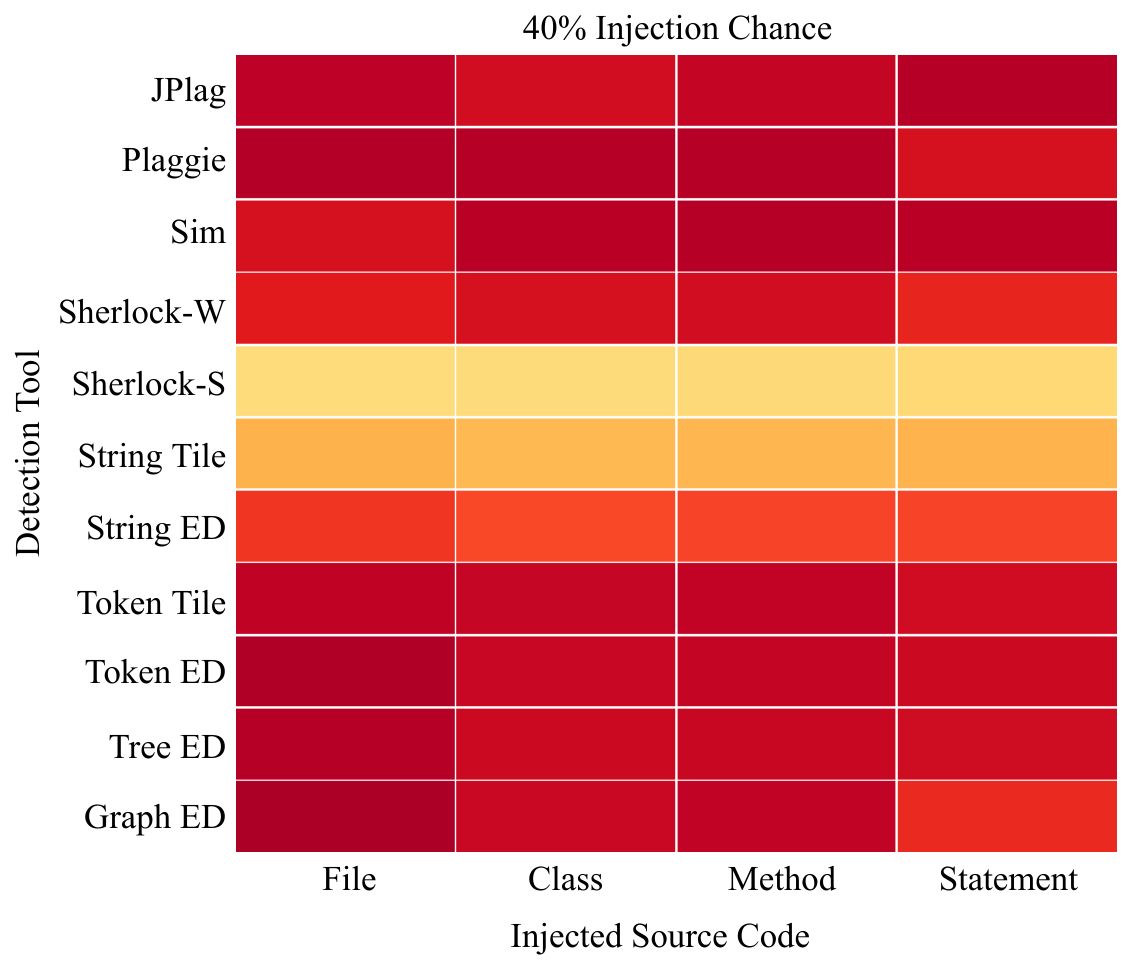}  
        } &
    
        \resizebox{0.45\textwidth}{!}{
            
            \includegraphics[width=1.0\linewidth]{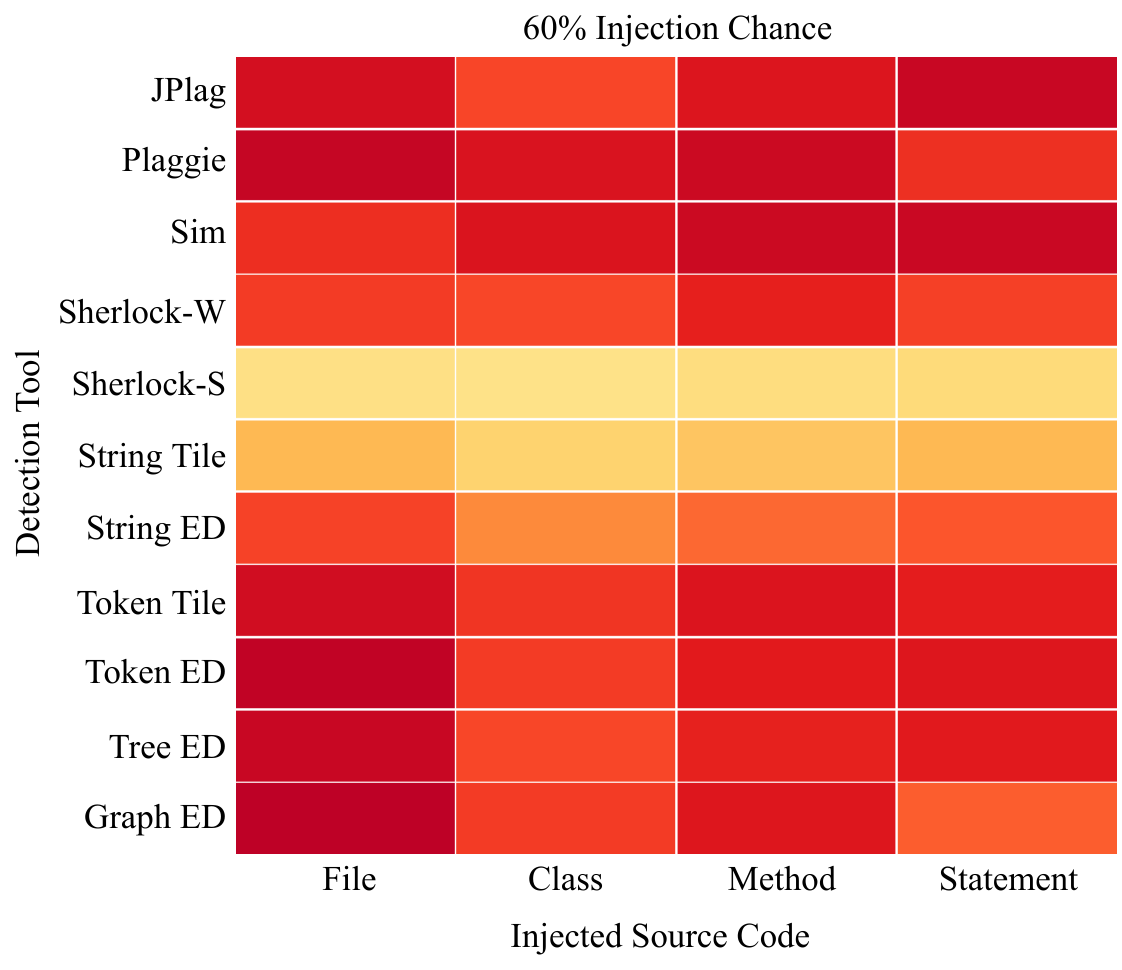}  
        } \\
        
        \resizebox{0.45\textwidth}{!}{
            
            \includegraphics[width=1.0\linewidth]{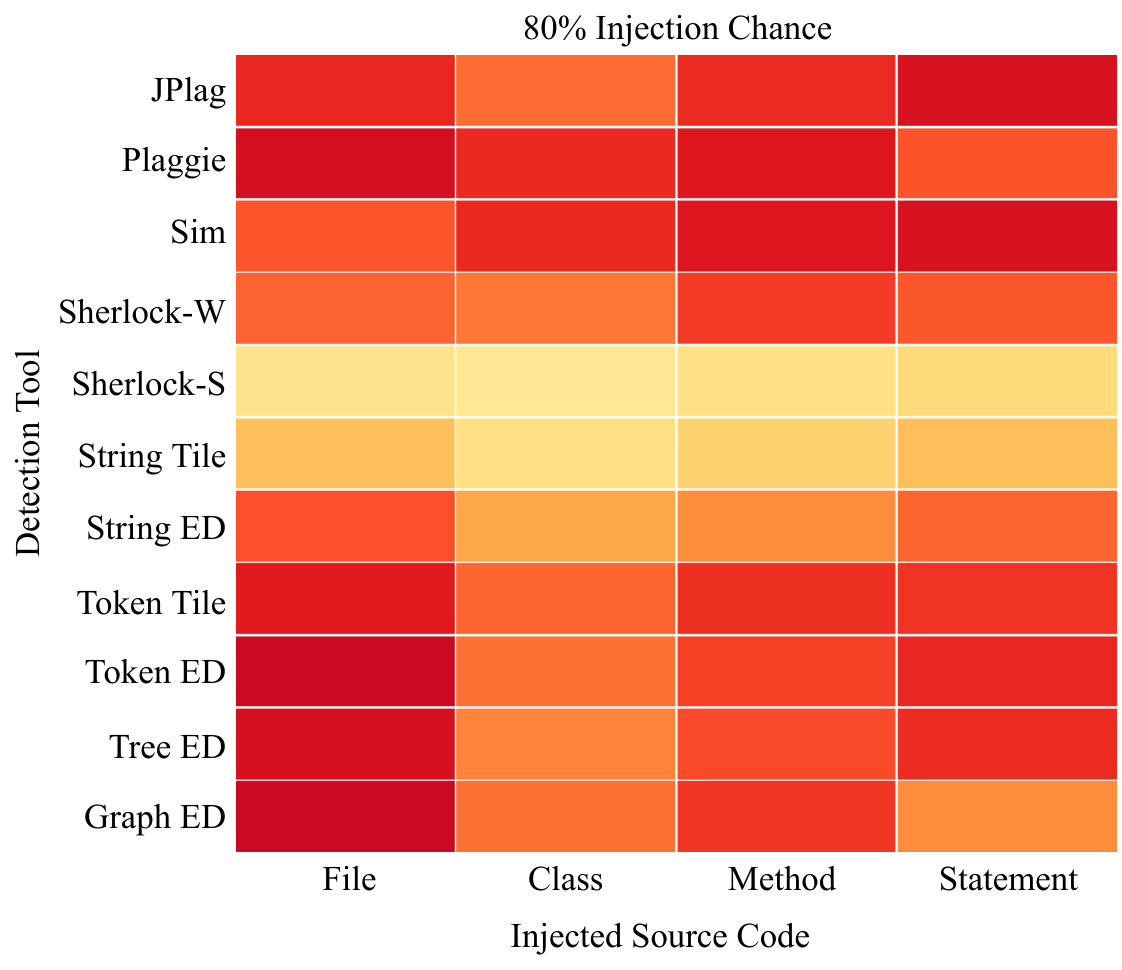}  
        } &
    
        \resizebox{0.45\textwidth}{!}{
            
            \includegraphics[width=1.0\linewidth]{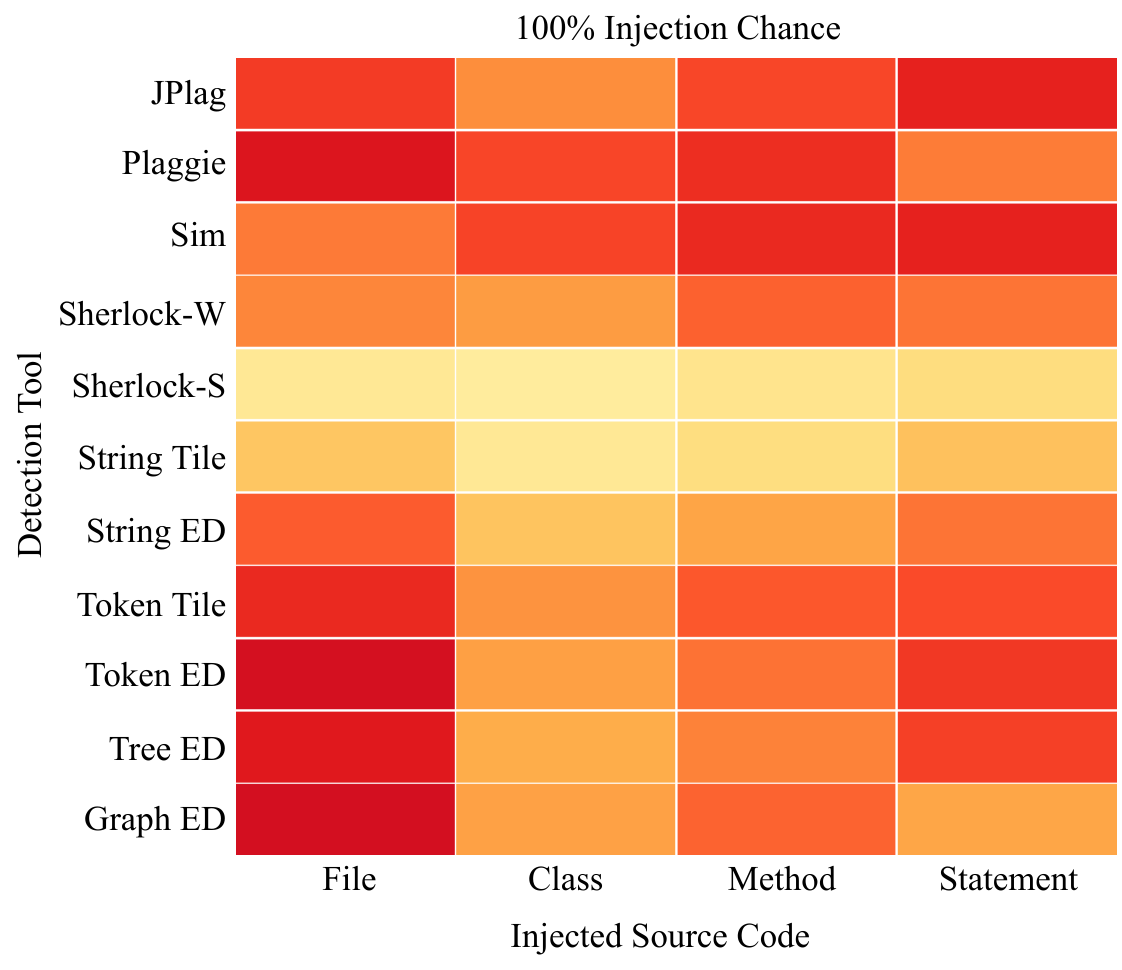}  
        } \\
        
        \multicolumn{2}{c}{
        \resizebox{0.45\textwidth}{!}{
            
            \includegraphics[width=1.0\linewidth]{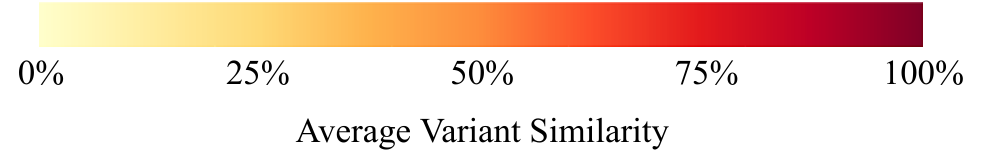} 
        }} \\
        
    \end{tabular}

    \caption{Heat maps representing the average similarity of variants generated with each individual source code fragments injected. Darker colours indicate a higher similarity scores, and hence higher robustness to transformation.}
    \label{fig:ev3aheatmaps}
\end{figure}

%% file: Fig14_ev2a_resilience.tex
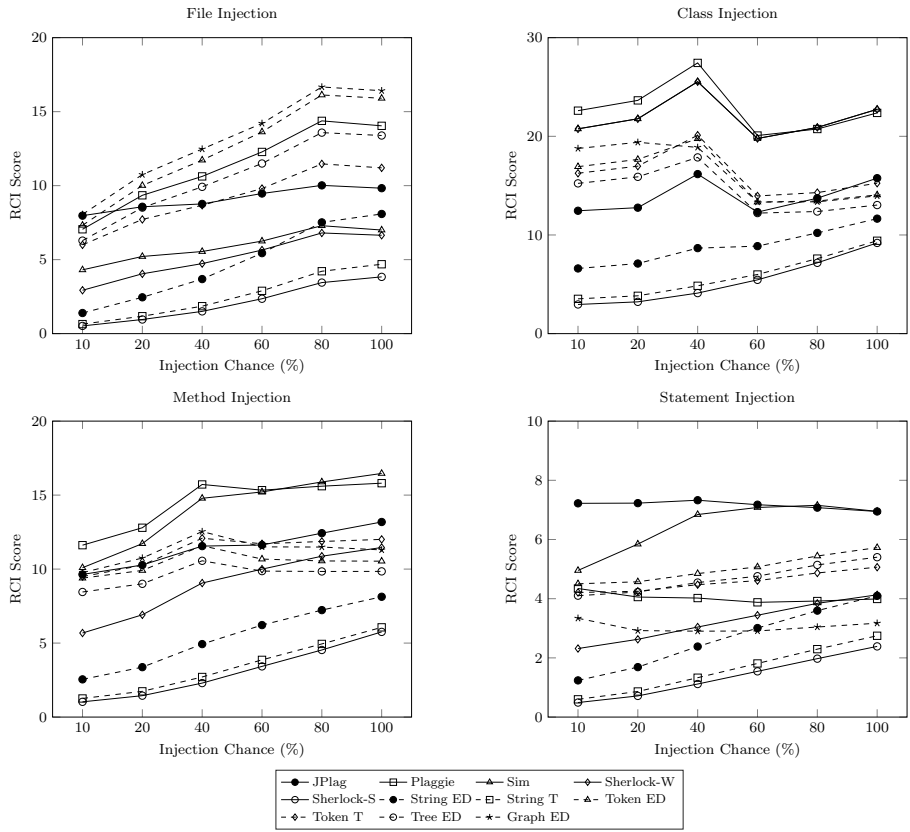
\begin{figure}[htbp]

    \begin{tabular}{p{.5\linewidth}p{.5\linewidth}}
         
        \resizebox{.45\textwidth}{!}{
            \begin{tikzpicture}
            \begin{axis}[
                title={File Injection},
                xlabel={Injection Chance (\%)},
                ylabel={RCI Score},
                xmin=0.5, xmax=6.5,
                ymin=0.0, ymax=20,
                xtick={0,1,2,3,4,5,6,7},
                xticklabels={
                    ,10,20,40,60,80,100
                }
            ]
            
	\addplot[mark=*,black] coordinates {
(1,7.972448393075384)
(2,8.572773240663395)
(3,8.764750425790623)
(4,9.469043840513283)
(5,10.018178731523138)
(6,9.828388530043322)
	};

	\addplot[dashed,color=black,mark=star,mark options={solid}] coordinates {
(1,8.056143352315832)
(2,10.745704515379497)
(3,12.47257134245094)
(4,14.207495735224258)
(5,16.667199412984285)
(6,16.416727002947095)
	};

	\addplot[dashed,color=black,mark=*,mark options={solid}] coordinates {
(1,1.3927540808018963)
(2,2.4546203384151055)
(3,3.686401346822905)
(4,5.439047433228655)
(5,7.517701394728157)
(6,8.086029783986993)
	};

	\addplot[dashed,color=black,mark=square,mark options={solid}] coordinates {
(1,0.6441104659609268)
(2,1.176365971302463)
(3,1.853784977055982)
(4,2.8923496321266757)
(5,4.215171781470306)
(6,4.6879611526065315)
	};

	\addplot[dashed,color=black,mark=triangle,mark options={solid}] coordinates {
(1,7.3202460777948675)
(2,10.010637446676402)
(3,11.724853745420173)
(4,13.629956412209431)
(5,16.126629704250377)
(6,15.903879321414653)
	};

	\addplot[dashed,color=black,mark=diamond,mark options={solid}] coordinates {
(1,5.998740875529775)
(2,7.716879279523596)
(3,8.672116872638641)
(4,9.80607658144396)
(5,11.46953179754451)
(6,11.202589066612811)
	};

	\addplot[dashed,color=black,mark=o,mark options={solid}] coordinates {
(1,6.28467842579881)
(2,8.515926558630904)
(3,9.919898310830385)
(4,11.496186022108741)
(5,13.582520322963422)
(6,13.38824819231853)
	};

	\addplot[color=black,mark=square] coordinates {
(1,7.050633015944544)
(2,9.346349361414948)
(3,10.624268872601608)
(4,12.27748501010132)
(5,14.378059775374892)
(6,14.042639062854214)
	};

	\addplot[color=black,mark=o] coordinates {
(1,0.521628395328869)
(2,0.9544201362542502)
(3,1.5032418796985718)
(4,2.356930452410192)
(5,3.4492675025130577)
(6,3.8359598501870993)
	};

	\addplot[color=black,mark=diamond] coordinates {
(1,2.93060310434875)
(2,4.042104207849287)
(3,4.731306375756678)
(4,5.645897577065052)
(5,6.8029008194298735)
(6,6.64687475424227)
	};

	\addplot[color=black,mark=triangle] coordinates {
(1,4.309451817270592)
(2,5.214074240537189)
(3,5.548747477526565)
(4,6.252905093692822)
(5,7.292988080975585)
(6,7.000401145787815)
	};

            \end{axis}
            \end{tikzpicture}  
        } &
    
    \resizebox{.45\textwidth}{!}{
    
        \begin{tikzpicture}
            \begin{axis}[
                title={Class Injection},
                xlabel={Injection Chance (\%)},
                ylabel={RCI Score},
                xmin=0.5, xmax=6.5,
                ymin=0.0, ymax=30,
                xtick={0,1,2,3,4,5,6,7},
                xticklabels={
                    ,10,20,40,60,80,100
                }
            ]
            
	\addplot[mark=*,black] coordinates {
(1,12.45477245870854)
(2,12.766860824582562)
(3,16.17088521421848)
(4,12.317637351933111)
(5,13.731046025992342)
(6,15.757075297543965)
	};

	\addplot[dashed,color=black,mark=star,mark options={solid}] coordinates {
(1,18.768585184748556)
(2,19.39285237965365)
(3,18.878411595175834)
(4,13.33298888075917)
(5,13.361863325333173)
(6,13.961996719841949)
	};

	\addplot[dashed,color=black,mark=*,mark options={solid}] coordinates {
(1,6.606191220726565)
(2,7.104456489527522)
(3,8.66684422582945)
(4,8.868119823540344)
(5,10.208476747360889)
(6,11.654992994128007)
	};

	\addplot[dashed,color=black,mark=square,mark options={solid}] coordinates {
(1,3.525374338049943)
(2,3.8334080023378605)
(3,4.848827795052061)
(4,5.993865450511903)
(5,7.6049771655565)
(6,9.413789223915577)
	};

	\addplot[dashed,color=black,mark=triangle,mark options={solid}] coordinates {
(1,16.923067827902788)
(2,17.658147038057898)
(3,19.758620813909772)
(4,13.33877038852843)
(5,13.43998821980785)
(6,14.059582302386957)
	};

	\addplot[dashed,color=black,mark=diamond,mark options={solid}] coordinates {
(1,16.256746910983892)
(2,16.97502143596935)
(3,20.11138061164745)
(4,13.954408251948912)
(5,14.296220100007531)
(6,15.24625406068487)
	};

	\addplot[dashed,color=black,mark=o,mark options={solid}] coordinates {
(1,15.230155544569435)
(2,15.893724332575301)
(3,17.86305723630805)
(4,12.218024892731485)
(5,12.377477414456491)
(6,13.032934559116159)
	};

\addplot[color=black,mark=square] coordinates {
(1,22.60010261906077)
(2,23.64249589922669)
(3,27.441685087570974)
(4,20.080590797305426)
(5,20.73963347136262)
(6,22.373612256011917)
};

	\addplot[color=black,mark=o] coordinates {
(1,2.956672700358297)
(2,3.221678872606904)
(3,4.108907097703395)
(4,5.451685779347146)
(5,7.175414032316558)
(6,9.193791203841078)
	};

	\addplot[color=black,mark=diamond] coordinates {
(1,20.756231316797606)
(2,21.77734022795332)
(3,25.534727094453025)
(4,19.780111782099755)
(5,20.899524216502034)
(6,22.73734048363163)
	};

\addplot[color=black,mark=triangle] coordinates {
(1,20.756231316797606)
(2,21.77734022795332)
(3,25.534727094453025)
(4,19.780111782099755)
(5,20.899524216502034)
(6,22.73734048363163)
};
            
            \end{axis}
            \end{tikzpicture}  
        } \\
        
                \resizebox{.45\textwidth}{!}{
            \begin{tikzpicture}
            \begin{axis}[
                title={Method Injection},
                xlabel={Injection Chance (\%)},
                ylabel={RCI Score},
                xmin=0.5, xmax=6.5,
                ymin=0.0, ymax=20,
                xtick={0,1,2,3,4,5,6,7},
                xticklabels={
                    ,10,20,40,60,80,100
                }
            ]
            
	\addplot[mark=*,black] coordinates {
(1,9.640779008015315)
(2,10.274838429844932)
(3,11.554009196928485)
(4,11.62857269283653)
(5,12.427632885097744)
(6,13.183015489107285)
	};

	\addplot[dashed,color=black,mark=star,mark options={solid}] coordinates {
(1,9.761457303968191)
(2,10.746281228636871)
(3,12.537507228595405)
(4,11.502390461998058)
(5,11.489693889859428)
(6,11.296257654076058)
	};

	\addplot[dashed,color=black,mark=*,mark options={solid}] coordinates {
(1,2.5482511011024895)
(2,3.373693382956884)
(3,4.931965823621549)
(4,6.217415588050683)
(5,7.227796482122055)
(6,8.130105970133691)
	};

	\addplot[dashed,color=black,mark=square,mark options={solid}] coordinates {
(1,1.25758577943446)
(2,1.7400509592588302)
(3,2.709147644371055)
(4,3.86627793296732)
(5,4.943914573938445)
(6,6.063781602489995)
	};

	\addplot[dashed,color=black,mark=triangle,mark options={solid}] coordinates {
(1,9.389383265228837)
(2,9.908744802171675)
(3,11.57913966366909)
(4,10.676843008677574)
(5,10.55125327583141)
(6,10.535170582513436)
	};

	\addplot[dashed,color=black,mark=diamond,mark options={solid}] coordinates {
(1,9.492902912216941)
(2,10.261007965543751)
(3,12.08469512508306)
(4,11.712334898567043)
(5,11.866452211136032)
(6,12.0074505918992)
	};

	\addplot[dashed,color=black,mark=o,mark options={solid}] coordinates {
(1,8.452834728192599)
(2,9.005469226250767)
(3,10.561588407595307)
(4,9.864173297711035)
(5,9.84031004863413)
(6,9.845218990985892)
	};

	\addplot[color=black,mark=square] coordinates {
(1,11.615807759107021)
(2,12.794596293100827)
(3,15.71928250555228)
(4,15.333024270614278)
(5,15.602577942268551)
(6,15.804577801330304)
	};

	\addplot[color=black,mark=o] coordinates {
(1,1.0307016424767004)
(2,1.4461437767947145)
(3,2.2973510817740155)
(4,3.4252767078050628)
(5,4.5340066576507825)
(6,5.756078535842296)
	};

	\addplot[color=black,mark=diamond] coordinates {
(1,5.6805856495817055)
(2,6.910466915021755)
(3,9.05744951373808)
(4,9.996494914575523)
(5,10.870899990320465)
(6,11.458857067174053)
	};

	\addplot[color=black,mark=triangle] coordinates {
(1,10.09434183241509)
(2,11.719770933345881)
(3,14.785088006469495)
(4,15.213327098606992)
(5,15.88789633135552)
(6,16.463453408485716)
	};
            
            \end{axis}
            \end{tikzpicture}  
        } &
    
    \resizebox{.45\textwidth}{!}{
            \begin{tikzpicture}
            \begin{axis}[
                title={Statement Injection},
                xlabel={Injection Chance (\%)},
                ylabel={RCI Score},
                xmin=0.5, xmax=6.5,
                ymin=0.0, ymax=10,
                xtick={0,1,2,3,4,5,6,7},
                xticklabels={
                    ,10,20,40,60,80,100
                }
            ]
            
	\addplot[mark=*,black] coordinates {
(1,7.222168505194542)
(2,7.228775270498439)
(3,7.327262663534416)
(4,7.17556032048329)
(5,7.076653865949122)
(6,6.946975413063985)
	};

	\addplot[dashed,color=black,mark=star,mark options={solid}] coordinates {
(1,3.3386993369136526)
(2,2.9219888771291984)
(3,2.9021556708530807)
(4,2.9094517063782384)
(5,3.0431388767549756)
(6,3.171570021571423)
	};

	\addplot[dashed,color=black,mark=*,mark options={solid}] coordinates {
(1,1.2391653506431768)
(2,1.6878554510824384)
(3,2.382927170341404)
(4,3.008172503391787)
(5,3.5967593714443025)
(6,4.097180637339229)
	};

	\addplot[dashed,color=black,mark=square,mark options={solid}] coordinates {
(1,0.5997464476607266)
(2,0.8649069549662346)
(3,1.3303167510721159)
(4,1.8117365551439526)
(5,2.294968111896425)
(6,2.7475812561218893)
	};

	\addplot[dashed,color=black,mark=triangle,mark options={solid}] coordinates {
(1,4.50364779701918)
(2,4.571198113129686)
(3,4.848971583692091)
(4,5.073921890111887)
(5,5.4427784473412615)
(6,5.725892806143147)
	};

	\addplot[dashed,color=black,mark=diamond,mark options={solid}] coordinates {
(1,4.21022175247358)
(2,4.250375392612006)
(3,4.473715212654152)
(4,4.60965390467565)
(5,4.869441318063519)
(6,5.0643431038883024)
	};

	\addplot[dashed,color=black,mark=o,mark options={solid}] coordinates {
(1,4.102053782546973)
(2,4.225825508899548)
(3,4.5452060016837565)
(4,4.759335539813928)
(5,5.141298621880104)
(6,5.40210269626193)
	};

	\addplot[color=black,mark=square] coordinates {
(1,4.34399436048439)
(2,4.055563813677371)
(3,4.02135492811796)
(4,3.876984631735579)
(5,3.9203892867362633)
(6,3.9910937184942465)
	};

	\addplot[color=black,mark=o] coordinates {
(1,0.49061255170959567)
(2,0.7149212564945236)
(3,1.1169280660386989)
(4,1.545059242848494)
(5,1.9738676381821778)
(6,2.388036028666929)
	};

	\addplot[color=black,mark=diamond] coordinates {
(1,2.315866778340737)
(2,2.627586368786445)
(3,3.0421779115807306)
(4,3.4410756571816203)
(5,3.842025834574987)
(6,4.130471877991679)
	};

	\addplot[color=black,mark=triangle] coordinates {	
(1,4.954595903053945)
(2,5.8448167146388705)
(3,6.843765491910521)
(4,7.0818712167790965)
(5,7.155299672840919)
(6,6.951219450378467)
	};
            
            \end{axis}
            \end{tikzpicture}  
        } \\
        
        \multicolumn{2}{c}{
        \resizebox{.45\linewidth}{!}{
        \begin{tikzpicture}
        \begin{axis}[
            hide axis,
            xmin=0,
            xmax=0,
            ymin=0,
            ymax=0,
            legend style={draw=white!15!black,legend cell align=left},
            legend columns = 4
        ]
        
        \addlegendimage{color=black,mark=*}
        \addlegendentry{JPlag}
        
       \addlegendimage{color=black,mark=square}
        \addlegendentry{Plaggie}
        
        \addlegendimage{color=black,mark=triangle}
        \addlegendentry{Sim}
        
        \addlegendimage{color=black,mark=diamond}
        \addlegendentry{Sherlock-W}
        
        \addlegendimage{color=black,mark=o}
        \addlegendentry{Sherlock-S}
        
        \addlegendimage{dashed,color=black,mark=*,mark options={solid}}
        \addlegendentry{String ED}
        
        \addlegendimage{dashed,color=black,mark=square,mark options={solid}}
        \addlegendentry{String T}
        
        \addlegendimage{dashed,color=black,mark=triangle,mark options={solid}}
        \addlegendentry{Token ED}
        
        \addlegendimage{dashed,color=black,mark=diamond,mark options={solid}}
        \addlegendentry{Token T}
        
        \addlegendimage{dashed,color=black,mark=o,mark options={solid}}
        \addlegendentry{Tree ED}
        
        \addlegendimage{dashed,color=black,mark=star,mark options={solid}}
        \addlegendentry{Graph ED}
    
        \end{axis}
        \end{tikzpicture}
        }
        } \\
        
    \end{tabular}

    \caption{Average RCI for each type of injected source code fragment at each chance of injection. Larger values indicate a tool is more robust to source code injection.}
    \label{fig:ev3aTransformationResilience}
\end{figure}

%% file: Tab10.tex
\begin{table}[htbp]
    \caption{Average logical lines of code injected into each variant generated with all types of injected source code fragments at each injection chance.}
    \label{tab:ev3bavglocinjected}
    \begin{tabular}{p{3cm}|cccccc}
        \hline\noalign{\smallskip}
        {Injection Chance} & {10\%} & {20\%} & {40\%} & {60\%} & {80\%} & {100\%} \\
        \noalign{\smallskip}
        \cline{2-7}
        \noalign{\smallskip}
        {Avg. LLOC Injected} & 189.50 & 361.96 & 643.80 & 969.95 & 1,249.63 & 1,631.64 \\
        \noalign{\smallskip}
        {LLOC Increase (\%)} & 56.39 & 107.72 & 191.59 & 288.65 & 385.27 & 485.56 \\
        \noalign{\smallskip}\hline
    \end{tabular}
\end{table}

%% file: Fig15_ev2b_avgsimilarity.tex
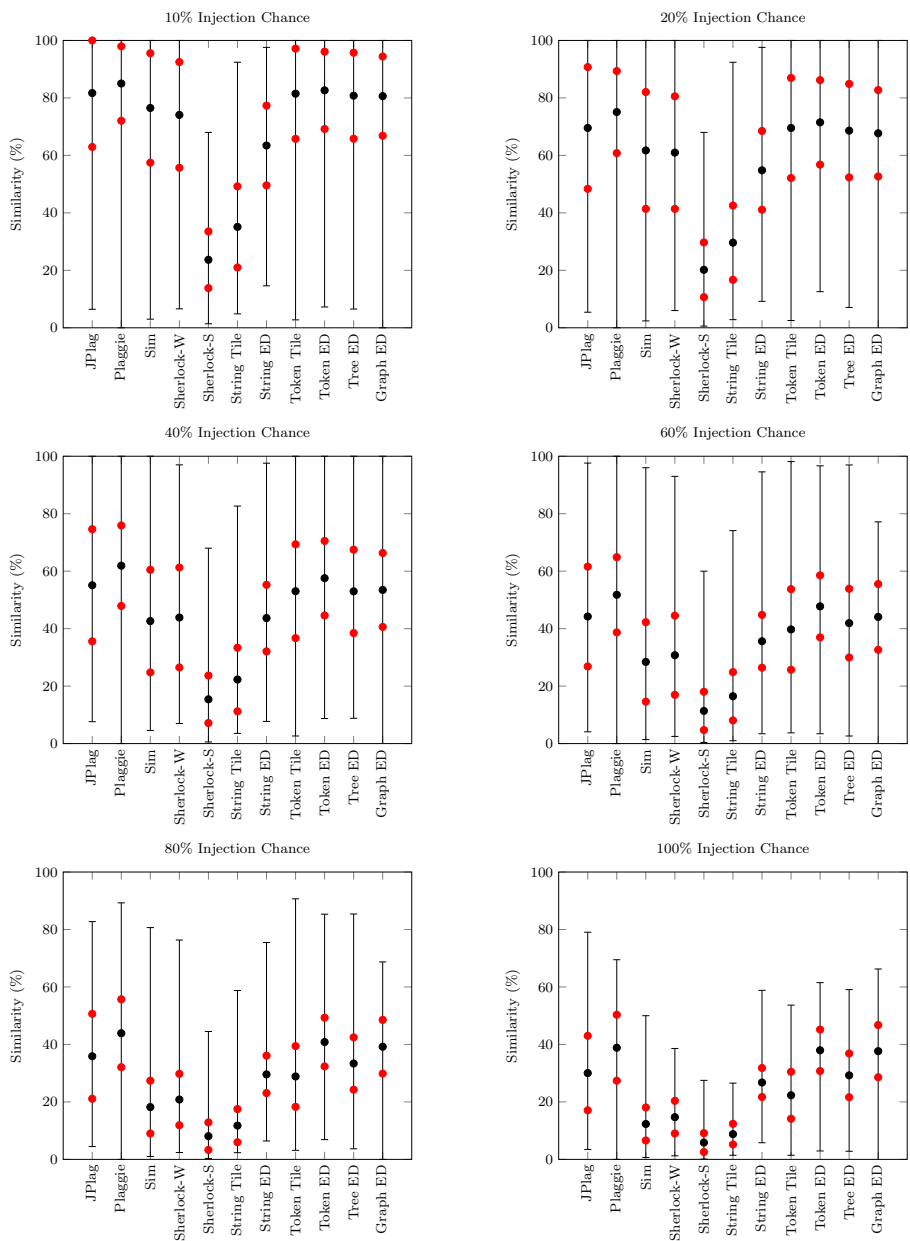
\begin{figure}[htbp]
    \begin{tabular}{p{.5\linewidth}p{.5\linewidth}}
         
        \resizebox{.45\textwidth}{!}{
            \begin{tikzpicture}
            \begin{axis}[
                title={10\% Injection Chance},
                ylabel={Similarity (\%)},
                xmin=-1, xmax=11,
                ymin=0.0, ymax=100,
                xtick={0,1,2,3,4,5,6,7,8,9,10},
                xticklabel style={rotate=90},
                xticklabels={
                    JPlag,Plaggie,Sim,Sherlock-W,Sherlock-S,String Tile,String ED,Token Tile,Token ED,Tree ED,Graph ED
                }
            ]

\addplot[mark=*,black] coordinates {(0, 81.6869355337209) };
\addplot[mark=-,black] coordinates { (0,6.4453125)(0,100.0) };
\addplot[mark=*,red] coordinates { (0,62.93379336294717) };
\addplot[mark=*,red] coordinates { (0,100.0) };

\addplot[mark=*,black] coordinates {(1, 85.00611986619394) };
\addplot[mark=-,black] coordinates { (1,0.0)(1,100.0) };
\addplot[mark=*,red] coordinates { (1,72.05468490544675) };
\addplot[mark=*,red] coordinates { (1,97.95755482694113) };

\addplot[mark=*,black] coordinates {(2, 76.51327064729594) };
\addplot[mark=-,black] coordinates { (2,3.0)(2,100.0) };
\addplot[mark=*,red] coordinates { (2,57.465439230532496) };
\addplot[mark=*,red] coordinates { (2,95.56110206405938) };

\addplot[mark=*,black] coordinates {(8, 82.63573126593309) };
\addplot[mark=-,black] coordinates { (8,7.24681504464432)(8,100.0) };
\addplot[mark=*,red] coordinates { (8,69.1883576754777) };
\addplot[mark=*,red] coordinates { (8,96.08310485638847) };

\addplot[mark=*,black] coordinates {(7, 81.46236880308638) };
\addplot[mark=-,black] coordinates { (7,2.763816206480038)(7,100.0) };
\addplot[mark=*,red] coordinates { (7,65.75876497429154) };
\addplot[mark=*,red] coordinates { (7,97.16597263188122) };

\addplot[mark=*,black] coordinates {(9, 80.7752417932455) };
\addplot[mark=-,black] coordinates { (9,6.528371647565301)(9,100.0) };
\addplot[mark=*,red] coordinates { (9,65.80288982965016) };
\addplot[mark=*,red] coordinates { (9,95.74759375684084) };

\addplot[mark=*,black] coordinates {(10, 80.6267267516796) };
\addplot[mark=-,black] coordinates { (10,0.0)(10,100.0) };
\addplot[mark=*,red] coordinates { (10,66.84267842944959) };
\addplot[mark=*,red] coordinates { (10,94.4107750739096) };

\addplot[mark=*,black] coordinates {(4, 23.67476384071244) };
\addplot[mark=-,black] coordinates { (4,1.4)(4,68.0) };
\addplot[mark=*,red] coordinates { (4,13.81580348001254) };
\addplot[mark=*,red] coordinates { (4,33.53372420141234) };

\addplot[mark=*,black] coordinates {(3, 74.08888461708348) };
\addplot[mark=-,black] coordinates { (3,6.6)(3,100.0) };
\addplot[mark=*,red] coordinates { (3,55.69101164525759) };
\addplot[mark=*,red] coordinates { (3,92.48675758890937) };

\addplot[mark=*,black] coordinates {(6, 63.44511865287968) };
\addplot[mark=-,black] coordinates { (6,14.608098705859174)(6,97.58320126782884) };
\addplot[mark=*,red] coordinates { (6,49.56558394428674) };
\addplot[mark=*,red] coordinates { (6,77.32465336147261) };

\addplot[mark=*,black] coordinates {(5, 35.11667840736081) };
\addplot[mark=-,black] coordinates { (5,4.855213255512501)(5,92.40624341957955) };
\addplot[mark=*,red] coordinates { (5,21.001572817908286) };
\addplot[mark=*,red] coordinates { (5,49.231783996813334) };

            \end{axis}
            \end{tikzpicture}  
        } &
    
    \resizebox{.45\textwidth}{!}{
            \begin{tikzpicture}
            \begin{axis}[
                title={20\% Injection Chance},
                ylabel={Similarity (\%)},
                xmin=-1, xmax=11,
                ymin=0.0, ymax=100,
                xtick={0,1,2,3,4,5,6,7,8,9,10},
                xticklabel style={rotate=90},
                xticklabels={
                    JPlag,Plaggie,Sim,Sherlock-W,Sherlock-S,String Tile,String ED,Token Tile,Token ED,Tree ED,Graph ED
                }
            ]
            
\addplot[mark=*,black] coordinates {(0, 69.55194594895352) };
\addplot[mark=-,black] coordinates { (0,5.4196463)(0,100.0) };
\addplot[mark=*,red] coordinates { (0,48.36695983656726) };
\addplot[mark=*,red] coordinates { (0,90.73693206133977) };

\addplot[mark=*,black] coordinates {(1, 75.07398089042083) };
\addplot[mark=-,black] coordinates { (1,0.0)(1,100.0) };
\addplot[mark=*,red] coordinates { (1,60.79780568933137) };
\addplot[mark=*,red] coordinates { (1,89.35015609151029) };

\addplot[mark=*,black] coordinates {(2, 61.723564788009895) };
\addplot[mark=-,black] coordinates { (2,2.4)(2,100.0) };
\addplot[mark=*,red] coordinates { (2,41.40199934943021) };
\addplot[mark=*,red] coordinates { (2,82.04513022658958) };

\addplot[mark=*,black] coordinates {(8, 71.48891484476478) };
\addplot[mark=-,black] coordinates { (8,12.571197409487118)(8,100.0) };
\addplot[mark=*,red] coordinates { (8,56.78926624494132) };
\addplot[mark=*,red] coordinates { (8,86.18856344458824) };

\addplot[mark=*,black] coordinates {(7, 69.5396468658085) };
\addplot[mark=-,black] coordinates { (7,2.5401540640484566)(7,100.0) };
\addplot[mark=*,red] coordinates { (7,52.125449421651936) };
\addplot[mark=*,red] coordinates { (7,86.95384430996508) };

\addplot[mark=*,black] coordinates {(9, 68.60839269010367) };
\addplot[mark=-,black] coordinates { (9,7.085288167237214)(9,100.0) };
\addplot[mark=*,red] coordinates { (9,52.342525654972235) };
\addplot[mark=*,red] coordinates { (9,84.8742597252351) };

\addplot[mark=*,black] coordinates {(10, 67.69257419981095) };
\addplot[mark=-,black] coordinates { (10,-117.75700934579439)(10,100.0) };
\addplot[mark=*,red] coordinates { (10,52.666068025081515) };
\addplot[mark=*,red] coordinates { (10,82.71908037454038) };

\addplot[mark=*,black] coordinates {(4, 20.18237935985198) };
\addplot[mark=-,black] coordinates { (4,0.6)(4,68.0) };
\addplot[mark=*,red] coordinates { (4,10.659604585052074) };
\addplot[mark=*,red] coordinates { (4,29.705154134651885) };

\addplot[mark=*,black] coordinates {(3, 60.969157137005006) };
\addplot[mark=-,black] coordinates { (3,6.0)(3,100.0) };
\addplot[mark=*,red] coordinates { (3,41.37944428301908) };
\addplot[mark=*,red] coordinates { (3,80.55886999099093) };

\addplot[mark=*,black] coordinates {(6, 54.791104793031835) };
\addplot[mark=-,black] coordinates { (6,9.212973413949651)(6,97.58320126782884) };
\addplot[mark=*,red] coordinates { (6,41.08939499488653) };
\addplot[mark=*,red] coordinates { (6,68.49281459117714) };

\addplot[mark=*,black] coordinates {(5, 29.623403045298993) };
\addplot[mark=-,black] coordinates { (5,2.829983087194027)(5,92.40624341957955) };
\addplot[mark=*,red] coordinates { (5,16.703426856662958) };
\addplot[mark=*,red] coordinates { (5,42.54337923393503) };
            
            \end{axis}
            \end{tikzpicture}  
        } \\
        
                \resizebox{.45\textwidth}{!}{
            \begin{tikzpicture}
            \begin{axis}[
                title={40\% Injection Chance},
                ylabel={Similarity (\%)},
                xmin=-1, xmax=11,
                ymin=0.0, ymax=100,
                xtick={0,1,2,3,4,5,6,7,8,9,10},
                xticklabel style={rotate=90},
                xticklabels={
                    JPlag,Plaggie,Sim,Sherlock-W,Sherlock-S,String Tile,String ED,Token Tile,Token ED,Tree ED,Graph ED
                }
            ]
            
\addplot[mark=*,black] coordinates {(0, 55.10303246926355) };
\addplot[mark=-,black] coordinates { (0,7.6208177)(0,100.0) };
\addplot[mark=*,red] coordinates { (0,35.596309267661596) };
\addplot[mark=*,red] coordinates { (0,74.60975567086551) };

\addplot[mark=*,black] coordinates {(1, 61.91009610139607) };
\addplot[mark=-,black] coordinates { (1,0.0)(1,100.0) };
\addplot[mark=*,red] coordinates { (1,47.9045951086905) };
\addplot[mark=*,red] coordinates { (1,75.91559709410164) };

\addplot[mark=*,black] coordinates {(2, 42.62834817834123) };
\addplot[mark=-,black] coordinates { (2,4.571428571428571)(2,100.0) };
\addplot[mark=*,red] coordinates { (2,24.778002523992054) };
\addplot[mark=*,red] coordinates { (2,60.4786938326904) };

\addplot[mark=*,black] coordinates {(8, 57.5582769830128) };
\addplot[mark=-,black] coordinates { (8,8.733367418117787)(8,100.0) };
\addplot[mark=*,red] coordinates { (8,44.59269918421846) };
\addplot[mark=*,red] coordinates { (8,70.52385478180715) };

\addplot[mark=*,black] coordinates {(7, 53.03211266317291) };
\addplot[mark=-,black] coordinates { (7,2.645809216246933)(7,100.0) };
\addplot[mark=*,red] coordinates { (7,36.71050750619171) };
\addplot[mark=*,red] coordinates { (7,69.35371782015412) };

\addplot[mark=*,black] coordinates {(9, 52.97602838415789) };
\addplot[mark=-,black] coordinates { (9,8.809726741546253)(9,100.0) };
\addplot[mark=*,red] coordinates { (9,38.46351896052306) };
\addplot[mark=*,red] coordinates { (9,67.48853780779272) };

\addplot[mark=*,black] coordinates {(10, 53.46370923089346) };
\addplot[mark=-,black] coordinates { (10,-73.08679196869588)(10,100.0) };
\addplot[mark=*,red] coordinates { (10,40.6070132957321) };
\addplot[mark=*,red] coordinates { (10,66.32040516605481) };

\addplot[mark=*,black] coordinates {(4, 15.412997537899598) };
\addplot[mark=-,black] coordinates { (4,0.5714285714285714)(4,68.0) };
\addplot[mark=*,red] coordinates { (4,7.158735769723336) };
\addplot[mark=*,red] coordinates { (4,23.66725930607586) };

\addplot[mark=*,black] coordinates {(3, 43.87481103016878) };
\addplot[mark=-,black] coordinates { (3,7.0)(3,97.0) };
\addplot[mark=*,red] coordinates { (3,26.48215908932699) };
\addplot[mark=*,red] coordinates { (3,61.267462971010566) };

\addplot[mark=*,black] coordinates {(6, 43.6766326463535) };
\addplot[mark=-,black] coordinates { (6,7.7238620012677615)(6,97.58320126782884) };
\addplot[mark=*,red] coordinates { (6,32.10081178813699) };
\addplot[mark=*,red] coordinates { (6,55.252453504570006) };

\addplot[mark=*,black] coordinates {(5, 22.295508951712907) };
\addplot[mark=-,black] coordinates { (5,3.5579713250892477)(5,82.65240670800303) };
\addplot[mark=*,red] coordinates { (5,11.236530014067133) };
\addplot[mark=*,red] coordinates { (5,33.35448788935868) };
            
            \end{axis}
            \end{tikzpicture}  
        } &
    
    \resizebox{.45\textwidth}{!}{
            \begin{tikzpicture}
            \begin{axis}[
                title={60\% Injection Chance},
                ylabel={Similarity (\%)},
                xmin=-1, xmax=11,
                ymin=0.0, ymax=100,
                xtick={0,1,2,3,4,5,6,7,8,9,10},
                xticklabel style={rotate=90},
                xticklabels={
                    JPlag,Plaggie,Sim,Sherlock-W,Sherlock-S,String Tile,String ED,Token Tile,Token ED,Tree ED,Graph ED
                }
            ]
            
\addplot[mark=*,black] coordinates {(0, 44.21909872131777) };
\addplot[mark=-,black] coordinates { (0,4.098124)(0,97.61905) };
\addplot[mark=*,red] coordinates { (0,26.849439585898033) };
\addplot[mark=*,red] coordinates { (0,61.58875785673751) };

\addplot[mark=*,black] coordinates {(1, 51.78103070106608) };
\addplot[mark=-,black] coordinates { (1,0.0)(1,100.0) };
\addplot[mark=*,red] coordinates { (1,38.69238741928685) };
\addplot[mark=*,red] coordinates { (1,64.86967398284531) };

\addplot[mark=*,black] coordinates {(2, 28.42121411015789) };
\addplot[mark=-,black] coordinates { (2,1.4)(2,96.0) };
\addplot[mark=*,red] coordinates { (2,14.597324601890708) };
\addplot[mark=*,red] coordinates { (2,42.24510361842508) };

\addplot[mark=*,black] coordinates {(8, 47.75689106326133) };
\addplot[mark=-,black] coordinates { (8,3.4267001243470596)(8,96.66153055983564) };
\addplot[mark=*,red] coordinates { (8,36.96475777644432) };
\addplot[mark=*,red] coordinates { (8,58.549024350078334) };

\addplot[mark=*,black] coordinates {(7, 39.73193764225075) };
\addplot[mark=-,black] coordinates { (7,3.7272168215737387)(7,98.13463098134632) };
\addplot[mark=*,red] coordinates { (7,25.720098329376988) };
\addplot[mark=*,red] coordinates { (7,53.74377695512451) };

\addplot[mark=*,black] coordinates {(9, 41.91627857588436) };
\addplot[mark=-,black] coordinates { (9,2.6405376808567302)(9,96.96467991169978) };
\addplot[mark=*,red] coordinates { (9,29.95562290768025) };
\addplot[mark=*,red] coordinates { (9,53.876934244088474) };

\addplot[mark=*,black] coordinates {(10, 44.074994022755604) };
\addplot[mark=-,black] coordinates { (10,-152.05992509363293)(10,77.16999433757617) };
\addplot[mark=*,red] coordinates { (10,32.63024735445563) };
\addplot[mark=*,red] coordinates { (10,55.519740691055574) };

\addplot[mark=*,black] coordinates {(4, 11.374313748414973) };
\addplot[mark=-,black] coordinates { (4,0.375)(4,60.0) };
\addplot[mark=*,red] coordinates { (4,4.7016122216127485) };
\addplot[mark=*,red] coordinates { (4,18.0470152752172) };

\addplot[mark=*,black] coordinates {(3, 30.750819075019887) };
\addplot[mark=-,black] coordinates { (3,2.5)(3,93.0) };
\addplot[mark=*,red] coordinates { (3,16.974882401207125) };
\addplot[mark=*,red] coordinates { (3,44.52675574883265) };

\addplot[mark=*,black] coordinates {(6, 35.61090803154532) };
\addplot[mark=-,black] coordinates { (6,3.417862757642423)(6,94.55968688845401) };
\addplot[mark=*,red] coordinates { (6,26.416106924081404) };
\addplot[mark=*,red] coordinates { (6,44.80570913900924) };

\addplot[mark=*,black] coordinates {(5, 16.47293446888834) };
\addplot[mark=-,black] coordinates { (5,0.997083999623742)(5,74.1068955967876) };
\addplot[mark=*,red] coordinates { (5,8.073460392770013) };
\addplot[mark=*,red] coordinates { (5,24.872408545006667) };
            
            \end{axis}
            \end{tikzpicture}  
        } \\
        
                \resizebox{.45\textwidth}{!}{
            \begin{tikzpicture}
            \begin{axis}[
                title={80\% Injection Chance},
                ylabel={Similarity (\%)},
                xmin=-1, xmax=11,
                ymin=0.0, ymax=100,
                xtick={0,1,2,3,4,5,6,7,8,9,10},
                xticklabel style={rotate=90},
                xticklabels={
                    JPlag,Plaggie,Sim,Sherlock-W,Sherlock-S,String Tile,String ED,Token Tile,Token ED,Tree ED,Graph ED
                }
            ]
            
\addplot[mark=*,black] coordinates {(0, 35.88563105182179) };
\addplot[mark=-,black] coordinates { (0,4.4549575)(0,82.74902) };
\addplot[mark=*,red] coordinates { (0,21.113717853635926) };
\addplot[mark=*,red] coordinates { (0,50.65754425000765) };

\addplot[mark=*,black] coordinates {(1, 43.897586171833446) };
\addplot[mark=-,black] coordinates { (1,0.0)(1,89.2770709077532) };
\addplot[mark=*,red] coordinates { (1,32.07831432646039) };
\addplot[mark=*,red] coordinates { (1,55.716858017206505) };

\addplot[mark=*,black] coordinates {(2, 18.20068625779255) };
\addplot[mark=-,black] coordinates { (2,1.0)(2,80.66666666666667) };
\addplot[mark=*,red] coordinates { (2,9.014266289969035) };
\addplot[mark=*,red] coordinates { (2,27.387106225616066) };

\addplot[mark=*,black] coordinates {(8, 40.83185001476619) };
\addplot[mark=-,black] coordinates { (8,6.87106601559927)(8,85.32836964501338) };
\addplot[mark=*,red] coordinates { (8,32.35625954943852) };
\addplot[mark=*,red] coordinates { (8,49.30744048009386) };

\addplot[mark=*,black] coordinates {(7, 28.857244033069776) };
\addplot[mark=-,black] coordinates { (7,3.1447585908696416)(7,90.6693924057974) };
\addplot[mark=*,red] coordinates { (7,18.285380712577663) };
\addplot[mark=*,red] coordinates { (7,39.42910735356189) };

\addplot[mark=*,black] coordinates {(9, 33.358325912047704) };
\addplot[mark=-,black] coordinates { (9,3.6741281550772777)(9,85.39818888656097) };
\addplot[mark=*,red] coordinates { (9,24.24959290493926) };
\addplot[mark=*,red] coordinates { (9,42.467058919156145) };

\addplot[mark=*,black] coordinates {(10, 39.19929717849854) };
\addplot[mark=-,black] coordinates { (10,0.0)(10,68.71563586433226) };
\addplot[mark=*,red] coordinates { (10,29.85329794146459) };
\addplot[mark=*,red] coordinates { (10,48.54529641553249) };

\addplot[mark=*,black] coordinates {(4, 8.081498202323417) };
\addplot[mark=-,black] coordinates { (4,0.3333333333333333)(4,44.5) };
\addplot[mark=*,red] coordinates { (4,3.285504049879453) };
\addplot[mark=*,red] coordinates { (4,12.87749235476738) };

\addplot[mark=*,black] coordinates {(3, 20.83083044060633) };
\addplot[mark=-,black] coordinates { (3,2.4)(3,76.33333333333333) };
\addplot[mark=*,red] coordinates { (3,11.881580985654734) };
\addplot[mark=*,red] coordinates { (3,29.780079895557925) };

\addplot[mark=*,black] coordinates {(6, 29.59187701888342) };
\addplot[mark=-,black] coordinates { (6,6.4041033635494475)(6,75.4604818880032) };
\addplot[mark=*,red] coordinates { (6,23.087390391079403) };
\addplot[mark=*,red] coordinates { (6,36.096363646687436) };

\addplot[mark=*,black] coordinates {(5, 11.749818848969962) };
\addplot[mark=-,black] coordinates { (5,2.340248388601827)(5,58.757207522648415) };
\addplot[mark=*,red] coordinates { (5,5.980749173207113) };
\addplot[mark=*,red] coordinates { (5,17.518888524732812) };
            
            \end{axis}
            \end{tikzpicture}  
        } &
    
    \resizebox{.45\textwidth}{!}{
            \begin{tikzpicture}
            \begin{axis}[
                title={100\% Injection Chance},
                ylabel={Similarity (\%)},
                xmin=-1, xmax=11,
                ymin=0.0, ymax=100,
                xtick={0,1,2,3,4,5,6,7,8,9,10},
                xticklabel style={rotate=90},
                xticklabels={
                    JPlag,Plaggie,Sim,Sherlock-W,Sherlock-S,String Tile,String ED,Token Tile,Token ED,Tree ED,Graph ED
                }
            ]
            
\addplot[mark=*,black] coordinates {(0, 30.0404590783333) };
\addplot[mark=-,black] coordinates { (0,3.4604905)(0,79.06015) };
\addplot[mark=*,red] coordinates { (0,17.076736530106388) };
\addplot[mark=*,red] coordinates { (0,43.00418162656021) };

\addplot[mark=*,black] coordinates {(1, 38.85870466637503) };
\addplot[mark=-,black] coordinates { (1,0.0)(1,69.48357163115809) };
\addplot[mark=*,red] coordinates { (1,27.38003615425361) };
\addplot[mark=*,red] coordinates { (1,50.337373178496456) };

\addplot[mark=*,black] coordinates {(2, 12.32314754168754) };
\addplot[mark=-,black] coordinates { (2,0.6666666666666666)(2,50.0) };
\addplot[mark=*,red] coordinates { (2,6.5782398256533705) };
\addplot[mark=*,red] coordinates { (2,18.06805525772171) };

\addplot[mark=*,black] coordinates {(8, 37.962226935894314) };
\addplot[mark=-,black] coordinates { (8,2.9220353234699803)(8,61.504661115686844) };
\addplot[mark=*,red] coordinates { (8,30.747684829331682) };
\addplot[mark=*,red] coordinates { (8,45.17676904245695) };

\addplot[mark=*,black] coordinates {(7, 22.308506239265057) };
\addplot[mark=-,black] coordinates { (7,1.4375385084358978)(7,53.68629042487484) };
\addplot[mark=*,red] coordinates { (7,14.145065000309156) };
\addplot[mark=*,red] coordinates { (7,30.471947478220958) };

\addplot[mark=*,black] coordinates {(9, 29.235207052914102) };
\addplot[mark=-,black] coordinates { (9,2.825956146845112)(9,59.08084745381471) };
\addplot[mark=*,red] coordinates { (9,21.632067001355168) };
\addplot[mark=*,red] coordinates { (9,36.83834710447304) };

\addplot[mark=*,black] coordinates {(10, 37.669983405942105) };
\addplot[mark=-,black] coordinates { (10,-5.6547273369966895)(10,66.26252251284838) };
\addplot[mark=*,red] coordinates { (10,28.583930317188013) };
\addplot[mark=*,red] coordinates { (10,46.7560364946962) };

\addplot[mark=*,black] coordinates {(4, 5.8553010124510125) };
\addplot[mark=-,black] coordinates { (4,0.1111111111111111)(4,27.5) };
\addplot[mark=*,red] coordinates { (4,2.567663491592712) };
\addplot[mark=*,red] coordinates { (4,9.142938533309312) };

\addplot[mark=*,black] coordinates {(3, 14.692569002080276) };
\addplot[mark=-,black] coordinates { (3,1.25)(3,38.6) };
\addplot[mark=*,red] coordinates { (3,9.018055221923252) };
\addplot[mark=*,red] coordinates { (3,20.3670827822373) };

\addplot[mark=*,black] coordinates {(6, 26.76151250830819) };
\addplot[mark=-,black] coordinates { (6,5.791522594685292)(6,58.821487204556014) };
\addplot[mark=*,red] coordinates { (6,21.71634067906786) };
\addplot[mark=*,red] coordinates { (6,31.80668433754852) };

\addplot[mark=*,black] coordinates {(5, 8.786981877719585) };
\addplot[mark=-,black] coordinates { (5,1.4350489374393598)(5,26.538858827500654) };
\addplot[mark=*,red] coordinates { (5,5.196422361787078) };
\addplot[mark=*,red] coordinates { (5,12.377541393652091) };
            
            \end{axis}
            \end{tikzpicture}  
        } \\
        
    \end{tabular}

    \caption{Average similarity of variants generated with all 4 source code fragments injected, evaluated using all 11 SCPDTs. Bars indicate range of similarity scores. Red marks indicate range of standard deviation around the average.}
    \label{fig:ev3bavgsimilarity}
\end{figure}

%% file: Fig16_ev2b_resilience.tex
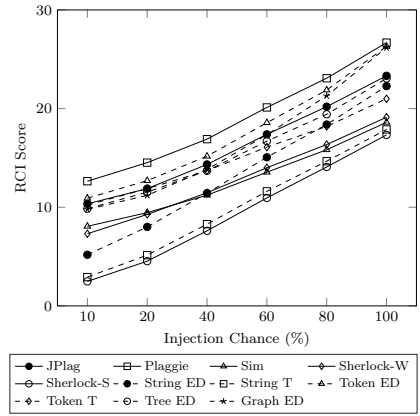
\begin{figure}[htbp]
\centering
\resizebox{.45\linewidth}{!}{
     \begin{tikzpicture}
        \begin{axis}[
            xlabel={Injection Chance (\%)},
            ylabel={RCI Score},
            xmin=0.5, xmax=6.5,
            ymin=0.0, ymax=30,
            xtick={0,1,2,3,4,5,6,7},
            xticklabels={
                ,10,20,40,60,80,100
            },
            legend=none
        ]
            
	\addplot[mark=*,black] coordinates {
(1,10.347803905181312)
(2,11.887787619963175)
(3,14.339498532039045)
(4,17.38856808989362)
(5,20.19250943647924)
(6,23.32262302617077)
	};
    
	\addplot[color=black,mark=square] coordinates {
(1,12.638489724400456)
(2,14.521372161706209)
(3,16.90211667936491)
(4,20.115527438730318)
(5,23.076190695916605)
(6,26.686382601100558)
	};
	
	\addplot[color=black,mark=triangle] coordinates {	
(1,8.068386072587948)
(2,9.456471011349977)
(3,11.221569879167301)
(4,13.550802628766684)
(5,15.8269053953198)
(6,18.609700898829516)
	};
	
	\addplot[color=black,mark=diamond] coordinates {
(1,7.313463631324007)
(2,9.273691610261714)
(3,11.470785431939651)
(4,14.006663862938371)
(5,16.352704053927877)
(6,19.126586991463718)
	};

	\addplot[color=black,mark=o] coordinates {
(1,2.482796117453492)
(2,4.534838261238963)
(3,7.611098410638887)
(4,10.944344027380133)
(5,14.08454200928593)
(6,17.331193551490255)
	};

	\addplot[dashed,color=black,mark=*,mark options={solid}] coordinates {
(1,5.183986187796181)
(2,8.006388971527226)
(3,11.430424533349901)
(4,15.063886915429638)
(5,18.387509071179462)
(6,22.278450250424594)
	};

	\addplot[dashed,color=black,mark=square,mark options={solid}] coordinates {
(1,2.9206272944802225)
(2,5.143187020437791)
(3,8.285235400357104)
(4,11.612403642250774)
(5,14.669998215464167)
(6,17.888236060916427)
	};
	
	\addplot[dashed,color=black,mark=triangle,mark options={solid}] coordinates {
(1,10.913215114450542)
(2,12.695412960580937)
(3,15.169035426349692)
(4,18.566084977341514)
(5,21.88052187406725)
(6,26.30075064612607)
	};

	\addplot[dashed,color=black,mark=diamond,mark options={solid}] coordinates {
(1,10.222449566886967)
(2,11.882987646446649)
(3,13.707237785319807)
(4,16.093930384594223)
(5,18.197636321564374)
(6,21.001526949976423)
	};

	\addplot[dashed,color=black,mark=o,mark options={solid}] coordinates {
(1,9.857081059850227)
(2,11.530470435194653)
(3,13.690889516084757)
(4,16.699171062363938)
(5,19.426732862253346)
(6,23.05722848959726)
	};
	
	\addplot[dashed,color=black,mark=star,mark options={solid}] coordinates {
(1,9.781516916168465)
(2,11.203616228622025)
(3,13.834364307079483)
(4,17.34376211590693)
(5,21.29301044102683)
(6,26.17743567479732)
	};

        \end{axis}
    \end{tikzpicture}  
    }
    \\
    \resizebox{.45\linewidth}{!}{
    \begin{tikzpicture}
        \begin{axis}[
            hide axis,
            xmin=0,
            xmax=0,
            ymin=0,
            ymax=0,
            legend style={draw=white!15!black,legend cell align=left},
            legend columns = 4
        ]
        
        \addlegendimage{color=black,mark=*}
        \addlegendentry{JPlag}
        
       \addlegendimage{color=black,mark=square}
        \addlegendentry{Plaggie}
        
        \addlegendimage{color=black,mark=triangle}
        \addlegendentry{Sim}
        
        \addlegendimage{color=black,mark=diamond}
        \addlegendentry{Sherlock-W}
        
        \addlegendimage{color=black,mark=o}
        \addlegendentry{Sherlock-S}
        
        \addlegendimage{dashed,color=black,mark=*,mark options={solid}}
        \addlegendentry{String ED}
        
        \addlegendimage{dashed,color=black,mark=square,mark options={solid}}
        \addlegendentry{String T}
        
        \addlegendimage{dashed,color=black,mark=triangle,mark options={solid}}
        \addlegendentry{Token ED}
        
        \addlegendimage{dashed,color=black,mark=diamond,mark options={solid}}
        \addlegendentry{Token T}
        
        \addlegendimage{dashed,color=black,mark=o,mark options={solid}}
        \addlegendentry{Tree ED}
        
        \addlegendimage{dashed,color=black,mark=star,mark options={solid}}
        \addlegendentry{Graph ED}
    
        \end{axis}
    \end{tikzpicture}
    }
    \caption{Average RCI Scores for variants generated with all source code fragments injected.}
    \label{fig:ev3brciranking}
\end{figure}

%% file: Tab11.tex
\begin{table}[htbp]
    \caption{Average SCPDT runtimes per program pair in seconds.}
    \label{tab:ev0normliasedruntimes}
    \begin{tabular}{r|cccccc|c}
        \hline\noalign{\smallskip}
        Tool & AS1 & AS2 & AS3 & AS4 & AS5 & AS6 & AVG \\
        \noalign{\smallskip}\hline\noalign{\smallskip}
        JPlag & 0.01 & 0.01 & 0.02 & 0.02 & 0.01 & 0.35 & 0.07 \\
        \noalign{\smallskip}
        Plaggie & 0.79 & 0.80 & 0.64 & 0.58 & 0.35 & 0.71 & 0.65 \\
        \noalign{\smallskip}
        Sim & 0.50 & 0.11 & 0.07 & 0.09 & 0.35 & 0.35 & 0.25 \\
        \noalign{\smallskip}
        Sherlock-W & 3.66 & 3.69 & 2.96 & 3.02 & 1.06 & 3.88 & 3.05 \\
        \noalign{\smallskip}
        Sherlock-S & 0.02 & 0.06 & 0.01 & 0.02 & 0.01 & 0.01 & 0.02 \\
        \noalign{\smallskip}
        String ED & 3.57 & 4.06 & 1.74 & 1.56 & 0.35 & 3.53 & 2.47 \\
        \noalign{\smallskip}
        String Tile & 1.35 & 1.21 & 1.41 & 1.37 & 0.35 & 3.18 & 1.48 \\
        \noalign{\smallskip}
        Token ED & 0.93 & 0.94 & 0.73 & 0.71 & 0.35 & 0.71 & 0.73 \\
        \noalign{\smallskip}
        Token Tile & 0.92 & 0.91 & 0.79 & 0.75 & 0.35 & 0.71 & 0.74 \\
        \noalign{\smallskip}
        Tree ED & 12.34 & 11.64 & 7.32 & 5.95 & 1.76 & 9.88 & 8.15 \\
        \noalign{\smallskip}
        Graph ED & 3.61 & 4.39 & 3.00 & 2.89 & 0.71 & 3.18 & 2.96 \\
        \noalign{\smallskip}\hline
    \end{tabular}
\end{table}

%% file: main.bbl
\begin{thebibliography}{69}
\providecommand{\natexlab}[1]{#1}
\providecommand{\url}[1]{{#1}}
\providecommand{\urlprefix}{URL }
\expandafter\ifx\csname urlstyle\endcsname\relax
  \providecommand{\doi}[1]{DOI~\discretionary{}{}{}#1}\else
  \providecommand{\doi}{DOI~\discretionary{}{}{}\begingroup
  \urlstyle{rm}\Url}\fi
\providecommand{\eprint}[2][]{\url{#2}}

\bibitem[{Ahadi and Mathieson(2019)}]{ahadi2019}
Ahadi A, Mathieson L (2019) A comparison of three popular source code
  similarity tools for detecting student plagiarism. In: Proceedings of the
  Twenty-First Australasian Computing Education Conference, Association for
  Computing Machinery, New York, NY, USA, ACE ’19, p 112–117,
  \doi{10.1145/3286960.3286974}

\bibitem[{Ahtiainen et~al.(2006)Ahtiainen, Surakka, and
  Rahikainen}]{ahtiainen2006}
Ahtiainen A, Surakka S, Rahikainen M (2006) Plaggie: Gnu-licensed source code
  plagiarism detection engine for java exercises. In: Proceedings of the 6th
  Baltic Sea Conference on Computing Education Research: Koli Calling 2006,
  Association for Computing Machinery, New York, NY, USA, Baltic Sea ’06, p
  141–142, \doi{10.1145/1315803.1315831}

\bibitem[{{Allyson} et~al.(2019){Allyson}, {Danilo}, {José}, and
  {Giovanni}}]{allyson2019}
{Allyson} FB, {Danilo} ML, {José} SM, {Giovanni} BC (2019) Sherlock n-overlap:
  Invasive normalization and overlap coefficient for the similarity analysis
  between source code. IEEE Transactions on Computers 68(5):740--751

\bibitem[{Anjali et~al.(2015)Anjali, Swapna, and Jayaraman}]{anjali2015}
Anjali V, Swapna T, Jayaraman B (2015) Plagiarism detection for java programs
  without source codes. Procedia Computer Science 46:749 -- 758,
  \doi{https://doi.org/10.1016/j.procs.2015.02.143}, proceedings of the
  International Conference on Information and Communication Technologies, ICICT
  2014, 3-5 December 2014 at Bolgatty Palace \& Island Resort, Kochi, India

\bibitem[{{Anzai} and {Watanobe}(2019)}]{anzai2019}
{Anzai} K, {Watanobe} Y (2019) Algorithm to determine extended edit distance
  between program codes. In: 2019 IEEE 13th International Symposium on Embedded
  Multicore/Many-core Systems-on-Chip (MCSoC), pp 180--186

\bibitem[{Baxter et~al.(1998)Baxter, Yahin, Moura, Sant'Anna, and
  Bier}]{baxter1998}
Baxter ID, Yahin A, Moura L, Sant'Anna M, Bier L (1998) Clone detection using
  abstract syntax trees. In: Proceedings of the International Conference on
  Software Maintenance, IEEE Computer Society, Washington, DC, USA, ICSM '98,
  pp 368--377

\bibitem[{{Bellon} et~al.(2007){Bellon}, {Koschke}, {Antoniol}, {Krinke}, and
  {Merlo}}]{bellon2007}
{Bellon} S, {Koschke} R, {Antoniol} G, {Krinke} J, {Merlo} E (2007) Comparison
  and evaluation of clone detection tools. IEEE Transactions on Software
  Engineering 33(9):577--591

\bibitem[{{Burd} and {Bailey}(2002)}]{burd2002}
{Burd} E, {Bailey} J (2002) Evaluating clone detection tools for use during
  preventative maintenance. In: Proceedings. Second IEEE International Workshop
  on Source Code Analysis and Manipulation, pp 36--43

\bibitem[{{Cebrian} et~al.(2009){Cebrian}, {Alfonseca}, and
  {Ortega}}]{cebrian2009}
{Cebrian} M, {Alfonseca} M, {Ortega} A (2009) Towards the validation of
  plagiarism detection tools by means of grammar evolution. IEEE Transactions
  on Evolutionary Computation 13(3):477--485

\bibitem[{Chae et~al.(2013)Chae, Ha, Kim, Kang, and Im}]{chae2013}
Chae DK, Ha J, Kim SW, Kang B, Im EG (2013) Software plagiarism detection: A
  graph-based approach. In: Proceedings of the 22nd ACM International
  Conference on Information \& Knowledge Management, Association for Computing
  Machinery, New York, NY, USA, CIKM ’13, p 1577–1580,
  \doi{10.1145/2505515.2507848}

\bibitem[{{Cheers} et~al.(2019){Cheers}, {Lin}, and {Smith}}]{cheers2019}
{Cheers} H, {Lin} Y, {Smith} SP (2019) Spplagiarise: A tool for generating
  simulated semantics-preserving plagiarism of java source code. In: 2019 IEEE
  10th International Conference on Software Engineering and Service Science
  (ICSESS), pp 617--622

\bibitem[{Cheers et~al.(2020)Cheers, Lin, and Smith}]{cheers2020}
Cheers H, Lin Y, Smith SP (2020) Detecting pervasive source code plagiarism
  through dynamic program behaviours. In: Proceedings of the Twenty-Second
  Australasian Computing Education Conference, Association for Computing
  Machinery, New York, NY, USA, ACE’20, p 21–30,
  \doi{10.1145/3373165.3373168}

\bibitem[{{Chen} et~al.(2010){Chen}, {Hong}, {Chunyan Lü}, and
  {Deng}}]{chen2010}
{Chen} R, {Hong} L, {Chunyan Lü} C, {Deng} W (2010) Author identification of
  software source code with program dependence graphs. In: 2010 IEEE 34th
  Annual Computer Software and Applications Conference Workshops, pp 281--286

\bibitem[{{Chen} et~al.(2004){Chen}, {Francia}, {Ming Li}, {McKinnon}, and
  {Seker}}]{chen2004}
{Chen} X, {Francia} B, {Ming Li}, {McKinnon} B, {Seker} A (2004) Shared
  information and program plagiarism detection. IEEE Transactions on
  Information Theory 50(7):1545--1551

\bibitem[{{Cosma} and {Joy}(2008)}]{cosma2008}
{Cosma} G, {Joy} M (2008) Towards a definition of source-code plagiarism. IEEE
  Transactions on Education 51(2):195--200

\bibitem[{{Cosma} and {Joy}(2012)}]{cosma2012}
{Cosma} G, {Joy} M (2012) An approach to source-code plagiarism detection and
  investigation using latent semantic analysis. IEEE Transactions on Computers
  61(3):379--394

\bibitem[{Curtis and Popal(2011)}]{curtis2011}
Curtis G, Popal R (2011) An examination of factors related to plagiarism and a
  five-year follow-up of plagiarism at an australian university. International
  Journal for Educational Integrity 7(1):30--42, \doi{10.21913/IJEI.v7i1.742}

\bibitem[{Faidhi and Robinson(1987)}]{faidhi1987}
Faidhi J, Robinson S (1987) An empirical approach for detecting program
  similarity and plagiarism within a university programming environment.
  Computers \& Education 11(1):11 -- 19,
  \doi{https://doi.org/10.1016/0360-1315(87)90042-X}

\bibitem[{Ferrante et~al.(1987)Ferrante, Ottenstein, and Warren}]{ferrante1987}
Ferrante J, Ottenstein KJ, Warren JD (1987) The program dependence graph and
  its use in optimization. ACM Trans Program Lang Syst 9(3):319–349,
  \doi{10.1145/24039.24041}

\bibitem[{Flores et~al.(2014)Flores, Rosso, Moreno, and
  Villatoro-Tello}]{flores2014}
Flores E, Rosso P, Moreno L, Villatoro-Tello E (2014) On the detection of
  source code re-use. In: Proceedings of the Forum for Information Retrieval
  Evaluation, Association for Computing Machinery, New York, NY, USA, FIRE
  ’14, p 21–30, \doi{10.1145/2824864.2824878}

\bibitem[{Freire et~al.(2007)Freire, Cebri{\'{a}}n, and del Rosal}]{freire2007}
Freire M, Cebri{\'{a}}n M, del Rosal E (2007) {AC:} an integrated source code
  plagiarism detection environment. CoRR abs/cs/0703136, \eprint{cs/0703136}

\bibitem[{Gitchell and Tran(1999{\natexlab{a}})}]{gitchell1999}
Gitchell D, Tran N (1999{\natexlab{a}}) Sim: A utility for detecting similarity
  in computer programs. In: The Proceedings of the Thirtieth SIGCSE Technical
  Symposium on Computer Science Education, Association for Computing Machinery,
  New York, NY, USA, SIGCSE ’99, p 266–270, \doi{10.1145/299649.299783}

\bibitem[{Gitchell and Tran(1999{\natexlab{b}})}]{gitchell1999a}
Gitchell D, Tran N (1999{\natexlab{b}}) Sim: A utility for detecting similarity
  in computer programs. SIGCSE Bull 31(1):266–270,
  \doi{10.1145/384266.299783}

\bibitem[{Granzer et~al.(2013)Granzer, Praus, and Balog}]{granzer2013}
Granzer W, Praus F, Balog P (2013) Source code plagiarism in computer
  engineering courses. Journal on Systemics, Cybernetics and Informatics
  11(6):22--26

\bibitem[{Grune and Huntjens(1989)}]{grune1989}
Grune D, Huntjens M (1989) Het detecteren van kopie\"{e}n bij
  informatica-practica. Informatie (in Dutch) 31(11):864--867

\bibitem[{Halstead(1977)}]{halstead1977}
Halstead MH (1977) Elements of Software Science (Operating and Programming
  Systems Series). Elsevier Science Inc., New York, NY, USA

\bibitem[{Jadalla and Elnagar(2008)}]{jadalla2008}
Jadalla A, Elnagar A (2008) Pde4java: Plagiarism detection engine for java
  source code: A clustering approach. Int J Bus Intell Data Min 3(2):121–135,
  \doi{10.1504/IJBIDM.2008.020514}

\bibitem[{{Jhi} et~al.(2011){Jhi}, {Wang}, {Jia}, {Zhu}, {Liu}, and
  {Wu}}]{jhi2011}
{Jhi} Y, {Wang} X, {Jia} X, {Zhu} S, {Liu} P, {Wu} D (2011) Value-based program
  characterization and its application to software plagiarism detection. In:
  2011 33rd International Conference on Software Engineering (ICSE), pp
  756--765

\bibitem[{Jones(2001)}]{jones2001}
Jones E (2001) Metrics based plagarism monitoring. Journal of Computing
  Sciences in Colleges 16:253--261

\bibitem[{{Joy} and {Luck}(1999)}]{joy1999}
{Joy} M, {Luck} M (1999) Plagiarism in programming assignments. IEEE
  Transactions on Education 42(2):129--133

\bibitem[{Kapser and Godfrey(2003)}]{kapser2003}
Kapser C, Godfrey MW (2003) Toward a taxonomy of clones in source code: A case
  study. In: ELISA ’03, pp 67--78

\bibitem[{{Karnalim}(2016)}]{karnalim2016}
{Karnalim} O (2016) Detecting source code plagiarism on introductory
  programming course assignments using a bytecode approach. In: 2016
  International Conference on Information Communication Technology and Systems
  (ICTS), pp 63--68

\bibitem[{{Ko} et~al.(2017){Ko}, {Choi}, and {Kim}}]{ko2017}
{Ko} S, {Choi} J, {Kim} H (2017) Coat: Code obfuscation tool to evaluate the
  performance of code plagiarism detection tools. In: 2017 International
  Conference on Software Security and Assurance (ICSSA), pp 32--37

\bibitem[{Kolmogorov(1998)}]{kolmogorov1998}
Kolmogorov A (1998) On tables of random numbers. Theoretical Computer Science
  207(2):387 -- 395, \doi{https://doi.org/10.1016/S0304-3975(98)00075-9}

\bibitem[{{Kustanto} and {Liem}(2009)}]{kustanto2009}
{Kustanto} C, {Liem} I (2009) Automatic source code plagiarism detection. In:
  2009 10th ACIS International Conference on Software Engineering, Artificial
  Intelligences, Networking and Parallel/Distributed Computing, pp 481--486

\bibitem[{Lancaster and Tetlow(2005)}]{lancaster2005}
Lancaster T, Tetlow M (2005) Does automated anti-plagiarism have to be complex?
  evaluating more appropriate software metrics for finding collusion

\bibitem[{{Li} and {Zhong}(2010)}]{li2010}
{Li} X, {Zhong} XJ (2010) The source code plagiarism detection using ast. In:
  2010 International Symposium on Intelligence Information Processing and
  Trusted Computing, pp 406--408

\bibitem[{Liu et~al.(2006)Liu, Chen, Han, and Yu}]{liu2006}
Liu C, Chen C, Han J, Yu PS (2006) Gplag: Detection of software plagiarism by
  program dependence graph analysis. In: Proceedings of the 12th ACM SIGKDD
  International Conference on Knowledge Discovery and Data Mining, Association
  for Computing Machinery, New York, NY, USA, KDD ’06, p 872–881,
  \doi{10.1145/1150402.1150522}

\bibitem[{{Luo} et~al.(2017){Luo}, {Ming}, {Wu}, {Liu}, and {Zhu}}]{luo2017}
{Luo} L, {Ming} J, {Wu} D, {Liu} P, {Zhu} S (2017) Semantics-based
  obfuscation-resilient binary code similarity comparison with applications to
  software and algorithm plagiarism detection. IEEE Transactions on Software
  Engineering 43(12):1157--1177

\bibitem[{Martins et~al.(2014)Martins, Fonte, Henriques, and
  da~Cruz}]{martins2014}
Martins VT, Fonte D, Henriques PR, da~Cruz D (2014) {Plagiarism Detection: A
  Tool Survey and Comparison}. In: Pereira MJV, Leal JP, Sim{\~o}es A (eds) 3rd
  Symposium on Languages, Applications and Technologies, Schloss
  Dagstuhl--Leibniz-Zentrum fuer Informatik, Dagstuhl, Germany, OpenAccess
  Series in Informatics (OASIcs), vol~38, pp 143--158,
  \doi{10.4230/OASIcs.SLATE.2014.143}

\bibitem[{Mozgovoy(2006)}]{mozgovoy2006}
Mozgovoy M (2006) Desktop tools for offline plagiarism detection in computer
  programs. Informatics in Education 5(1):97–112

\bibitem[{{Novak}(2016)}]{novak2016}
{Novak} M (2016) Review of source-code plagiarism detection in academia. In:
  2016 39th International Convention on Information and Communication
  Technology, Electronics and Microelectronics (MIPRO), pp 796--801

\bibitem[{Novak et~al.(2019)Novak, Joy, and Kermek}]{novak2019}
Novak M, Joy M, Kermek D (2019) Source-code similarity detection and detection
  tools used in academia: A systematic review. ACM Trans Comput Educ 19(3),
  \doi{10.1145/3313290}

\bibitem[{Ottenstein(1976)}]{ottenstein1976}
Ottenstein KJ (1976) An algorithmic approach to the detection and prevention of
  plagiarism. SIGCSE Bull 8(4):30--41, \doi{10.1145/382222.382462}

\bibitem[{{Parker} and {Hamblen}(1989)}]{parker1989}
{Parker} A, {Hamblen} JO (1989) Computer algorithms for plagiarism detection.
  IEEE Transactions on Education 32(2):94--99

\bibitem[{Pawlik and Augsten(2015)}]{pawlik2015}
Pawlik M, Augsten N (2015) Efficient computation of the tree edit distance. ACM
  Trans Database Syst 40(1), \doi{10.1145/2699485}

\bibitem[{Pawlik and Augsten(2016)}]{pawlik2016}
Pawlik M, Augsten N (2016) Tree edit distance: Robust and memory-efficient.
  Information Systems 56:157 -- 173,
  \doi{https://doi.org/10.1016/j.is.2015.08.004}

\bibitem[{Pierce and Zilles(2017)}]{pierce2017}
Pierce J, Zilles C (2017) Investigating student plagiarism patterns and
  correlations to grades. In: Proceedings of the 2017 ACM SIGCSE Technical
  Symposium on Computer Science Education, Association for Computing Machinery,
  New York, NY, USA, SIGCSE ’17, p 471–476, \doi{10.1145/3017680.3017797}

\bibitem[{Pike(n.d.)}]{pikeunknown}
Pike R (n.d.) Sherlock Plagiarism Detector.
  \urlprefix\url{\url{https://web.archive.org/web/20150323030146/http://rp-www.cs.usyd.edu.au/~scilect/sherlock/}}

\bibitem[{Prechelt et~al.(2002)Prechelt, Malpohl, and
  Philippsen}]{prechelt2002}
Prechelt L, Malpohl G, Philippsen M (2002) Finding plagiarisms among a set of
  programs with jplag 8(11):1016--1038

\bibitem[{Ragkhitwetsagul et~al.(2018)Ragkhitwetsagul, Krinke, and
  Clark}]{ragkhitwetsagul2018}
Ragkhitwetsagul C, Krinke J, Clark D (2018) A comparison of code similarity
  analysers. Empirical Software Engineering 23(4):2464--2519

\bibitem[{Rani and Singh(2018)}]{rani2018}
Rani S, Singh J (2018) Enhancing levenshtein's edit distance algorithm for
  evaluating document similarity. In: Sharma R, Mantri A, Dua S (eds)
  Computing, Analytics and Networks, Springer Singapore, Singapore, pp 72--80

\bibitem[{Roy and Cordy(2007)}]{roy2007}
Roy C, Cordy J (2007) A survey on software clone detection research. School of
  Computing TR 2007-541

\bibitem[{Roy et~al.(2009)Roy, Cordy, and Koschke}]{roy2009}
Roy CK, Cordy JR, Koschke R (2009) Comparison and evaluation of code clone
  detection techniques and tools: A qualitative approach. Science of Computer
  Programming 74(7):470 -- 495,
  \doi{https://doi.org/10.1016/j.scico.2009.02.007}

\bibitem[{Schleimer et~al.(2003)Schleimer, Wilkerson, and
  Aiken}]{schleimer2003}
Schleimer S, Wilkerson DS, Aiken A (2003) Winnowing: Local algorithms for
  document fingerprinting. In: Proceedings of the 2003 ACM SIGMOD International
  Conference on Management of Data, ACM, New York, NY, USA, SIGMOD '03, pp
  76--85, \doi{10.1145/872757.872770}

\bibitem[{{Schulze} and {Meyer}(2013)}]{schulze2013}
{Schulze} S, {Meyer} D (2013) On the robustness of clone detection to code
  obfuscation. In: 2013 7th International Workshop on Software Clones (IWSC),
  pp 62--68

\bibitem[{Shan et~al.(2014)Shan, Tian, Guo, and Ren}]{shan2014}
Shan SQ, Tian ZG, Guo FJ, Ren JX (2014) Similarity detection’s application
  using chi-square test in the property of counting method. In: Advances in
  Computers, Electronics and Mechatronics, Trans Tech Publications Ltd, Applied
  Mechanics and Materials, vol 667, pp 32--35,
  \doi{10.4028/www.scientific.net/AMM.667.32}

\bibitem[{Sheard et~al.(2003)Sheard, Markham, and Dick}]{sheard2003}
Sheard J, Markham S, Dick M (2003) Investigating differences in cheating
  behaviours of it undergraduate and graduate students: The maturity and
  motivation factors. Higher Education Research \& Development 22(1):91--108,
  \doi{10.1080/0729436032000056526}

\bibitem[{{Sraka} and {Kaucic}(2009)}]{sraka2009}
{Sraka} D, {Kaucic} B (2009) Source code plagiarism. In: Proceedings of the ITI
  2009 31st International Conference on Information Technology Interfaces, pp
  461--466

\bibitem[{{Svajlenko} and {Roy}(2015)}]{svajlenko2015}
{Svajlenko} J, {Roy} CK (2015) Evaluating clone detection tools with
  bigclonebench. In: 2015 IEEE International Conference on Software Maintenance
  and Evolution (ICSME), pp 131--140

\bibitem[{{Svajlenko} et~al.(2013){Svajlenko}, {Roy}, and
  {Duszynski}}]{svajlenko2013}
{Svajlenko} J, {Roy} CK, {Duszynski} S (2013) Forksim: Generating software
  forks for evaluating cross-project similarity analysis tools. In: 2013 IEEE
  13th International Working Conference on Source Code Analysis and
  Manipulation (SCAM), pp 37--42

\bibitem[{Verco and Wise(1996)}]{verco1996}
Verco KL, Wise MJ (1996) {Plagiarism à la Mode: A Comparison of Automated
  Systems for Detecting Suspected Plagiarism}. The Computer Journal
  39(9):741--750, \doi{10.1093/comjnl/39.9.741},
  \eprint{https://academic.oup.com/comjnl/article-pdf/39/9/741/993714/390741.pdf}

\bibitem[{Walker et~al.(2020)Walker, Cerny, and Song}]{walker2020}
Walker A, Cerny T, Song E (2020) Open-source tools and benchmarks for
  code-clone detection: Past, present, and future trends. SIGAPP Appl Comput
  Rev 19(4):28–39, \doi{10.1145/3381307.3381310}

\bibitem[{Whale(1990{\natexlab{a}})}]{whale1990}
Whale G (1990{\natexlab{a}}) Identification of program similarity in large
  populations. Comput J 33(2):140–146, \doi{10.1093/comjnl/33.2.140}

\bibitem[{Whale(1990{\natexlab{b}})}]{whale1990a}
Whale G (1990{\natexlab{b}}) Software metrics and plagiarism detection. Journal
  of Systems and Software 13(2):131 -- 138,
  \doi{https://doi.org/10.1016/0164-1212(90)90118-6}, special Issue on Using
  Software Metrics

\bibitem[{Wise(1996)}]{wise1996}
Wise MJ (1996) Yap3: Improved detection of similarities in computer program and
  other texts. SIGCSE Bull 28(1):130–134, \doi{10.1145/236462.236525}

\bibitem[{Yeo(2007)}]{yeo2007}
Yeo S (2007) First‐year university science and engineering students’
  understanding of plagiarism. Higher Education Research \& Development
  26(2):199--216, \doi{10.1080/07294360701310813}

\bibitem[{{Zhang} et~al.(2014){Zhang}, {Wu}, {Liu}, and {Zhu}}]{zhang2014}
{Zhang} F, {Wu} D, {Liu} P, {Zhu} S (2014) Program logic based software
  plagiarism detection. In: 2014 IEEE 25th International Symposium on Software
  Reliability Engineering, pp 66--77

\bibitem[{{Zhao} et~al.(2015){Zhao}, {Xia}, {Fu}, and {Cui}}]{zhao2015}
{Zhao} J, {Xia} K, {Fu} Y, {Cui} B (2015) An ast-based code plagiarism
  detection algorithm. In: 2015 10th International Conference on Broadband and
  Wireless Computing, Communication and Applications (BWCCA), pp 178--182

\end{thebibliography}
